\documentclass[a4paper,11pt]{article}
\pdfoutput=1 
\usepackage{jheppub}
\usepackage[T1]{fontenc}
\usepackage{amssymb,amsmath,amsfonts,amsthm}
\usepackage{mathtools}
\usepackage{amsmath}
\usepackage{hyperref}
\usepackage{IEEEtrantools}
\usepackage{graphicx}
\usepackage[utf8]{inputenc}
\usepackage{bm,bbm,bbold}	
\usepackage{centernot}
\usepackage{csquotes}
\usepackage{scalerel}[2016/12/29]

\def\dd{\mathrm{d}}
\newcommand{\pd}{\partial}

\newcommand{\M}{\mathcal{M}}

\newcommand{\R}{\mathcal{R}}
\newcommand{\T}{\mathcal{T}}

\newcommand{\bbQ}{\mathbb{Q}}
\newcommand{\bbR}{\mathbb{R}}
\newcommand{\bbZ}{\mathbb{Z}}
\newcommand{\x}{\boldsymbol{x}}
\renewcommand{\k}{\boldsymbol{k}}
\newcommand{\khat}{\hat{\boldsymbol{k}}}
\newcommand{\n}{\boldsymbol{n}}
\renewcommand{\v}{\boldsymbol{v}}
\newcommand{\ee}{\boldsymbol{e}}

\newcommand{\ud}[2]{^{#1}{}_{#2}}
\newcommand{\du}[2]{_{#1}{}^{#2}}

\newcommand{\ce}{\coloneqq}
\newcommand{\ec}{\eqqcolon}

\DeclareMathOperator{\e}{e}
\DeclareMathOperator{\diag}{diag}
\newcommand{\lie}{\mathcal{L}}

\newcommand{\PD}[2]{\ensuremath{\frac{\partial #1}{\partial #2}}}

\title{A Re-Examination Of Foundational Elements Of Cosmology}
\author{Lavinia Heisenberg}
\affiliation{Institut f\"{u}r Theoretische Physik, Philosophenweg 16, 69120 Heidelberg, Germany}

\emailAdd{heisenberg@thphys.uni-heidelberg.de}

\abstract{This paper undertakes a conceptual re-examination of several foundational elements of cosmology through the lens of spacetime symmetries. A new derivation of the Friedmann–Lema\^itre–Robertson–Walker metric is obtained by a careful conceptual examination of rotations and translations on generic manifolds, followed by solving the rotational and translational Killing equations, yielding both the metric \emph{and} its translational generators for $k\in\{-1,0,1\}$ without any further assumptions. We then analyze how continuous symmetries are inherited by the Einstein tensor and the Hilbert energy-momentum tensor, proving two general propositions. Furthermore, we use the Maxwell and Kalb-Ramond fields to show that a homogeneous and isotropic energy-momentum tensor, in general, does \emph{not} give rise to field configurations which share these symmetries. In particular, the Kalb-Ramond field we derive is significantly more general than what is usually encountered in the cosmological context.
Finally, we provide a rigorous but accessible, elementary, and transparent derivation of the scalar–vector–tensor decomposition from the linearized Einstein equations. Together, these results highlight the value of multiple complementary formulations of the same cosmological physics.}


\begin{document}
	\allowdisplaybreaks[1]
	\maketitle
	\flushbottom

\newtheorem{proposition}{Proposition}[section]


\section{Introduction}\label{sec:Introduction}
Modern cosmology is based on two pillars: The Cosmological Principle and General Relativity (GR). The former asserts that on large enough scales the Universe is \emph{spatially homogeneous and isotropic}. Informally, the requirement of spatial homogeneity means that at any instant of time the Universe looks the same, no matter where an observer is located, while spatial isotropy means that the Universe looks the same in all directions. This principle has far reaching consequences for the geometry of spacetime. Indeed, without ever invoking the Einstein field equations, one can deduce that the metric has to be of the Friedmann–Lema\^itre–Robertson–Walker (FLRW) type. Several different derivations of this fact can be found in the literature (see for instance~\cite{EinsteinBook, WeinbergBook, HawkingEllisBook, MisnerThorneWheelerBook, WaldBook, WeinbergCosmologyBook, SchutzBook, MalamentBook, CarrollBook, RovelliBook, BaumannBook, Blau}).

These references all have in common that the main focus of the argument is the homogeneity and isotropy of \emph{space}. In this paper we present a new derivation of the FLRW metric which, to the best of our knowledge, has not been discussed in the literature. What sets this derivation apart from the usual methods employed in~\cite{EinsteinBook, WeinbergBook, HawkingEllisBook, MisnerThorneWheelerBook, WaldBook, WeinbergCosmologyBook, SchutzBook, MalamentBook, CarrollBook, RovelliBook, BaumannBook, Blau} is the focus on symmetries of the full  \emph{spacetime} metric, rather than just its spatial part, and their description in terms of diffeomorphisms and generating vector fields. The strategy is to define homogeneity and isotropy in terms of diffeomorphisms, to derive generating vector fields from that description, and to then impose the Killing equation on the metric. By solving the Killing equations for the metric components, one finds the FLRW metric.

However, as it stands, this strategy faces an immediate obstacle: In order to know which generating vector fields constitute translations on a generic manifold requires us to know the metric. Thus, the above strategy seems to be circular and fall apart upon closer scrutiny.

We overcome this obstacle by a careful conceptual discussion of the difference between rotations and translations on generic manifolds. This allows us then to adapt the strategy such that we do not need to make any assumptions on the metric or the generating vector fields. Rather, we find that the Killing equations can be simultaneously solved for (a) some of the components of a generic spherically symmetric metric and (b) a function which parametrizes all possible generating vector fields of translations on a spherically symmetric manifold. The solutions we obtain uniquely recover the FLRW metric. Our method is interesting for didactical and pedagogical purposes. It is distinct from other known derivations of the FLRW metric and it requires the careful examination of concepts and assumptions which are either taken for granted or are never thought about. It thus provides a deeper understanding of symmetries and their implementation than other approaches. 

The conceptual discussions and the derivation of the FLRW metric can be found in sections~\ref{sec:Symmetries} and~\ref{sec:Metric}. Once the Killing vectors of the metric have been derived, they can be used to impose homogeneity and isotropy also on other fields. In section~\ref{sec:MatterFields} we apply this technique to perform a symmetry-reduction of scalar fields, vector fields, $2$-forms, $3$-forms, and a generic symmetric energy-momentum tensor $T_{\mu\nu}$. In doing so, we recover well-known results for the form of homogeneous and isotropic matter fields, as well as the perfect-fluid-form of the energy-momentum tensor. However, all these results are purely kinematical in nature and they serve as preparation for the explorations of section~\ref{sec:SymmetryReducedT}. Since matter fields in GR influence the metric through the Einstein field equations, and these fields only enter the equation through the energy-momentum tensor, it is reasonable to impose homogeneity and isotropy on that tensor, rather than on the matter fields themselves. We explore this possibility in section~\ref{sec:SymmetryReducedT}.

We open section~\ref{sec:SymmetryReducedT} with the proof of two propositions on symmetry inheritance. First, we show that if the metric has certain symmetries, the Einstein tensor must have the same symmetries. Then we prove an analogous statement for the energy-momentum tensor: If the metric and the matter fields share a common set of symmetries, then the Hilbert energy-momentum tensor must possess the same symmetries. 
It is natural to wonder whether the converse is true: If the Einstein tensor or the energy-momentum tensor possess certain symmetries, does this imply that the metric and the matter fields which make up these tensors also share these symmetries? 

In the case of the energy-momentum tensor we provide a negative answer by constructing two counter-examples. Specifically, in subsections~\ref{ssec:KleinGordon},~\ref{ssec:MaxwellEMT}, and~\ref{ssec:T2Form} we construct the Hilbert energy-momentum tensors for the Klein-Gordon field, Maxwell's electromagnetism in curved spacetime, and the Kalb-Ramond $2$-form. We then impose that these tensors are homogeneous and isotropic, and we work out what this implies for the matter fields themselves. 
In the case of the Maxwell and Kalb-Ramond fields, we find that they are neither homogeneous nor isotropic. In particular, the form of Kalb-Ramond field strongly differs from the expressions one commonly finds in the literature~\cite{Stein-Schabes:1986owe,Copeland:1994km, Matsuo:2021xas, Piazza:2017bsd, Aoki:2022ylc}. 
This opens up the possibility for future studies of the cosmological implications of such fields. 

In section~\ref{sec:SVT} we turn our attention to cosmological perturbation theory and the scalar-vector-tensor (SVT) decomposition. We provide a simple but rigorous derivation of the SVT decomposition, where every step is explained and made intuitively clear. Our goal is to provide a demystified, self-contained, and novel exposition of this notoriously difficult subject which is of high pedagogical value. 
We present our conclusions in section~\ref{sec:Conclusion} and we provide two appendices. In appendix~\ref{app:A} we derive geometric identities involving the Lie derivative, which were used in the main text. The goal of this appendix is to render the paper as a whole self-contained.
Finally, in appendix~\ref{app:B} we discuss the energy-momentum tensor of an imperfect fluid and its relation to cosmological perturbation theory. The purpose of this discussion is again to make the paper self-contained. \vfill

\section{Homogeneity and Isotropy as Symmetries}\label{sec:Symmetries}
In GR spacetime is described as a smooth four-dimensional, connected manifold $\M$ equipped with a non-degenerate metric $g_{\mu\nu}$ of Lorentzian signature\footnote{Our convention for the signature of $g_{\mu\nu}$ is the ``mostly plus'' convention: $(-,+,+,+)$.}. Matter fields are represented by tensorial fields $\Psi\ud{\bullet}{\circ}$, where the superscript $\bullet$ and the subscript $\circ$ are placeholders for an arbitrary number of contravariant and covariant indices, respectively. We call $(\M, g_{\mu\nu}, \Psi\ud{\bullet}{\circ})$ a \emph{spacetime model}.

Every spacetime model is required to satisfy the Einstein field equations. In the presence of a cosmological constant $\Lambda$, these equations, which govern the dynamics of the metric, can be written as
\begin{align}
    \underbrace{R_{\mu\nu} - \frac12 R\, g_{\mu\nu}}_{\ec G_{\mu\nu}}  + \Lambda\, g_{\mu\nu} = 8\pi\, T_{\mu\nu}\,.
\end{align}
We use a geometrized unit system, in which $c=G=1$. Here, $R_{\mu\nu}$ is the Ricci tensor, constructed from the metric $g_{\mu\nu}$ and its first and second order derivatives, $R \ce g^{\mu\nu} R_{\mu\nu}$ is the Ricci scalar, $G_{\mu\nu}$ denotes the Einstein tensor, and $T_{\mu\nu}$ is the energy-momentum tensor associated with the matter field $\Psi\ud{\bullet}{\circ}$.

In the present context, the spacetime model $(\M, g_{\mu\nu}, \Psi\ud{\bullet}{\circ})$ is also required to satisfy the Cosmological Principle. In the absence of matter fields, this principle demands that $g_{\mu\nu}$ is spatially homogeneous and isotropic. We provide a mathematically precise statement of this requirement further below. Before doing so, however, we note in the presence of matter there is a choice to be made. Either the requirements of homogeneity and isotropy are directly imposed on the metric and the matter fields $\Psi\ud{\bullet}{\circ}$. It is intuitively clear that if $g_{\mu\nu}$ and $\Psi\ud{\bullet}{\circ}$ possess these symmetries, then the Einstein tensor $G_{\mu\nu}$ and $T_{\mu\nu}$ will inherit them. In subsection~\ref{ssec:TProposition} we will mathematically confirm these intuitions.

An equally valid option is to impose homogeneity and isotropy on the metric and the energy-momentum tensor, since this is the only form in which matter fields manifest themselves and influence the metric in the field equations. Because the Einstein tensor inherits the symmetries of the metric, we know that the resulting field equations will be self-consistent, in the sense that both sides possess the same symmetries. This second option amounts to regarding $(\M, g_{\mu\nu}, T_{\mu\nu})$ as spacetime model, rather than $(\M, g_{\mu\nu}, \Psi\ud{\bullet}{\circ})$. In section~\ref{sec:SymmetryReducedT} we show that imposing symmetry-conditions on $T_{\mu\nu}$ produces valid cosmological models. 

It follows that both options are viable, because they lead to consistent cosmological models. Which option to choose therefore seems like a matter of taste. However, it should be noted that these options are not equivalent! 

The discrepancy in the models $(\M, g_{\mu\nu}, \Psi\ud{\bullet}{\circ})$ and $(\M, g_{\mu\nu}, T_{\mu\nu})$ lies in the precise form of the matter content of the Universe. While it is true that a tuple $(g_{\mu\nu}, \Psi\ud{\bullet}{\circ})$ which abides by the Cosmological Principle gives rise to a $T_{\mu\nu}$ which respects homogeneity and isotropy, the converse is not true. In subsections~\ref{ssec:MaxwellEMT} and~\ref{ssec:T2Form} we construct two explicit counterexamples, starting from a tuple $(g_{\mu\nu}, T_{\mu\nu})$ which is homogeneous and isotropic and showing that the underlying matter fields $\Psi\ud{\bullet}{\circ}$ do not share either one of these symmetries. This opens interesting new avenues in the construction of cosmological models. Even more intriguing, but technically challenging, would be to impose symmetries directly on the Einstein and energy-momentum tensors; $(\M, G_{\mu\nu}, T_{\mu\nu})$. If possible, this could allow inhomogeneous and anisotropic metric and matter field configurations, which nevertheless do not manifest themselves in the Einstein field equations. Here we are only able to prove that anisotropic and inhomogeneous matter configurations can give rise to energy-momentum tensors which respect the Cosmological Principle. Whether this is also true for the Einstein tensor remains to be investigated. There could be fundametally different implications for the models $(\M, g_{\mu\nu}, \Psi\ud{\bullet}{\circ})$ and $(\M, G_{\mu\nu}, T_{\mu\nu})$. We regard this as an important open question of potentially high impact. 

To make the claims of section~\ref{sec:Introduction} and in the introduction to section~\ref{sec:Symmetries} precise, we now provide a clear and practical definition of what it means for a field to be homogeneous and isotropic. We begin with the notion of homogeneity, focusing first on the metric; the extension to matter fields will follow naturally.

\subsection{Definition of Homogeneity and Isotropy through Diffeomorphisms}
\begin{figure}[!ht]
    \centering
    \includegraphics[width=0.75\linewidth]{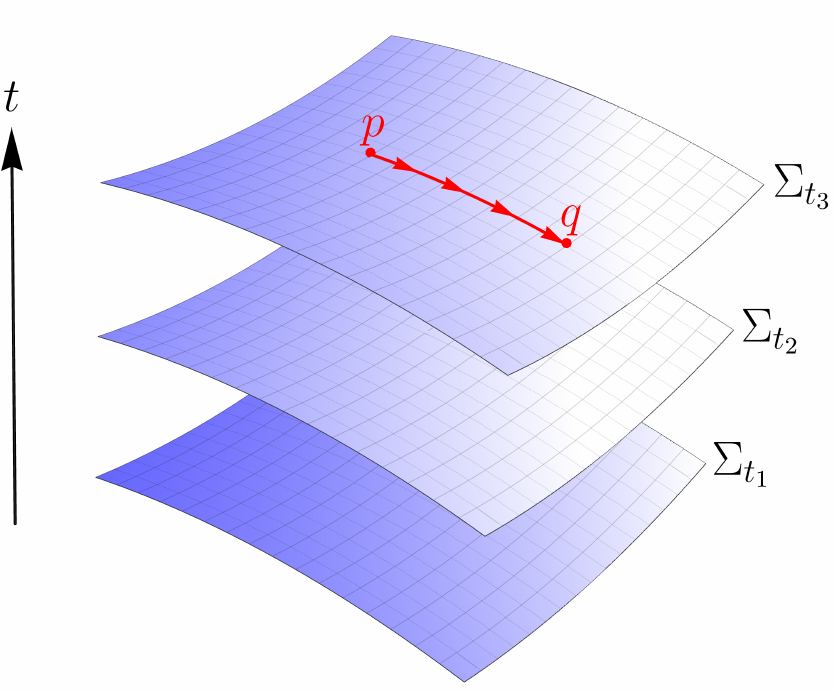}
    \caption{The foliation of spacetime into spacelike leaves $\Sigma_t$, threaded along the time line $t$. The requirement of spatial homogeneity means that any spatial point $p$ can be mapped to any other spatial point $q$ on the same leave, while leaving the metric invariant.}
    \label{fig:Foliation}
\end{figure}

Recall our informal definition of homogeneity: \emph{at any given instant, the Universe looks the same, regardless of an observer's location}. To formalize this idea, we must distinguish between space and time by slicing spacetime into spacelike hypersurfaces, organized along a time axis. More precisely, we assume that the spacetime manifold $\M$ can be written as $\bbR \times \Sigma$, where $\bbR$ represents the temporal dimension, and for any fixed time $t$, the corresponding spatial section is a three-dimensional spacelike hypersurface $\Sigma_t$. We refer to $\Sigma_t$ as the leaves of the foliation, which we can picture as being stacked along the time direction (see Figure~\ref{fig:Foliation}).

This foliation allows us to discuss spatial points rather than merely spacetime points. For any $t \in \bbR$, a point $p \in \Sigma_t$ represents a spatial location, while the corresponding spacetime point is $(t, p) \in \M \simeq \bbR \times \Sigma$. Given such a spatial point $p$, we may imagine an observer located there at time $t$. Homogeneity then requires that if we consider another observer at a different spatial point $q \in \Sigma_t$, both observers perceive the same spatial geometry. In other words, transforming from the perspective of the first observer to that of the second should leave the spatial part of the metric unchanged.

Following Wald~\cite{WaldBook}, we can formalize this idea as follows. A spacetime model $(\M, g_{\mu\nu})$ is said to be \emph{spatially homogeneous} if there exists a one-parameter family of spacelike hypersurfaces $\Sigma_t$ foliating spacetime such that, for any $t$ and for any two points $p, q \in \Sigma_t$, there exists a diffeomorphism $\phi: \{t\} \times \Sigma_t \to \{t\} \times \Sigma_t$ that maps $p$ to $q$ while leaving the metric invariant:
\begin{align}
(\phi^* g)_{\mu\nu} = g_{\mu\nu} \,.
\end{align}
Here, $\phi^* g$ denotes the pullback of $g$ by $\phi$, i.e., the metric obtained when transforming coordinates according to $\phi$. The requirement that a diffeomorphism exists mapping any point $p$ to any other point $q$ implies that all spatial points in $\Sigma_t$ are indistinguishable. Hence, this definition precisely captures the intuitive meaning of spatial homogeneity.

Notice that the definition above only requires that a diffeomorphism mapping $p$ to $q$ \emph{exists}; it does not specify which one. In general, there may be more than one such diffeomorphism.

\begin{figure}[!ht]
    \centering
    \includegraphics[width=0.75\linewidth]{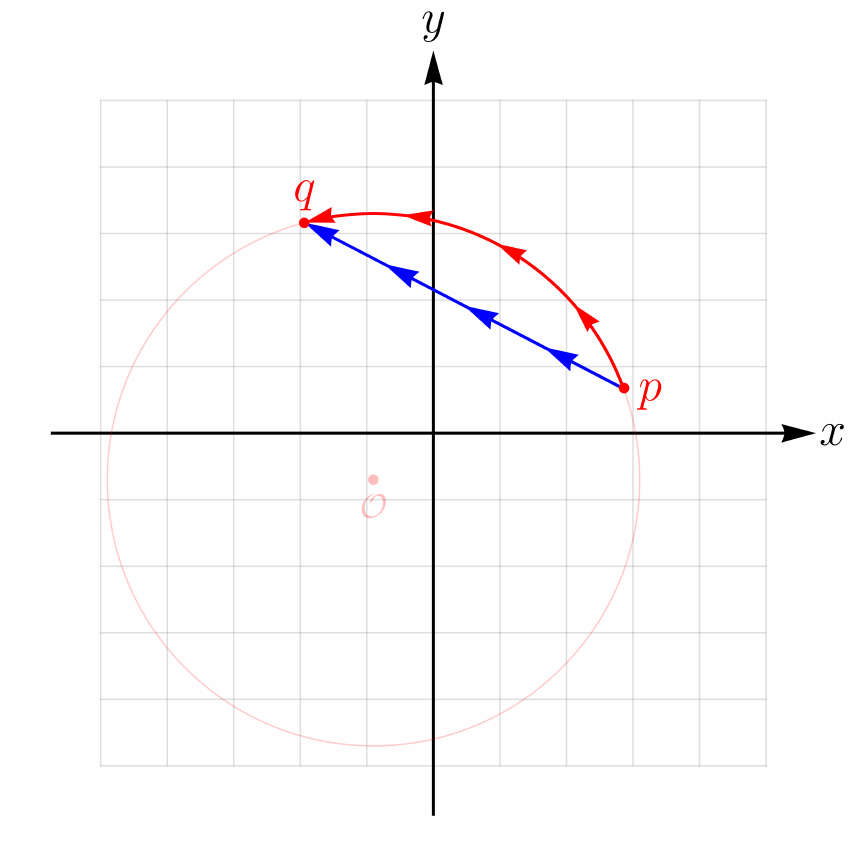}
    \caption{There are many different diffeomorphisms which map $p$ to $q$. Here, a rotation about $\begin{smallmatrix}\mathcal{O}\end{smallmatrix}$ (red arc) and a translation (blue line) are depicted.}
    \label{fig:RotationTranslation}
\end{figure}

As a simple example, consider two points in the Euclidean plane (see Figure~\ref{fig:RotationTranslation}). It is always possible to draw a circle passing through both points. One possible diffeomorphism mapping $p$ to $q$ is a rotation about the circle's center $\begin{smallmatrix}\mathcal{O}\end{smallmatrix}$ (illustrated by the red arc in Figure~\ref{fig:RotationTranslation}). Another equally valid diffeomorphism is a translation taking $p$ to $q$ (shown as the blue line in Figure~\ref{fig:RotationTranslation}).
In general, stipulating that the diffeomorphism in the definition of a homogeneous spacetime be a \emph{translation} is more convenient than other choices. For instance, if one were to use rotations instead, one would always need to refer to a third point, namely the center of the circle passing through $p$ and $q$. No such auxiliary structure is required when using translations. Therefore, from now on, we identify homogeneity with invariance under spatial translations.

To define \emph{spatial isotropy}, we need a notion of direction. Recall that our informal definition stated that \emph{the Universe looks the same in all directions}. Consider an observer located at $p \in \Sigma_t$. We can specify a direction at $p$ by attaching a spatial vector $\boldsymbol{v}$ to that point. We may further imagine the observer rotating around an axis passing through $p$, carrying the vector $\boldsymbol{v}$ along. Given a reference vector $\boldsymbol{v}$, any other direction in space can then be obtained by rotating $\boldsymbol{v}$.

If space is isotropic, the spatial part of the metric remains unchanged under such rotations, regardless of the observer's position $p \in \Sigma_t$. Formally, for any $t$ and any point $p \in \Sigma_t$, a rotation $\phi_p : \{t\} \times \Sigma_t \to \{t\} \times \Sigma_t$ leaves the metric invariant:
\begin{align}
(\phi_p^* g)_{\mu\nu} = g_{\mu\nu} \,.
\end{align}
A spacetime model $(\M, g_{\mu\nu})$ satisfying the Cosmological Principle is thus one that is invariant under spatial translations and under rotations about \emph{any} point on the spatial hypersurfaces.

\subsection{Continuous Symmetries and Infinitesimal Symmetry Conditions}
The characterization of translational and rotational invariance in terms of diffeomorphisms~$\phi$ is not practical in its current form. It would require verifying, for every point $p$, whether the metric remains invariant under all possible rotations about that point. Similarly, for every pair of points $p$ and $q$, one would need to check that translating $p$ to $q$ leaves the metric unchanged.

However, since rotations and translations are smooth transformations, we can instead derive \emph{local} conditions that are far more convenient to work with. Conceptually, we may imagine three axes emanating from any point $p$. Testing rotational and translational invariance then reduces to checking whether the metric is invariant under infinitesimal rotations about these three axes and infinitesimal translations along them. This suffices to ensure invariance under finite rotations and translations as well.

To derive these local symmetry conditions, we make use of the continuous nature of rotations and translations. Let us introduce a one-parameter family of diffeomorphisms 
\begin{align}
    \phi_s:I\times\M \to \M\,,
\end{align}
where the parameter $s$ labels the members of the family and takes values in the interval~$I$. For rotations, this interval is $I = [0, 2\pi)$, while for translations it is the entire real line, $I = \bbR$. We further require that this one-parameter family satisfies:
\begin{enumerate}
    \item $\phi_{s=0} = \mathrm{id}$
    \item $\phi_s \circ \phi_t = \phi_{s+t}$.
\end{enumerate}
The first condition ensures that for $s = 0$, the diffeomorphism reduces to the identity transformation, i.e., $\phi_{s=0}$ leaves all tensor fields unchanged. The second expresses the group property: applying $\phi_t$ followed by $\phi_s$ is equivalent to applying $\phi_{s+t}$ directly. This is indeed the expected behavior for rotations and translations. For example, performing a rotation by $t$ degrees about a given axis and then by an additional $s$ degrees around the same axis is equivalent to a single rotation by $s + t$ degrees.

We can now define a \emph{continuous symmetry} $\phi_s$ of the metric as a diffeomorphism that leaves the metric invariant for all values of $s$ in $I$:
\begin{align}
    (\phi^*_s g)_{\mu\nu} = g_{\mu\nu}\,.
\end{align}
This definition extends naturally to include matter fields. We say that a spacetime model $(\M, g_{\mu\nu}, \Psi\ud{\bullet}{\circ})$ possesses the continuous symmetry $\phi_s$ if
\begin{align}\label{eq:ContSymmCond}
    (\phi^*_s g)_{\mu\nu} &= g_{\mu\nu} &\text{and} && (\phi^*_s \Psi)\ud{\bullet}{\circ} &= \Psi\ud{\bullet}{\circ}
\end{align}
for all values of $s$ in $I$. In the case of rotations and translations, $\phi_s$ is not only continuous in~$s$ but also \emph{differentiable}. This differentiability allows us to introduce the \emph{generating vector field} $\xi$ associated with $\phi_s$:
\begin{align}
    \xi \ce \left.\frac{\dd \phi_s}{\dd s}\right|_{s=0}\,.
\end{align}
The local, or infinitesimal, version of~\eqref{eq:ContSymmCond} follows by performing a Taylor expansion of the left-hand side up to first order in $s$:
\begin{align}
    (\phi^*_sg)_{\mu\nu} = \underbrace{(\phi^*_0 g)_{\mu\nu}}_{=g_{\mu\nu}} + s\left(\left.\frac{\dd}{\dd s}(\phi^*_sg)_{\mu\nu}\right|_{s=0}\right) + \mathcal{O}(s^2) &= g_{\mu\nu} \notag\\
     (\phi^*_s\Psi)\ud{\bullet}{\circ} = \underbrace{(\phi^*_0 \Psi)\ud{\bullet}{\circ}}_{=\Psi\ud{\bullet}{\circ}} + s\left(\left.\frac{\dd}{\dd s}(\phi^*_s\Psi)\ud{\bullet}{\circ}\right|_{s=0}\right) + \mathcal{O}(s^2) &= \Psi\ud{\bullet}{\circ}\,,
\end{align}
where we have used $\phi_{s=0} = \mathrm{id}$. By subtracting the zeroth order term from both sides, dividing by $s$, and taking the $s\to0$ limit, we obtain the infinitesimal symmetry conditions
\begin{align}\label{eq:InfinitSymmCond}
    (\lie_\xi g)_{\mu\nu} &= 0 &\text{and} && (\lie_\xi\Psi)\ud{\bullet}{\circ} &= 0\,,
\end{align}
where $\lie_\xi$ denotes the Lie derivative along the generating vector field $\xi$:
\begin{align}
    (\lie_\xi\Psi)\ud{\bullet}{\circ} &\ce \left.\frac{\dd}{\dd s}(\phi^*_s\Psi)\ud{\bullet}{\circ}\right|_{s=0} = \lim_{s\to0} \frac{(\phi^*_s\Psi)\ud{\bullet}{\circ}-\Psi\ud{\bullet}{\circ}}{s}\,.
\end{align}

Homogeneity and isotropy of space can now be understood as continuous symmetries of both the metric and the matter fields,\footnote{Since the metric and matter fields are related via Einstein’s field equations, some symmetry assumptions on the matter fields are necessary. These assumptions are explored in detail in Sections~\ref{sec:MatterFields} and~\ref{sec:SymmetryReducedT}.} in the sense of~\eqref{eq:InfinitSymmCond}.

If we can provide three independent generators of rotations and three independent generators of translations, we can test whether a given metric is spherically symmetric and translationally invariant. Alternatively, these generators can be used to impose conditions on a generic metric, reducing it to a form consistent with the desired symmetries. We refer to the process of imposing symmetry conditions on a generic tensor field as \emph{symmetry-reduction}. This is precisely the strategy we pursue in the next section to determine the most general metric compatible with homogeneity and isotropy. To this end, we must first identify the generating vector fields corresponding to rotations and translations.

\subsection{The Generating Vector Fields of Rotations and Translations}\label{ssec:GenVecFRotAndTransl}
Let us begin by deriving the generators of rotations. Given a three-dimensional vector $\x \ce x\,\boldsymbol{\hat{e}}_1 + y\,\boldsymbol{\hat{e}}_2 + z\,\boldsymbol{\hat{e}}_3$, with unit basis vectors $\boldsymbol{\hat{e}}_i$ and Cartesian coordinates $(x,y,z)$, rotations act as
\begin{align}
    \x\quad \mapsto\quad \tilde{\x} = R_{\hat{\n}}(s)\x\,,
\end{align}
where $\hat{\n}$ denotes the rotation axis and $s \in I = [0, 2\pi)$ is the rotation angle. There are three distinct rotation matrices,
\begin{align}
    R_{\boldsymbol{\hat{e}}_1}(s) &= \begin{pmatrix}
        1 & 0 & 0 \\
        0 & \cos s & -\sin s \\
        0 & \sin s & \cos s
    \end{pmatrix}
    & 
    R_{\boldsymbol{\hat{e}}_2}(s) &= \begin{pmatrix}
        \cos s & 0 & \sin s\\
        0 & 1 & 0 \\
        -\sin s & 0 & \cos s
    \end{pmatrix} \\
    R_{\boldsymbol{\hat{e}}_3}(s) &= \begin{pmatrix}
        \cos s & -\sin s & 0 \\
        \sin s & \cos s & 0 \\
        0 & 0 & 1
    \end{pmatrix}\,.
\end{align}
Each matrix defines a one-parameter family of diffeomorphisms acting on the coordinates $x^\mu = (t, x, y, z)$. For instance, using the first rotation matrix we obtain
\begin{align}
    \phi_s(x^\mu) = (t, R_{\boldsymbol{\hat{e}}_1}(s)\x) = (t, x, y\,\cos s - z\, \sin s, z\,\cos s + y\,\sin s)\,.
\end{align}
The corresponding generating vector field, denoted $\R_1$, is
\begin{align}
    \R_1 = \left.\frac{\dd \phi_s(x^\mu)}{\dd s}\right|_{s=0} = (0, 0, -z, y)\,.
\end{align}
Proceeding analogously for the remaining rotation matrices and transforming from Cartesian to spherical coordinates $(t, r, \theta, \phi)$, we find
\begin{align}
    \R_1 &= \left(0, 0, -\sin\phi, - \frac{\cos\phi}{\tan \theta}\right)\,,   & \R_2 &= \left(0,0, \cos\phi, - \frac{\sin\phi}{\tan \theta}\right)\,, & \R_3 &= (0,0,0,1)\,.
\end{align}
Expressed in the basis $\{\pd_t, \pd_r, \pd_\theta, \pd_\phi\}$ of the tangent bundle~$T\M$, the generating vector fields of rotations can be written in the equivalent but more compact form
\begin{align}\label{eq:GenOfRotationsSpherical}
    \R_1 &= -\sin\phi\, \pd_\theta - \frac{\cos\phi}{\tan \theta}\, \pd_\phi \,, & \R_2 &= \cos\phi\,\pd_\theta - \frac{\sin\phi}{\tan \theta}\, \pd_\phi\,, & \R_3 &= \pd_\phi\,.
\end{align}

In the case of translations, we would like to proceed analogously, but we must be more careful because translations are sensitive to curvature. To see this crucial distinction, recall that a rotation acts on a vector $v \in T_p\M$ at a fixed point $p \in \M$: the point $p$ remains unchanged, while $v$ is mapped to another vector $v'$ in the same tangent space~$T_p\M$.

Translations behave differently. Given a point $p \in \M$ and a vector $v \in T_p\M$, a translation moves $v$ to a new vector $v'$ in a different tangent space $T_{p'}\M$ at some point $p' \neq p$. Thus, translations ``feel'' the curvature of the underlying space.

To illustrate this, consider $\M$ to be the two-sphere $\mathbb{S}^2$. For any point $p \in \mathbb{S}^2$, a vector $v \in T_p\mathbb{S}^2$ lies in the plane tangent to the sphere at~$p$. Rotating $v$ about an axis through $p$ merely rotates the vector within that plane.
By contrast, translating $v$ from $p$ to another point $p'$ drags it along the surface of $\mathbb{S}^2$ so that it always remains tangent to the sphere; it does not pass through the interior of the sphere (when viewed as embedded in~$\bbR^3$). Hence, translations inherently carry information about the curvature of~$\mathbb{S}^2$.

Because of this curvature sensitivity, translations pose a potential problem. On a pseudo-Riemannian manifold $(\M, g_{\mu\nu})$, curvature is encoded in the Riemann tensor, which itself is entirely determined by the metric. However, when the metric is still unknown, as it is in our case before solving Einstein's field equations, the precise form of the curvature tensor, and thus the action of translations on vectors and tensors, cannot be known.

We can overcome this obstacle by first considering the special case of translations in flat space and then introducing a deformation of the generating vector fields. By imposing that these deformed generators be Killing vectors of a spherically symmetric metric, we can determine their exact form.
This procedure also reveals that the spatial part of the metric must correspond to one of three possible geometries: flat Euclidean space, a positively curved three-sphere, or a negatively curved hyperbolic space~$H^3$.
The explicit derivation of this result is presented in Section~\ref{sec:Metric}.

In preparation for this, let us first consider translations in flat Euclidean space. In Cartesian coordinates, a translation can be written as
\begin{align}
    \x\quad \mapsto\quad \tilde{\x} = \x + s\,\boldsymbol{\hat{e}}_i \,,
\end{align}
where $\boldsymbol{\hat{e}}_i$ denotes the axis along which the translation occurs and $s \in I = \bbR$ represents the magnitude of the translation. The corresponding one-parameter family of diffeomorphisms acting on $x^\mu = (t, \x)$ is
\begin{align}
    \phi_s(x^\mu) = (t, \x) + s\,(0, \boldsymbol{\hat{e}}_i)\,.
\end{align}
From this, we deduce that the generating vector fields take the simple form
\begin{align}
    \T_i = \left.\frac{\dd \phi_s(x^\mu)}{\dd s}\right|_{s=0} = \boldsymbol{\hat{e}}_i\,,
\end{align}
for $i = 1,2,3$. In the basis $\{\pd_t, \pd_x, \pd_y, \pd_z\}$ of the tangent bundle $T\M$, these generators can equivalently be written as
\begin{align}\label{eq:GenOfTranslationsCartesian}
    \T_1 &= \pd_x\,, & \T_2 &= \pd_y\,, & \T_3 &= \pd_z\,.
\end{align}
For later convenience, we express these generators in spherical coordinates. With respect to the basis $\{\pd_t, \pd_r, \pd_\theta, \pd_\phi\}$ of the tangent bundle $T\M$, the generators of translation take the form
\begin{align}\label{eq:GenOfTranslationsSpherical}
    \T_1 &= \sin\theta \cos\phi\, \pd_r + \frac1r \cos\theta \cos\phi \, \pd_\theta - \frac1r \frac{\sin\phi}{\sin\theta}\,\pd_\phi\notag\\
    \T_2 &= \sin\theta \sin\phi\, \pd_r + \frac1r \cos\theta \sin\phi \, \pd_\theta + \frac1r \frac{\cos\phi}{\sin\theta}\,\pd_\phi\notag\\
    \T_3 &= \cos\theta\,\pd_r - \frac1r \sin\theta\, \pd_\theta\,.
\end{align}
As emphasized earlier, these vector fields represent translation generators only in flat Euclidean space.
To obtain suitable candidates for translation generators on a curved spatial manifold~$\Sigma_t$, we must deform~\eqref{eq:GenOfTranslationsSpherical} by introducing a set of unknown functions of $(r,\theta,\phi)$.

A completely general deformation would multiply each component of each $\T_i$ by a distinct function of $(r, \theta, \phi)$, and in the case of $\T_3$, add an additional component:
\begin{align}\label{eq:DeformedTGeneral}
    \T_1 &= A(r,\theta,\phi)\sin\theta \cos\phi\, \pd_r + B(r,\theta,\phi)\frac1r \cos\theta \cos\phi \, \pd_\theta - C(r,\theta,\phi)\frac1r \frac{\sin\phi}{\sin\theta}\,\pd_\phi\notag\\
    \T_2 &= D(r,\theta,\phi)\sin\theta \sin\phi\, \pd_r + E(r,\theta,\phi)\frac1r \cos\theta \sin\phi \, \pd_\theta + F(r,\theta,\phi)\frac1r \frac{\cos\phi}{\sin\theta}\,\pd_\phi\notag\\
    \T_3 &= G(r,\theta,\phi)\cos\theta\,\pd_r - H(r,\theta,\phi)\frac1r \sin\theta\, \pd_\theta + I(r,\theta,\phi)\, \pd_\phi\,.
\end{align}
Such a procedure introduces nine arbitrary functions. However, most choices of these functions would not correspond to anything we would reasonably interpret as translations. Indeed, with inappropriate choices, one could even tune the $\T_i$ so that they become equivalent to the rotation generators~$\R_i$, which is clearly nonsensical.

We therefore need a more restrictive deformation procedure; one that produces viable candidates for translation generators in curved space while preserving their geometric meaning. There are two natural approaches to achieve this.
\begin{enumerate}
    \item \textbf{Geodesic-based definition}\\ We may start by asking what we mean by a ``translation'' in a curved space.
    Intuitively, given a point $p \in \Sigma_t$, moving along a geodesic $\gamma$ that passes through $p$ with tangent vector $\boldsymbol{v}$ constitutes a translation from $p$ in the direction of~$\boldsymbol{v}$.
    This definition agrees with our earlier discussion of translations on the two-sphere~$\mathbb{S}^2$, and it is both coordinate-independent and sensitive to curvature.
    Indeed, geodesics are determined by the metric and its Christoffel symbols, which encode curvature.
    Hence, one could in principle construct the generators of translations by studying the geodesics of $(\Sigma_t, h_{ab})$, where $h_{ab}$ is the spatial metric of signature~$(+,+,+)$.
    Since $(\Sigma_t, h_{ab})$ must be spherically symmetric, we could start from the most general spherically symmetric three-dimensional metric and derive the geodesic equations.
    Although viable, this strategy is labor-intensive and not particularly transparent.
    \item \textbf{Algebraic and symmetry-based construction}\\ A more efficient and conceptually clear route is to reflect on how translations should transform under rotations. Suppose we pick a point $p \in \Sigma_t$ and move along a geodesic $\gamma$ with tangent vector $\boldsymbol{v}$ at~$p$.
    This describes a translation from $p$ in the direction~$\boldsymbol{v}$.
    If we now perform a rotation about $p$, the vector $\boldsymbol{v}$ is mapped to a new vector $\boldsymbol{v}'$ within the same tangent space~$T_p\Sigma_t$, and the geodesic~$\gamma$ is mapped to another geodesic~$\gamma'$ that still passes through $p$ with tangent~$\boldsymbol{v}'$.
    Hence, under rotations, the generators of translations transform as vectors.
\end{enumerate}
Notice that in the second approach we did not assume $(\Sigma_t, h_{ab})$ to be spherically symmetric. Thus, the statement that translation generators transform as vectors under rotations holds on any manifold. However, once we assume spherical symmetry, this transformation property can be exploited to constrain possible deformations of the generators of translation. 

To see this, we demand that under an infinitesimal rotation generated by the Lie derivative $\lie_{\R_i}$ the generator of translations $\T_j$ transforms as
\begin{align}
    \lie_{\R_i}\T_j \overset{!}{=} c\du{ij}{k}\T_k\,.
\end{align}
The right hand side is simply a linear combination of vectors $\T_k$, which expresses the fact that $\T_j$ transform as components of a vector under rotations. Since the Lie derivative acting on vector fields is equivalent to their commutator, this condition can be rewritten as
\begin{align}\label{eq:TransformationCondition}
    [\R_i, \T_j] = c\du{ij}{k}\T_k\,.
\end{align}
The commutator is antisymmetric in its arguments, implying that $c\du{ij}{k}$ must be anti-symmetric in $i$ and~$j$. In principle, $c\du{ij}{k}$ could depend on $(r, \theta, \phi)$, but spherical symmetry implies that the relation~\eqref{eq:TransformationCondition} must hold identically at every point $p \in \Sigma_t$. Thus, $c\du{ij}{k}$ cannot depend on position and must be a constant rank-three tensor invariant under rotations.
Up to normalization, the only such tensor is the Levi–Civita symbol~$\epsilon_{ijk}$, so that
\begin{align}
    c_{ijk} = \lambda\, \epsilon_{ijk}\,,
\end{align}
where $\lambda$ is a constant. Without loss of generality, we may set $\lambda = 1$, since any other choice can be absorbed by a rescaling $\T_i \mapsto \frac{1}{\lambda} \T_i$. Hence, only those vector fields $\T_i$ satisfying
\begin{align}\label{eq:TransformationConditionFinalForm}
    [\R_i, \T_j] = \epsilon\du{ij}{k}\T_k
\end{align}
are admissible as generators of translations on a curved spatial manifold~$\Sigma_t$.

A direct computation confirms that the flat-space generators~\eqref{eq:GenOfTranslationsSpherical} satisfy this condition. Moreover, observe that the rotation generators $\R_i$ contain the derivative operators $\pd_\theta$ and $\pd_\phi$, but not the operator $\pd_r$. Consequently, if we consider generators of the form~\eqref{eq:DeformedTGeneral}, the only admissible deformation functions are those which depend on $r$; any dependence on $\theta$ or $\phi$ would introduce additional derivatives on the left-hand side of~\eqref{eq:TransformationConditionFinalForm}, while no such derivatives appear on the right-hand side.

Let us therefore assume that the nine function $A-I$ only depend on $r$. However, condition~\eqref{eq:TransformationConditionFinalForm} now implies that we do not need nine independent functions and that we do not need to add a third component to $\T_3$. Three such functions suffice. To see this, note that if, for instance, $\T_1$ contained functions differing from those in~$\T_3$, the relation
\begin{align}
    [\R_2, \T_1] = \epsilon\du{21}{3}\T_3
\end{align}
could not be satisfied. Also, $\T_3$ cannot contain a third component, since this would again spoil this equation. Similar arguments hold for other choices of $\R_i$ and $\T_j$. Therefore, the most general admissible ansatz for the deformed translation generators is
\begin{align}\label{eq:DeformedGenerators}
    \T_1 &=  F_r(r)\,\sin\theta \cos\phi\, \pd_r + F_\theta(r)\,\frac1r \cos\theta \cos\phi \, \pd_\theta -F_\phi(r)\, \frac1r \frac{\sin\phi}{\sin\theta}\,\pd_\phi\notag\\
    \T_2 &= F_r(r)\,\sin\theta \sin\phi\, \pd_r + F_\theta(r)\,\frac1r \cos\theta \sin\phi \, \pd_\theta + F_\phi(r)\, \frac1r \frac{\cos\phi}{\sin\theta}\,\pd_\phi\notag\\
    \T_3 &= F_r(r)\,\cos\theta\,\pd_r - F_\theta(r)\,\frac1r \sin\theta\, \pd_\theta\,,
\end{align}
where $F_r(r)$, $F_\theta(r)$, and $F_\phi(r)$ are arbitrary functions of the radial coordinate~$r$.

In the next section, we will use the generating vector fields~\eqref{eq:GenOfRotationsSpherical} and~\eqref{eq:DeformedGenerators} to perform a symmetry-reduction of the metric. Before doing so, however, note that the commutation relation~\eqref{eq:TransformationConditionFinalForm} can also be written in terms of the Lie derivative:
\begin{align}
    [\lie_{\R_i}, \lie_{\T_j}] = \epsilon\du{ij}{k}\lie_{\T_k}\,.
\end{align}
A useful conclusion can be drawn from these commutation relations. Suppose we are given a spherically symmetric tensor field, i.e., one that satisfies
\begin{align}\label{eq:SphericalSymmetryIndividual}
    \lie_{\R_i}\Psi\ud{\bullet}{\circ} &= 0 & \text{for } i=1,2,3\,.
\end{align}
Assume further that $\Psi\ud{\bullet}{\circ}$ is also invariant under translations generated by $\T_3$, so that $\lie_{\T_3}\Psi\ud{\bullet}{\circ} = 0$. From the commutation relation, we then find
\begin{align}
   \lie_{\T_1}\Psi\ud{\bullet}{\circ} &= \lie_{\R_2}\left(\lie_{\T_3}\Psi\ud{\bullet}{\circ}\right) - \lie_{\T_3}\left(\lie_{\R_2}\Psi\ud{\bullet}{\circ}\right) = 0 \notag\\
   \lie_{\T_2}\Psi\ud{\bullet}{\circ} &= \lie_{\R_3}\left(\lie_{\T_1}\Psi\ud{\bullet}{\circ}\right) - \lie_{\T_1}\left(\lie_{\R_3}\Psi\ud{\bullet}{\circ}\right) =  \lie_{\R_3}\left(\lie_{\T_1}\Psi\ud{\bullet}{\circ}\right)\,.
\end{align}
This shows that spherical symmetry, together with invariance under translations generated by $\T_3$, implies invariance under translations generated by $\T_1$. In turn, this leads to invariance under $\T_2$, since $\lie_{\R_3}\left(\lie_{\T_1}\Psi\ud{\bullet}{\circ}\right)=0$. Therefore, to obtain a tensor that is both spherically symmetric and translationally invariant, it suffices to impose only four independent conditions instead of six. These are
\begin{align}\label{eq:SymmetryConditionsFullSet}
    &\text{(i)} & \lie_{\R_3} \Psi\ud{\bullet}{\circ} &\overset{!}{=} 0 \notag\\
    &\text{(ii)} & \cos\phi \lie_{\R_1}\Psi\ud{\bullet}{\circ} + \sin\phi \lie_{\R_2}\Psi\ud{\bullet}{\circ} &\overset{!}{=} 0   \notag\\
    &\text{(iii)} & \sin\phi \lie_{\R_1}\Psi\ud{\bullet}{\circ} - \cos\phi \lie_{\R_2}\Psi\ud{\bullet}{\circ} &\overset{!}{=} 0 \notag\\
    &\text{(iv)} & \lie_{\T_3}\Psi\ud{\bullet}{\circ} &\overset{!}{=} 0\,.
\end{align}
As we will see later, we chose linear combinations of $\lie_{\R_1}$ and $\lie_{\R_2}$ in conditions (ii) and (iii) rather than imposing $\lie_{\R_1}\Psi\ud{\bullet}{\circ} = 0$ and $\lie_{\R_2}\Psi\ud{\bullet}{\circ} = 0$ individually, because this formulation yields simple algebraic conditions from (ii) and first-order differential equations from (iii). Taken together, conditions (i)–(iii) are equivalent to the spherical symmetry condition~\eqref{eq:SphericalSymmetryIndividual}.

\section{Symmetry-Reduced Form of the Metric}\label{sec:Metric}
The most general metric that respects spatial homogeneity and isotropy is the well-known Friedmann–Lema\^{i}tre–Robertson–Walker (FLRW) metric. Its line element in $(t, r, \theta, \phi)$ coordinates reads
\begin{align}
    \dd s^2 = - \dd t^2 + a(t) \left(\frac{\dd r^2}{1-k r^2} + r^2\,\dd\theta^2 + r^2\,\sin^2 \theta\, \dd \phi^2\right)\,,
\end{align}
where $a(t)$ is the scale factor and $k \in \{-1, 0, 1\}$ encodes the constant curvature of the spatial slices. When $k = 0$, the spatial sections are flat three-dimensional Euclidean space~$\bbR^3$; $k = 1$ corresponds to spatial slices with the geometry of the three-sphere $\mathbb{S}^3$; and $k = -1$ describes three-dimensional hyperbolic space $\mathbb{H}^3$.

Different derivations of this result exist in the literature, each with varying degrees of rigor. Some rely on physical intuition~\cite{RovelliBook}, others on coordinate-based arguments~\cite{SchutzBook, BaumannBook}. More commonly, one appeals to known properties of maximally symmetric spaces, which are either derived elsewhere in the same monograph~\cite{WeinbergBook, Blau, CarrollBook}, taken from earlier works~\cite{HawkingEllisBook}, or proven directly in context~\cite{WaldBook}. Another frequent approach is to argue that homogeneity and isotropy impose strong restrictions on the \emph{spatial} Riemann tensor. Such arguments can be qualitative and coordinate-dependent~\cite{MisnerThorneWheelerBook, SchutzBook}, refer to external results~\cite{EinsteinBook}, assume prior knowledge of special metrics~\cite{WeinbergCosmologyBook}, or be presented specifically for cosmological applications~\cite{MalamentBook}.

Both of these approaches focus primarily on \emph{space} rather than on \emph{spacetime}. Indeed, it is the spatial Riemann tensor that takes an algebraically special form, which reflects that the spatial hypersurfaces are maximally symmetric. However, homogeneity and isotropy are, strictly speaking, symmetries of the full \emph{spacetime} metric, not merely of the spatial slices. To the best of our knowledge, there is no derivation in the literature that treats homogeneity and isotropy as spacetime symmetries of $g_{\mu\nu}$ and then proceeds to solve the Killing equations for all metrics consistent with these symmetries.

Such an approach is for instance employed to derive the Schwarzschild metric~\cite{CarrollBook}. By imposing time-translation invariance and spherical symmetry, one can use the resulting Killing equations to reduce the metric to a simple ansatz, which then serves as the starting point for solving Einstein’s field equations.

In the cosmological context, the approach of performing a symmetry-reduction was, for instance, employed in~\cite{Hohmann:2021, DAmbrosio:2021b} to study $f(\bbQ)$ cosmology. There, the FLRW metric was assumed from the outset so that its Killing vectors could be used to define homogeneity and isotropy. The Killing equations were then imposed on the affine connection, which, in non-metricity theories such as $f(\bbQ)$, differs from the Levi-Civita connection of the metric. These equations were then solved to find the most general connections compatible with these symmetries. This method allows for the study of cosmological models with non-trivial connections.

Extending this procedure to derive the metric itself from its Killing vectors, however, presents a major challenge. Imposing spherical symmetry or time-translation invariance, as it is done when deriving the Schwarzschild metric, poses no difficulty, as neither requires prior knowledge of curvature or of the metric. The same is not true for spatial translations because of their susceptibility to curvature, as we have discussed in the previous section. Postulating the Killing vectors associated with spatial translations effectively amounts to postulating the metric itself. On the other hand, attempting to derive these translation Killing vectors inevitably involves the metric, since curvature enters through its Christoffel symbols.

This difficulty may explain why, to the best of our knowledge, no derivation of the FLRW metric based directly on Killing vectors has appeared in the literature. In what follows, we demonstrate that this obstacle can, in fact, be overcome. After imposing spherical symmetry and solving the Killing equations for the most general spherically symmetric metric, we use the most general ansatz for generators of translations compatible with this symmetry (cf. equation~\eqref{eq:DeformedGenerators}). Remarkably, this allows us to obtain simultaneously the most general homogeneous and isotropic metric together with its corresponding Killing vectors.

\subsection{Imposing Rotational-Invariance}
Let $g_{\mu\nu}$ be a non-degenerate metric ($\det g_{\mu\nu}\neq0$) of Lorentzian signature $(-,+,+,+)$. Furthermore, we work in coordinates $(t,r,\theta,\phi)$. A priori, $g_{\mu\nu}$ consists of ten independent components, which depend on $(t, r, \theta, \phi)$. In what follows, we perform the symmetry-reduction of $g_{\mu\nu}$ with respect to spherical symmetry. This procedure will eliminate certain components and restrict the functional dependence of others. We begin by imposing symmetry condition (i) from~\eqref{eq:SymmetryConditionsFullSet}:
\begin{align}
    \lie_{\R_3}g_{\mu\nu} = -\pd_\phi g_{\mu\nu} \overset{!}{=}0 \,.
\end{align}
This condition requires that all ten components of $g_{\mu\nu}$ be independent of the coordinate $\phi$. We can use this fact to simplify the evaluation of symmetry condition (ii), which gives
\begin{align}
    \cos\phi\, \lie_{\R_1}g_{\mu\nu} + \sin\phi\, \lie_{\R_2}g_{\mu\nu} &= \notag\\
    -\frac{1}{\sin^2\theta}
    &\begin{pmatrix}
        0 & 0 & g_{t\phi} & - \sin^2\theta g_{t\theta} \\
        0 & 0 & g_{r\phi} & - \sin^2\theta g_{r\theta} \\
        g_{t\phi} & g_{r\phi} & 2 g_{\theta\phi} & g_{\phi\phi} - \sin^2\theta g_{\theta\theta} \\
        - \sin^2\theta g_{t\theta} & - \sin^2\theta g_{r\theta} &  g_{\phi\phi} - \sin^2\theta g_{\theta\theta} & -2 \sin^2\theta g_{\theta\phi} 
    \end{pmatrix}
    \overset{!}{=} 0\,.
\end{align}
These are purely algebraic conditions on the components of $g_{\mu\nu}$, as no derivatives with respect to $\theta$ appear. We can immediately read off the solution:
\begin{align}
    g_{t\theta} &= 0\,, & g_{t\phi} &= 0\,, & g_{r\theta} &= 0\,, & g_{r\phi} &= 0\,, & g_{\theta\phi} &= 0\,, & g_{\phi\phi} &= \sin^2\theta g_{\theta\theta}\,.
\end{align}
Symmetry condition (ii) thus eliminates five of the ten independent components of $g_{\mu\nu}$ and expresses $g_{\phi\phi}$ in terms of $g_{\theta\theta}$. We are therefore left with four independent components, $\{g_{tt}, g_{tr}, g_{rr}, g_{\theta\theta}\}$, each a function of $t$, $r$, and $\theta$.

The third symmetry condition removes the remaining $\theta$-dependence:
\begin{align}
    \sin\phi \lie_{\R_1}g_{\mu\nu} - \cos\phi \lie_{\R_2}g_{\mu\nu} &= -\cot\theta \left(\delta\du{\mu}{\phi} g_{\phi\nu} + \delta\du{\nu}{\phi} g_{\phi\mu}\right) + \pd_\theta g_{\mu\nu} \notag\\
    &=
    \begin{pmatrix}
        \pd_\theta g_{tt} & \pd_\theta g_{tr} & 0 & 0 \\
        \pd_\theta g_{tr} & \pd_\theta g_{rr} & 0 & 0 \\
        0 & 0 & \pd_\theta g_{\theta\theta} & 0\\
        0 & 0 & 0 & \sin^2\theta \pd_\theta g_{\theta\theta}
    \end{pmatrix}
    \overset{!}{=} 0\,.
\end{align}
To obtain this result, we used $g_{\phi\phi} = \sin^2\theta\, g_{\theta\theta}$. As anticipated, these equations imply that $\{g_{tt}, g_{tr}, g_{rr}, g_{\theta\theta}\}$ depend only on $t$ and $r$. We thus conclude that the most general metric compatible with rotational symmetry has the form
\begin{align}
    g_{\mu\nu} = 
    \begin{pmatrix}
        g_{tt}(t,r) & g_{tr}(t, r) & 0 & 0 \\
        g_{tr}(t, r) & g_{rr}(t, r) & 0 & 0 \\
        0 & 0 & g_{\theta\theta}(t, r) & 0 \\
        0 & 0 & 0 & \sin^2\theta g_{\theta\theta}(t, r)
    \end{pmatrix}\,.
\end{align}
This form of the metric serves as the starting point for our analysis of translational invariance in the next subsection.

\subsection{Imposing Translational-Invariance}\label{ssec:ImposingTranslInv}
We now turn to the final symmetry condition, which enforces translational invariance. From
\begin{align}
    \lie_{\T_3} g_{\mu\nu} \overset{!}{=} 0\,
\end{align}
and using the deformed generating vector field~\eqref{eq:DeformedGenerators}, we obtain the following six equations:
\begin{align}\label{eq:SystemOfEqs}
    \text{(i)}& & F_r\, \pd_r g_{tt} &= 0 \notag\\
    \text{(ii)}& & \pd_r \left(F_r\, g_{tr}\right) &= 0 \notag\\
    \text{(iii)}& & F_r\,g_{tr} &= 0 \notag\\
    \text{(iv)} & & F_r\, \pd_r g_{rr} + 2 \pd_r F_r\, g_{rr}  &= 0 \notag\\
    \text{(v)} & &  - F_r\, g_{rr} + \frac{1}{r^2} \left(F_\theta - r \, \pd_r F_\theta\right)g_{\theta\theta} &= 0 \notag\\
    \text{(vi)} & & r \,F_r\, \pd_r g_{\theta\theta} - 2 F_\theta\, g_{\theta\theta} &= 0\,.
\end{align}
Equation (i) can be satisfied either by $F_r = 0$ or by $\pd_r g_{tt} = 0$. However, the choice $F_r = 0$ leads to a contradiction. Indeed, if $F_r = 0$, then equations (i)–(v) are trivially satisfied, while equation (vi) reduces to $F_\theta\, g_{\theta\theta} = 0$. This gives two possibilities: either $F_\theta = 0$, which together with $F_r = 0$ implies $\T_3 = 0$, or $g_{\theta\theta} = 0$. The former is clearly unacceptable, since we require\footnote{If we allowed $\T_3=0$, we would not be able to satisfy the commutation relation~\eqref{eq:TransformationCondition}, which we established to hold in full generality. Thus we obtain a contradiction.} $\T_3 \neq 0$, and the latter would make the metric degenerate ($\det g_{\mu\nu} = 0$), in violation of our assumptions. Hence, the only viable option is $\pd_r g_{tt} = 0$ with $F_r \neq 0$.

Given $F_r \neq 0$, equation (iii) immediately implies that the off-diagonal component $g_{tr}$ vanishes. Equation (ii) is then automatically satisfied. We are now left with the three equations (iv), (v), and (vi) for the four unknown functions $g_{rr}$, $g_{\theta\theta}$, $F_r$, and $F_\theta$.

Equation (iv) can be integrated by separation of variables, yielding
\begin{align}\label{eq:grr}
    \frac{\pd_r g_{rr}(t,r)}{g_{rr}(t,r)} &= -2 \frac{\pd_r F_r(r)}{F_r(r)} &\Longleftrightarrow && g_{rr}(t,r) &= \frac{c_1(t)}{F_r(r)^2}\,,
\end{align}
where $c_1(t)$ is an arbitrary function of time. Next, consider equation (vi), which can be rewritten as
\begin{align}\label{eq:ArgumentDependences}
    \frac{\pd g_{\theta\theta}(t, r)}{g_{\theta\theta}(t, r)} &= 2 \frac{F_\theta(r)}{r\, F_r(r)}\,.
\end{align}
Note that all functions on the right-hand side depend only on $r$, while $g_{\theta\theta}$ depends on both $t$ and $r$. For this equation to hold consistently, $g_{\theta\theta}$ must factorize as
\begin{align}\label{eq:gthetatheta}
    g_{\theta\theta}(t, r) = c_2(t) f(r)\,,
\end{align}
for some nonzero functions $c_2(t)$ and $f(r)$. This factorization removes the $t$-dependence from the left-hand side of~\eqref{eq:ArgumentDependences}. Substituting~\eqref{eq:grr} and~\eqref{eq:gthetatheta} into equations (v) and (vi) yields two ordinary differential equations for the three functions of $r$:
\begin{align}
    &\text{(v)} & - \frac{c_1(t)}{F_r(r)} + \frac{1}{r^2} \left(F_\theta(r) - r\, F'_\theta(r)\right) c_2(t) f(r) &= 0\notag\\
    &\text{(vi)} & r\, F_r\,  f'(r) - 2 F_\theta(r)\, f(r) &= 0\,,
\end{align}
where a prime denotes a derivative with respect to $r$.

Since the number of unknown functions exceeds the number of available equations, we must impose one additional condition on either $F_r$, $F_\theta$, or $f$. The most natural choice is to exploit the freedom of coordinate transformations. It is well-known that one can always perform a coordinate transformation such that the function $g_{\theta\theta}(t,r)$ of a spherically symmetric metric takes the form $r^2\, c_2(t)$ (see, for instance,~\cite{CarrollBook, DAmbrosio:2021}). This choice implies $f(r) = r^2$, and equation (vi) then reduces to
\begin{align}
    F_\theta(r) = F_r(r)\,.
\end{align}
Substituting this result into equation (v) yields
\begin{align}
    &\text{(v)} & - \frac{c_1(t)}{F_r(r)} + \left(F_r(r) - r\, F'_r(r)\right) c_2(t) &= 0\,.
\end{align}
This equation contains both $t$- and $r$-dependent terms. The $t$-dependence can be isolated as
\begin{align}
    &\text{(v)} & \frac{c_1(t)}{c_2(t)}   &=  F_r(r)\left(F_r(r) - r\, F'_r(r)\right)\,.
\end{align}
Since the left-hand side depends only on $t$ and the right-hand side only on $r$, both must be equal to a constant:
\begin{align}\label{eq:SeparationOfEquations}
    F_r\,\left(F_r - r \,F_r'\right) &= c &\text{and} && \frac{c_1}{c_2} &= c\,,
\end{align}
where $c$ is a nonzero constant. If we would allow $c=0$, then $c_1=0$ and consequently $g_{rr}=0$, rendering the metric degenerate. Moreover, since we assumed the metric to have Lorentzian signature $(-,+,+,+)$, both $c_1(t)$ and $c_2(t)$ must be positive, implying that $c>0$. To emphasize this, we set $c_1(t) = a(t)^2$ and denote $c = |c|$.

The first-order differential equation from~\eqref{eq:SeparationOfEquations} can be solved by separation of variables. Relabeling $F_r$ as $\chi$, we find
\begin{align}
   F_r\,\left(F_r - r \,\frac{\dd F_r}{\dd r}\right) = |c| \Longleftrightarrow  \frac{\chi\,\dd \chi}{\chi^2-|c|} = \frac{\dd r}{r} \quad  &\Longrightarrow\quad \int \frac{\chi\,\dd \chi}{\chi^2-|c|} = \int \frac{\dd r}{r}\notag\\
    &\Longrightarrow\quad\frac12 \log\left(\chi^2 - |c|\right) = \log(\sqrt{K}\,r) \notag\\
    &\Longrightarrow\quad \chi(r) = \pm \sqrt{|c|+K r^2}\,,
\end{align}
where $K\in\bbR$ is an integration constant. The overall sign is irrelevant, since $\chi$ appears quadratically in the metric (see~\eqref{eq:grr}). For convenience, we choose the plus sign.

At this point, all equations have been solved. We find that $g_{tt}$ depends only on time, $g_{tr}$ vanishes, and the remaining components are determined up to an arbitrary function of time $a(t)$ and two real constants $|c|\neq 0$ and $K$:
\begin{align}
    g_{\mu\nu} = 
    \begin{pmatrix}
        g_{tt}(t) & 0 & 0 & 0 \\
        0 & \frac{a(t)^2 |c|}{|c|+K r^2} & 0 & 0 \\
        0 & 0 & a(t)^2 r^2 & 0 \\
        0 & 0 & 0 &  a(t)^2 r^2\,\sin^2\theta
    \end{pmatrix}
\end{align}
Since the metric has signature $(-,+,+,+)$, $g_{tt}$ must be strictly negative. We therefore define $g_{tt} = -N(t)^2$, where $N(t)$ is the lapse function. Finally, we define $k\ce -\frac{K}{|c|}$, which casts the $rr$-component into the standard form found in most textbooks:
\begin{align}
    g_{\mu\nu} = 
    \begin{pmatrix}
        -N(t)^2 & 0 & 0 & 0 \\
        0 & \frac{a(t)^2}{1 - k r^2} & 0 & 0 \\
        0 & 0 & a(t)^2 r^2 & 0 \\
        0 & 0 & 0 &  a(t)^2 r^2\,\sin^2\theta
    \end{pmatrix}
\end{align}
This is the well-known FLRW metric, obtained by imposing the symmetry conditions~\eqref{eq:SymmetryConditionsFullSet}. Recall that the vector field $\T_3$, appearing in symmetry condition (iv), depends on two functions $F_r(r)$ and $F_\theta(r)$. In our derivation we found that $F_\theta(r) = F_r(r) = \chi(r) = \sqrt{1 - k\, r^2}$.

As discussed at the end of section~\ref{sec:Symmetries}, if $\T_3$ is a symmetry of a spherically symmetric metric, then $\T_1$ and $\T_2$ must also be symmetries. Both $\T_1$ and $\T_2$ depend on an additional function $F_\phi(r)$, which can be fixed by imposing $\lie_{\T_1} g_{\mu\nu} = 0$. This yields $F_\phi = \chi$. The complete set of generators of translations is therefore
\begin{align}
    \T_1 &= \chi(r)\left(\sin\theta \cos\phi\, \pd_r + \frac1r \cos\theta \cos\phi \, \pd_\theta - \frac1r \frac{\sin\phi}{\sin\theta}\,\pd_\phi \right)\notag\\
    \T_2 &= \chi(r)\left(\sin\theta \sin\phi\, \pd_r + \frac1r \cos\theta \sin\phi \, \pd_\theta + \frac1r \frac{\cos\phi}{\sin\theta}\,\pd_\phi \right) \notag\\
    \T_3 &= \chi(r)\left(\cos\theta\,\pd_r - \frac1r \sin\theta\, \pd_\theta\right)\,.
\end{align}
In one fell swoop, we have derived the most general metric compatible with rotational and translational invariance, together with its translational Killing vector fields. We can now use the generators of rotations and translations to \emph{define} homogeneity and isotropy of other fields than the metric. To do so, we need to impose the symmetry conditions~\eqref{eq:SymmetryConditionsFullSet}. In the next section, we derive the homogeneous and isotropic form of various tensor fields.


\section{Symmetry-Reduced Matter Fields}\label{sec:MatterFields}
In this section, we perform a step-by-step symmetry-reduction of scalar fields, vector fields, $2$-forms, $3$-forms, and symmetric rank-two tensor fields. Hence, we impose the symmetries directly on the fields themselves. In each case, we determine which field configurations are compatible with homogeneity and isotropy.

\subsection{Scalar Fields}\label{ssec:ScalarFields}
We consider a real scalar field $\Phi:\M \to \bbR$ and begin by imposing spherical symmetry. Since condition (i) in~\eqref{eq:SymmetryConditionsFullSet} applies to any tensor field, including scalars, we immediately find that $\Phi$ is independent of the coordinate $\phi$:
\begin{align}\label{eq:PhiIndepOfphi}
    \lie_{\R_3} \Phi = - \pd_\phi \Phi \overset{!}{=} 0\,.
\end{align}
Likewise, the Lie derivatives along the generators $\R_1$ and $\R_2$ imply that $\Phi$ is also independent of $\theta$:
\begin{align}
    \lie_{\R_1} \Phi &= \sin\phi\, \pd_\theta \Phi \overset{!}{=} 0 \notag\\
    \lie_{\R_2} \Phi &= - \cos\phi\, \pd_\theta \Phi \overset{!}{=} 0\,,
\end{align}
where we have used~\eqref{eq:PhiIndepOfphi}. Hence, a spherically symmetric scalar field can depend only on $t$ and $r$.

Next, we impose translational invariance with respect to the generator $\T_3$. The corresponding equation reads
\begin{align}
    \lie_{\T_3} \Phi &= \chi(r) \left(\cos\theta\, \pd_r \Phi - \frac{\sin\theta}{r} \pd_\theta \Phi\right)
    = \chi(r)\, \cos\theta\, \pd_r \Phi \overset{!}{=} 0\,.
\end{align}
Translational invariance therefore requires $\Phi$ to be independent of the radial coordinate as well. Consequently, a scalar field compatible with both homogeneity and isotropy can depend only on time:
\begin{align}
\Phi = \Phi(t)\,.
\end{align}
In other words, in a homogeneous and isotropic Universe, a scalar field must be spatially constant and can vary only with cosmic time.

\subsection{Vector Fields}\label{ssec:VectorFields}
We now turn our attention to real-valued vector fields $A^\mu$. From the first condition in~\eqref{eq:SymmetryConditionsFullSet}, we immediately see that all components of a spherically symmetric vector field must be independent of the coordinate $\phi$:
\begin{align}
    \lie_{\R_3} A^\mu = -\pd_\phi A^\mu \overset{!}{=} 0\,.
\end{align}
The second condition for spherical symmetry reads
\begin{align}
    \cos\phi\,\lie_{\R_1} A^\mu + \sin\phi\, \lie_{\R_2}A^\mu = 
    \begin{pmatrix}
        0 \\
        0 \\
        -A^\phi \\
        \frac{1}{\sin^2\theta} A^\theta
    \end{pmatrix}
    \overset{!}{=} 0\,.
\end{align}
Thus, in addition to being independent of $\phi$, the components $A^t$ and $A^r$ must also be independent of $\theta$. A spherically symmetric vector field therefore takes the general form
\begin{align}
    A^\mu =
    \begin{pmatrix}
        A^t (t,r) \\
        A^r (t, r) \\
        0 \\
        0
    \end{pmatrix}\,.
\end{align}
We now impose translational invariance. As before, it suffices to require that the Lie derivative along the generator $\T_3$ vanishes. Using
\begin{align}
   r\,\chi'(r) - \chi(r) &= -\frac{1}{\chi(r)}\,,
\end{align}
we obtain
\begin{align}
    \lie_{\T_3} A^\mu = 
    \begin{pmatrix}
        \cos\theta\, \chi(r)\, \pd_r A^t \\
        \cos\theta\, \pd_r\left\{\chi(r)\,A^r\right\} \\
        -\frac{\sin\theta}{\chi(r)\,r^2}  A^r \\
        0
    \end{pmatrix}
    \overset{!}{=} 0\,.
\end{align}
The first component implies that $A^t$ must be independent of $r$, while the third component requires $A^r$ to vanish. The second equation is then automatically satisfied. We therefore conclude that a spatially homogeneous and isotropic vector field must have the form
\begin{align}
    A^\mu = 
    \begin{pmatrix}
        A^t(t) \\
        0 \\
        0 \\
        0
    \end{pmatrix}\,.
\end{align}
In other words, homogeneity and isotropy restrict a vector field to have only a nonzero temporal component, which may depend solely on cosmic time.

\subsection{\texorpdfstring{$2$}{2}-Forms}\label{ssec:2Forms}
In this subsection, we perform the symmetry-reduction of the $2$-form $B_{\mu\nu} = -B_{\nu\mu}$, beginning with spherical symmetry. Because of its anti-symmetry, $B_{\mu\nu}$ has only $\frac{4\cdot 3}{2!} = 6$ independent nonzero components in four dimensions. As before, condition (i) in~\eqref{eq:SymmetryConditionsFullSet} implies that all six components must be independent of the coordinate~$\phi$:
\begin{align}
    \lie_{\R_3} B_{\mu\nu} = - \pd_{\phi} B_{\mu\nu} \overset{!}{=} 0\,.
\end{align}
From the second symmetry condition, we learn that four of these six components must vanish for the $2$-form to be compatible with spherical symmetry:
\begin{align}
    \cos\phi\,\lie_{\R_1} B_{\mu\nu} + \sin\phi\, \lie_{\R_2} B_{\mu\nu} =
    \begin{pmatrix}
        0 & 0 & -\frac{1}{\sin^2\theta} B_{t\phi} & B_{t\theta} \\
        0 & 0 & -\frac{1}{\sin^2\theta} B_{r\phi} & B_{r\theta} \\
        \frac{1}{\sin^2\theta} B_{t\phi} & \frac{1}{\sin^2\theta} B_{r\phi} & 0 & 0 \\
        -B_{t\theta} & -B_{r\theta} & 0 & 0
    \end{pmatrix}
    \overset{!}{=} 0\,.
\end{align}
This condition is uniquely solved by
\begin{align}
    B_{t\theta} &= 0\,, & B_{t\phi} &= 0\,, & B_{r\theta} &= 0\,, & B_{r\phi} &= 0\,,
\end{align}
leaving only $B_{tr}$ and $B_{\theta\phi}$ as potentially non-vanishing components. The third condition for spherical symmetry imposes differential constraints on these remaining components:
\begin{align}
    \sin\phi\,\lie_{\R_1} B_{\mu\nu} - \cos\phi \, \lie_{\R_2} B_{\mu\nu} =
    \begin{pmatrix}
        0 & \pd_\theta B_{tr} & 0 & 0 \\
        -\pd_\theta B_{tr} & 0 & 0 & 0 \\
        0 & 0 & 0 & \pd_\theta B_{\theta\phi} - \cot\theta B_{\theta\phi} \\
        0 & 0 & -\pd_\theta B_{\theta\phi} + \cot\theta B_{\theta\phi} & 0
    \end{pmatrix}
    \overset{!}{=} 0\,.
\end{align}
We find that $B_{tr}$ must be independent of $\theta$, while the equation for $B_{\theta\phi}$ can be solved by separation of variables. A straightforward computation gives
\begin{align}
    B_{\theta\phi} = \sin\theta\, c(t, r)\,,
\end{align}
where $c(t, r)$ is an arbitrary function of $t$ and $r$, since $B_{\theta\phi}$ is already known to be independent of~$\phi$.

Having imposed all conditions for spherical symmetry, we conclude that a spherically symmetric $2$-form must take the form
\begin{align}
    B_{\mu\nu} = 
    \begin{pmatrix}
        0 & B_{tr}(t, r) & 0 & 0 \\
        -B_{tr}(t, r) & 0 & 0 & 0 \\
        0 & 0 & 0 & \sin\theta\, c(t, r) \\
        0 & 0 & -\sin\theta\, c(t, r) & 0
    \end{pmatrix}\,.
\end{align}
We now impose translational invariance, which yields four independent conditions:
\begin{align}
    \lie_{\T_3} B_{\mu\nu} =
    \begin{pmatrix}
        0 &   \cos\theta\,\pd_r \left[\chi B_{tr}\right]  & -\sin\theta \,\chi \, B_{tr} & 0 \\
        -\cos\theta\,\pd_r \left[\chi B_{tr}\right] & 0 & 0 & \frac{\sin^2\theta}{r^2\,\chi(r)} c(t,r) \\
        \sin\theta \,\chi \, B_{tr} & 0 & 0 & - \frac{2 \sin\theta \cos\theta}{r}c(t, r) \\
        0 & -\frac{\sin^2\theta}{r^2\,\chi(r)} c(t,r) & \frac{2 \sin\theta \cos\theta}{r}c(t, r) & 0
    \end{pmatrix}
    \overset{!}{=} 0\,.
\end{align}
The second equation in the first row requires $B_{tr}$ to vanish, which also renders the first equation in that row automatically satisfied. The second equation in the second row then implies $c(t, r) = 0$.

Therefore, the only $2$-form compatible with both homogeneity and isotropy is the trivial one:
\begin{align}
    B_{\mu\nu} = 0\,.
\end{align}

\subsection{\texorpdfstring{$3$-Forms}{3-Forms}}\label{ssec:3Forms}
After finding that the only $2$-form compatible with homogeneity and isotropy is the trivial one, it may seem futile to search for nonzero $3$-forms satisfying the same symmetries. Surprisingly, however, nontrivial homogeneous and isotropic $3$-forms do exist.

We consider the totally anti-symmetric tensor $K_{\alpha\mu\nu}$ as our $3$-form. Owing to its anti-symmetry, a $3$-form in four dimensions has only $\frac{4\cdot3\cdot2}{3!} = 4$ independent components. In $(t, r, \theta, \phi)$ coordinates, these are
\begin{align}
    &K_{tr\theta}\,, & &K_{tr\phi}\,, & &K_{t\theta\phi}\,, & &K_{r\theta\phi}\,.
\end{align}
The first symmetry condition in~\eqref{eq:SymmetryConditionsFullSet} requires that all components be independent of the coordinate~$\phi$. From the second condition we find that
\begin{align}
    K_{tr\theta} &= 0 &\text{and} && K_{tr\phi} &= 0\,.
\end{align}
leaving only two potentially non-vanishing components. The third symmetry condition provides two first-order differential equations:
\begin{align}
    \pd_\theta K_{t\theta\phi} - \cot\theta\, K_{t\theta\phi} &= 0 &\text{and} && \pd_\theta K_{r\theta\phi} - \cot\theta\, K_{r\theta\phi} &= 0\,.
\end{align}
These equations integrate straightforwardly to
\begin{align}\label{eq:KSols}
    K_{t\theta\phi} &= \sin\theta\, c_1(t, r) &\text{and} && K_{r\theta\phi} &= \sin\theta\, c_2(t, r)\,,
\end{align}
where $c_1$ and $c_2$ are arbitrary functions of $t$ and $r$. Hence, a spherically symmetric $3$-form can have only two nonzero components, given by~\eqref{eq:KSols}.

We now impose translational invariance. This yields two independent conditions for $c_1$ and~$c_2$:
\begin{align}
    \frac{\sin^2\theta}{r^2} \left(\chi(r) - r\, \chi'(r)\right) c_1(t,r) &= 0 \notag\\
    \frac{\cos\theta\sin\theta}{r} \left(r\, c_2(t,r)\,\chi'(r) + \chi(r)\left\{r\,\pd_r c_2(t,r) - 2 c_2(t,r)\right\}\right) &= 0\,.
\end{align}
The first equation is uniquely solved by $c_1(t,r) = 0$, which sets $K_{t\theta\phi}$ to zero. The second equation, however, admits a nontrivial solution. Solving for $\pd_r c_2(t,r)$ and integrating gives
\begin{align}
    \pd_r c_2(t,r) &= c_2(t,r)\frac{2 \chi(r)- r\, \chi'(r)}{r \chi(r)} &\Longrightarrow && c_2(t,r) &= \frac{r^2\, c_3(t)}{\chi(r)}\,,
\end{align}
where $c_3(t)$ is an arbitrary integration function. We therefore conclude that a homogeneous and isotropic $3$-form has a single non-vanishing component of the form
\begin{align}
    K_{r\theta\phi} = \sin\theta\,\frac{r^2\, c_3(t)}{\chi(r)}\,.
\end{align}
Unlike the scalar field, the vector field, and the $2$-form, a homogeneous and isotropic $3$-form depends on cosmic time $t$, the radial coordinate $r$, and the angle $\theta$.\\

This approach can be applied to any symmetric or antisymmetric tensor of arbitrary rank (see for instance its successful application to the connection within metric affine theories \cite{DAmbrosio:2021b}). In this way, one obtains restricted field configurations, since the symmetries are imposed directly on the fields themselves rather than, for example, on their energy–momentum tensors, as illustrated in section \ref{sec:SymmetryReducedT}.

\subsection{Homogeneous and Isotropic Form of the Energy-Momentum Tensor}
To determine the most general form of an energy-momentum tensor compatible with homogeneity and isotropy, it is not necessary to impose the full set of symmetry conditions~\eqref{eq:SymmetryConditionsFullSet}. The derivation can be simplified by noting that, when we obtained the spherically symmetric form of the metric, we never used any of its special properties other than its symmetry $g_{\mu\nu} = g_{\nu\mu}$. Consequently, a spherically symmetric energy-momentum tensor must take the same general form:
\begin{align}
    T_{\mu\nu} = 
    \begin{pmatrix}
        T_{tt}(t,r) & T_{tr}(t,r) & 0 & 0 \\
        T_{tr}(t,r) & T_{rr}(t,r) & 0 & 0 \\
        0 & 0 & T_{\theta\theta}(t,r) & 0 \\
        0 & 0 & 0 & T_{\theta\theta}(t,r)\sin^2\theta 
    \end{pmatrix}\,.
\end{align}
With this ansatz, the first three symmetry conditions (i)--(iii) of~\eqref{eq:SymmetryConditionsFullSet} are automatically satisfied. The remaining step is to impose invariance under translations generated by $\T_3$. A straightforward computation shows that this requirement restricts the tensor to the form
\begin{align}
    T_{\mu\nu} = 
    \begin{pmatrix}
        T_{tt}(t) & 0 & 0 & 0 \\
        0 & \frac{b(t)}{\chi(r)^2} & 0 & 0 \\
        0 & 0 & r^2\, b(t) & 0 \\
        0 & 0 & 0 & r^2 \sin^2\theta\, b(t)
    \end{pmatrix}\,,
\end{align}
where $b(t)$ is an arbitrary function of time and $\chi(r) \ce \sqrt{1 - k r^2}$. For later convenience, it is useful to raise one index using the FLRW metric. This yields
\begin{align}
    T\ud{\mu}{\nu} =
    \begin{pmatrix}
        - T_{tt}(t) & 0 & 0 & 0 \\
        0 & \frac{b(t)}{a(t)^2} & 0 & 0 \\
        0 & 0 & \frac{b(t)}{a(t)^2} & 0 \\
        0 & 0 & 0 & \frac{b(t)}{a(t)^2}
    \end{pmatrix}\,.
\end{align}
This tensor is structurally identical to that of a perfect fluid. To make this correspondence explicit, we relabel $T_{tt}(t)$ as the energy density $\rho(t)$ and define
\begin{align}
    b(t) = a(t)^2\, p(t)\,,
\end{align}
where $p(t)$ is an arbitrary function of time representing the pressure of an ideal fluid. With these relabelings, the most general homogeneous and isotropic energy-momentum tensor takes the familiar form
\begin{align}\label{eq:PerfectFluidEMT}
    T\ud{\mu}{\nu} =
    \begin{pmatrix}
        -\rho(t) & 0 & 0 & 0 \\
        0 & p(t) & 0 & 0 \\
        0 & 0 & p(t) & 0 \\
        0 & 0 & 0 & p(t)
    \end{pmatrix}\,.
\end{align}
Thus, starting purely from the symmetry conditions~\eqref{eq:SymmetryConditionsFullSet}, we recover the standard cosmological energy-momentum tensor of a perfect fluid, characterized by a time-dependent energy density and isotropic pressure.

\section{Imposing Symmetries on the Energy-Momentum Tensor}\label{sec:SymmetryReducedT}
In Section~\ref{sec:Symmetries} we introduced two intuitive notions. First, if the metric is homogeneous and isotropic, then the Einstein tensor must share these symmetries. Second, if both the metric $g_{\mu\nu}$ and the matter field $\Psi\ud{\bullet}{\circ}$ are homogeneous and isotropic, then the same must hold for the energy-momentum tensor $T_{\mu\nu}$.

In subsection~\ref{ssec:TProposition} we prove a more general version of these intuitive claims. In fact, it is true that the Einstein tensor inherits \emph{any} symmetry of the metric. Similarly, the Hilbert energy-momentum tensor inherits any \emph{shared} symmetry of the metric and the matter fields from which it is constructed. 

The converse statements, however, are not necessarily true: the metric and the matter fields do not necessarily share the symmetries of the Einstein tensor and the Hilbert energy-momentum tensor. In fact, in subsections~\ref{ssec:MaxwellEMT} and~\ref{ssec:T2Form} we construct two explicit counter-examples, where $\Psi\ud{\bullet}{\circ}$ is neither homogeneous nor isotropic, despite $T_{\mu\nu}$ and $g_{\mu\nu}$ having these properties.

We establish this result by deriving the Hilbert energy-momentum tensor of three different field theories of increasing complexity in the subsection~\ref{ssec:KleinGordon}, \ref{ssec:MaxwellEMT}, and~\ref{ssec:T2Form}. After imposing that the resulting $T_{\mu\nu}$ be homogeneous and isotropic, we investigate what the consequences are for the underlying matter fields. In two out of three cases, we can prove that the matter fields are neither homogeneous nor isotropic.

Notice the conceptual difference to our analysis in section~\ref{sec:MatterFields}. There, we investigated the constraints that homogeneity and isotropy impose on matter fields, completely independently of any dynamics. In that sense, the results of section~\ref{sec:MatterFields} are purely kinematical. In this section, however, we need to specify a matter Lagrangian, from which the Hilbert energy-momentum tensor can be derived. Thus, dynamical aspects of matter fields enter indirectly into the considerations. 

What these results suggest, is that there are two inequivalent ways of implementing the Cosmological Principle. It can either be imposed on the spacetime model $(\M, g_{\mu\nu}, \Psi\ud{\bullet}{\circ})$, as we did in~\ref{sec:MatterFields}, or it can be imposed on $(\M, g_{\mu\nu}, T_{\mu\nu})$. The latter alternative is equally natural as the first one.

\subsection{A Proposition on the Energy-Momentum Tensor of Symmetry-Reduced Matter Fields}\label{ssec:TProposition}
Let $g_{\mu\nu}$ be a non-degenerate metric of Lorentzian signature. The Einstein tensor $G_{\mu\nu}$ is constructed from this metric, its inverse, and its first- and second-order derivatives. We now show that if $g_{\mu\nu}$ possesses any kind of continuous symmetry, generated by a vector field $\xi$, the Einstein tensor necessarily possesses the same symmetry.
\begin{proposition}\label{prop:Gmunu}
    If the Lie derivative of the metric $g_{\mu\nu}$ vanishes along a generator $\xi$, i.e.
    \begin{align}
        \lie_\xi g_{\mu\nu} = 0\,,
    \end{align}
    then the Einstein tensor $G_{\mu\nu} \ce R_{\mu\nu} - \frac{1}{2} R g_{\mu\nu}$ also satisfies
    \begin{align}
        \lie_\xi G_{\mu\nu} = 0\,.
    \end{align}
    In other words, the Einstein tensor inherits the symmetries of the metric.
\end{proposition}

\paragraph{Proof.}
We begin with the Palatini identity, which relates the Lie derivative of the Ricci tensor to that of the Christoffel symbols\footnote{For a derivation of the Palatini identity and the Lie derivative of the Christoffel symbols of the metric see appendices~\ref{app:A_LieDChristoffel} and~\ref{app:A_PalatiniIdentity}.}:
\begin{align}\label{eq:PalatiniId}
    \lie_\xi R_{\mu\nu} = \nabla_\lambda \left(\lie_\xi \Gamma\ud{\lambda}{\mu\nu}\right) - \nabla_\nu\left(\lie_\xi \Gamma\ud{\lambda}{\lambda\mu}\right)\,.
\end{align}
The Lie derivative of the Christoffel symbols can in turn be expressed as
\begin{align}\label{eq:LieGamma}
    \lie_\xi \Gamma\ud{\alpha}{\mu\nu} = \frac12 g^{\alpha\lambda} \left(\nabla_\mu \left(\lie_\xi g_{\nu\lambda}\right) + \nabla_\nu \left(\lie_\xi g_{\mu\lambda}\right) - \nabla_\lambda\left(\lie_\xi g_{\mu\nu}\right)\right)\,.
\end{align}
Combining~\eqref{eq:PalatiniId} and~\eqref{eq:LieGamma} yields
\begin{align}\label{eq:LieDRmunu}
    \lie_\xi R_{\mu\nu} = \frac12 g^{\alpha\beta}\left[\nabla_\beta\nabla_\mu \left(\lie_\xi g_{\alpha\nu}\right) + \nabla_\beta\nabla_\nu \left(\lie_\xi g_{\alpha\mu}\right) - \nabla_\beta\nabla_\alpha\left(\lie_\xi g_{\mu\nu}\right) - \nabla_\nu \nabla_\mu\left(\lie_\xi g_{\alpha\beta}\right)\right]\,.
\end{align}
Hence, if $\lie_\xi g_{\mu\nu} = 0$, it immediately follows that $\lie_\xi R_{\mu\nu} = 0$. The Ricci scalar also satisfies $\lie_\xi R = 0$, since
\begin{align}
    \lie_\xi R = \lie_\xi \left(g^{\mu\nu} R_{\mu\nu}\right) = \left(\lie_\xi g^{\mu\nu}\right) R_{\mu\nu} + g^{\mu\nu}\left(\lie_\xi R_{\mu\nu}\right)\,,
\end{align}
Finally, taking the Lie derivative of $G_{\mu\nu} = R_{\mu\nu} - \tfrac{1}{2} R g_{\mu\nu}$ gives
\begin{align}
    \lie_\xi G_{\mu\nu} = \lie_\xi R_{\mu\nu} - \frac12 R \lie_\xi g_{\mu\nu} - \frac12 g_{\mu\nu} \lie_\xi R\,,
\end{align}
which vanishes if $\lie_\xi g_{\mu\nu} = 0$. Thus, the Einstein tensor indeed inherits the symmetries of the metric.
\hfill $\Box$ \bigskip

We now turn to the analogous result for the energy-momentum tensor of a matter field.
\begin{proposition}\label{prop:Tmunu}
    Let $\Psi\ud{\bullet}{\circ}$ be a minimally coupled matter field with Lagrangian
    \begin{align*}
        L = L(\Psi, \nabla\Psi, g)\,,
    \end{align*}
    where $\nabla_\alpha$ is the unique torsion-free and metric-compatible covariant derivative associated with $g_{\mu\nu}$. If both the metric and the field share a continuous symmetry generated by $\xi$, that is,
    \begin{align*}
        \lie_\xi g_{\mu\nu} &= 0 &\text{and} && \lie_\xi \Psi\ud{\bullet}{\circ} = 0\,,
    \end{align*}
    then the Hilbert energy-momentum tensor satisfies
    \begin{align*}
        \lie_\xi T_{\mu\nu} = 0\,.
    \end{align*}
    In other words, the energy-momentum tensor inherits all \emph{common} symmetries of the metric and the matter field.
\end{proposition}

\paragraph{Proof.}
To verify the claim of the proposition, we need the commutator of the Lie derivative with the covariant derivative. As we show in appendix~\ref{app:A_Commutator}, given any torsion-free connection $\bar{\nabla}_\alpha$ (which is not necessarily metric-compatible) with Christoffel symbols $\bar{\Gamma}\ud{\alpha}{\mu\nu}$, the commutator of $\lie_\xi$ and $\bar{\nabla}_\alpha$ is given by
\begin{align}\label{eq:CommutatorLieNabla}
    [\lie_\xi, \bar{\nabla}_\alpha]\Psi\ud{\bullet}{\circ} &= \PD{\left(\bar{\nabla}_\alpha \Psi\ud{\bullet}{\circ}\right)}{\bar{\Gamma}\ud{\lambda}{\mu\nu}}\lie_\xi \bar{\Gamma}\ud{\lambda}{\mu\nu}\,.
\end{align}
If $\bar{\nabla}_\alpha$ is chosen to be the Levi-Civita connection of $g_{\mu\nu}$, then $\lie_\xi g_{\mu\nu}=0$ implies $\lie_\xi \Gamma\ud{\alpha}{\mu\nu}=0$, and the commutator reduces to
\begin{align}
    [\lie_\xi, \nabla_\alpha]\Psi\ud{\bullet}{\circ} &=  0\,,
\end{align}
Expanding the commutator and using $\lie_\xi \Psi\ud{\bullet}{\circ} = 0$ gives
\begin{align}\label{eq:LieDCovDEq0}
    \lie_\xi \left(\nabla_\alpha \Psi\ud{\bullet}{\circ}\right) = 0\,.
\end{align}
Thus, if $g_{\mu\nu}$ and $\Psi\ud{\bullet}{\circ}$ share a symmetry generated by $\xi$, then $\nabla_\alpha \Psi\ud{\bullet}{\circ}$ also shares it. 

The Hilbert energy-momentum tensor is defined by
\begin{align}\label{eq:HilbertSEM}
    T_{\mu\nu} = - \frac{2}{\sqrt{-g}}\frac{\delta\left(\sqrt{-g} L\right)}{\delta g^{\mu\nu}} = -2 \PD{L}{g^{\mu\nu}} + g_{\mu\nu}\,L\,.
\end{align}
Taking the Lie derivative and applying the chain rule gives
\begin{align}
    \lie_\xi T_{\mu\nu} &= -2\left(\PD{^2L}{g^{\mu\nu} \pd\Psi\ud{\bullet}{\circ}} \lie_\xi \Psi\ud{\bullet}{\circ} + \PD{^2 L}{g^{\mu\nu} \pd(\nabla_\alpha\Psi\ud{\bullet}{\circ})}\lie_\xi (\nabla_\alpha \Psi\ud{\bullet}{\circ}) + \PD{^2 L}{g^{\mu\nu}\pd g^{\alpha\beta}}\lie_\xi g_{\alpha\beta}\right) \notag\\
    &\phantom{=-}+(\lie_\xi g_{\mu\nu}) L + g_{\mu\nu}\left(\PD{L}{\Psi\ud{\bullet}{\circ}}(\lie_\xi \Psi\ud{\bullet}{\circ}) + \PD{L}{(\nabla_\alpha\Psi\ud{\bullet}{\circ})}(\lie_\xi(\nabla_\alpha\Psi\ud{\bullet}{\circ})) + \PD{L}{g^{\alpha\beta}} \lie_\xi g_{\alpha\beta}\right)\notag\\
    &= 0\,.
\end{align}
Each term vanishes because $\lie_\xi g_{\mu\nu}=0$, $\lie_\xi\Psi\ud{\bullet}{\circ}=0$, and $\lie_\xi(\nabla_\alpha\Psi\ud{\bullet}{\circ})=0$ by~\eqref{eq:LieDCovDEq0}. Hence, $\lie_\xi T_{\mu\nu}=0$, as claimed. \hfill $\Box$\bigskip

In the next section, we will impose homogeneity and isotropy on both the metric and the energy-momentum tensor and explicitly demonstrate that matter fields can nevertheless violate these symmetries. Formally, we will show that
\begin{align}
    \lie_\xi T_{\mu\nu} &= 0 & \text{and} && \lie_\xi g_{\mu\nu} &= 0 & \centernot\Longrightarrow && \lie_\xi \Psi\ud{\bullet}{\circ} &= 0\,.
\end{align}
Examining the proof of Proposition~\ref{prop:Tmunu}, one is naturally led to the converse consideration: even if the energy-momentum tensor and the matter field share a symmetry, this does not guarantee that the metric inherits it. In formal terms,
\begin{align}
    \lie_\xi T_{\mu\nu} &= 0 & \text{and} && \lie_\xi \Psi\ud{\bullet}{\circ} &= 0 & \centernot\Longrightarrow && \lie_\xi g_{\mu\nu} &= 0\,.
\end{align}
We will not pursue this possibility further here. However, we regard the following scenario as conceptually important and worthy of detailed future investigation:

Rather than imposing homogeneity and isotropy on the metric and the matter fields, one could instead impose these symmetries directly on the energy-momentum tensor and on the Einstein tensor. The rationale behind this alternative approach is rooted in the Cosmological Principle, which is fundamentally a statement about the large-scale matter distribution of the Universe. Information about this distribution enters the Einstein field equations only through the energy-momentum tensor $T_{\mu\nu}$, not through the matter fields themselves. It is therefore natural to require homogeneity and isotropy of $T_{\mu\nu}$ rather than of $\Psi\ud{\bullet}{\circ}$.

This approach has the additional advantage that the symmetry conditions are imposed on a gauge-\emph{independent} tensor.\footnote{This assumes that one uses the Hilbert energy-momentum tensor or, equivalently, applies the Belinfante–Rosenfeld procedure to symmetrize and improve the canonical (Noether) tensor.} The matter fields $\Psi\ud{\bullet}{\circ}$, by contrast, are typically gauge-\emph{dependent} quantities. Imposing symmetries directly on them is therefore questionable: a gauge transformation can, in general, destroy these symmetries without altering the underlying physics.

The Cosmological Principle states that the matter distribution on large scales is homogeneous and isotropic. It therefore allows us to disregard microscopic details and focus on the coarse-grained, averaged picture. The energy-momentum tensor provides precisely such a picture. In fact, a homogeneous and isotropic energy-momentum tensor is described by an energy-density and an isotropic pressure--both are macroscopic quantities! At the microscopic level, the matter field do not need to satisfy these symmetries. In fact, matter forms observable clumps, if we restrict ourselves to small scales. In the next subsection, we will indeed show that matter fields can be inhomogeneous and anisotropic, while still giving rise to an energy-momentum tensor with these properties. 
If we now choose to impose the Cosmological Principle on $T_{\mu\nu}$, we have to impose it on the Einstein tensor as well. If we do not demand the same symmetries for $G_{\mu\nu}$, we obtain an inconsistent theory because
\begin{align}
    G_{\mu\nu} &= 8\pi\, T_{\mu\nu} &\text{but}&&  \lie_\xi G_{\mu\nu} \neq 0 = 8\pi\, \lie_\xi T_{\mu\nu}\,.
\end{align}
An interesting and important question now arises: If $T_{\mu\nu}$ and $G_{\mu\nu}$ are both homogeneous and isotropic, can the underlying matter degrees of freedom and the metric break these symmetries? If this is possible, we could not dynamically distinguish between the FLRW metric and a more general metric which produces a homogeneous and isotropic Einstein tensor. The field equations are blind to such modifications. However, kinematically there would be observable or measurable differences. The metric defines quasi-local observables such as proper time, for instance. It also affects anything moving through spacetime. The geodesic equations as well as the proper time computed from the FLRW metric would differ from their counterparts determined by an anisotropic and inhomogeneous metric. 
Solving the condition that the Lie derivative of the Einstein tensor vanishes in order to determine the allowed anisotropic and inhomogeneous metrics is computationally very cumbersome. A natural starting point is to consider the pure vacuum case, 
$T_{\mu\nu}=0$, and impose homogeneity and isotropy directly on $G_{\mu\nu}$. The remaining question is whether the metric 
$g_{\mu\nu}$ necessarily shares these symmetries or may violate them. Due to the algebraic complexity of the Einstein tensor, we were not able to construct a concrete example in which such a violation occurs. We view the construction of such an example as an interesting open problem for future work.

In the following sections, we impose the homogeneity and isotropy conditions at the level of 
$T_{\mu\nu}$, rather than on the fields themselves, and investigate which matter field configurations are admissible. For the spacetime geometry, we assume the FLRW metric.

\subsection{Energy-Momentum Tensor for the Klein-Gordon Field}\label{ssec:KleinGordon}
The Klein-Gordon Lagrangian density is given by
\begin{equation}
    L = - \sqrt{-g}\left(\frac12 g^{\mu\nu}\pd_\mu \Phi \pd_\nu \Phi + V(\Phi)\right)\,,
\end{equation}
where $V(\Phi)$ denotes an arbitrary potential. Using the definition~\eqref{eq:HilbertSEM} and raising one index, we obtain the corresponding Hilbert energy-momentum tensor,
\begin{equation}
    T\ud{\mu}{\nu} = -\frac12 \delta\ud{\mu}{\nu} \left(g^{\alpha\beta}\pd_\alpha\Phi \pd_\beta\Phi + 2 V(\Phi)\right)  + g^{\mu\alpha}  \pd_\alpha\Phi \pd_\nu\Phi\,.
\end{equation}
Substituting the FLRW metric, the components of this tensor can be written explicitly as
\begin{align}
    T\ud{\mu}{\nu} &=
    \begin{pmatrix}
        -E & -\tau_{tr} & -\tau_{t\theta} & -\tau_{t\phi} \\
        \left(\frac{\chi(r)}{a(t)}\right)^2 \tau_{tr} &   \left(\frac{\chi(r)}{a(t)}\right)^2 (\pd_r\Phi)^2 -E & \left(\frac{\chi(r)}{a(t)}\right)^2 \tau_{r\theta} & \left(\frac{\chi(r)}{a(t)}\right)^2 \tau_{r\phi} \\
        \frac{1}{a(t)^2 r^2} \tau_{t\theta} & \frac{1}{a(t)^2 r^2} \tau_{r\theta} &  \frac{1}{a(t)^2 r^2} (\pd_\theta\Phi)^2 - E & \frac{1}{a(t)^2 r^2} \tau_{\theta\phi} \\
        \frac{1}{a(t)^2 r^2\sin^2\theta} \tau_{t\phi} & \frac{1}{a(t)^2 r^2\sin^2\theta} \tau_{r\phi} & \frac{1}{a(t)^2 r^2\sin^2\theta} \tau_{\theta\phi} &  \frac{1}{a(t)^2 r^2 \sin^2\theta} (\pd_\phi\Phi)^2 -E
    \end{pmatrix}\,.
\end{align}
where, for compactness, we defined
\begin{align}
    E &\ce \frac12(\pd_t\Phi)^2 + \frac{1}{2a(t)^2} \left(\chi(r)^2(\pd_r \Phi)^2 + \frac{(\pd_\theta \Phi)^2}{r^2} + \frac{(\pd_\phi \Phi)^2}{r^2 \sin^2\theta}\right) + V(\Phi)
\end{align}
and introduced the shorthand notation
\begin{align}
    \tau_{tr} &\ce \pd_t\Phi \pd_r\Phi & \tau_{r\theta} &\ce \pd_{r}\Phi \pd_{\theta} \Phi & \tau_{\theta\phi} &\ce \pd_\theta\Phi \pd_\phi \Phi  \notag\\
    \tau_{t\theta} &\ce \pd_t\Phi \pd_\theta\Phi & \tau_{r\phi} &\ce  \pd_{r}\Phi \pd_{\phi} \Phi\notag\\
    \tau_{t\phi} &\ce \pd_t\Phi \pd_\phi\Phi\,.
\end{align}
We now demand that $T\ud{\mu}{\nu}$ take the form of a perfect-fluid energy-momentum tensor. This requires all off-diagonal components to vanish and the spatial diagonal components to be equal, i.e., $T\ud{r}{r} = T\ud{\theta}{\theta} = T\ud{\phi}{\phi}$.

Observe that each $\tau_{t i}$ ($i = r, \theta, \phi$) contains $\pd_t\Phi$ as a factor. To eliminate these terms, we first assume $\pd_t\Phi = 0$. The tensor then simplifies to
\begin{align}
    T\ud{\mu}{\nu} &= 
    \begingroup
    \setlength\arraycolsep{7pt}
    \begin{pmatrix}
        -E & 0 & 0 & 0 \\
        0 & \left(\frac{\chi(r)}{a(t)}\right)^2 (\pd_r\Phi)^2 -E & \left(\frac{\chi(r)}{a(t)}\right)^2 \tau_{r\theta} & \left(\frac{\chi(r)}{a(t)}\right)^2 \tau_{r\phi} \\
        0 & \frac{1}{a(t)^2 r^2} \tau_{r\theta} &  \frac{1}{a(t)^2 r^2} (\pd_\theta\Phi)^2 - E & \frac{1}{a(t)^2 r^2} \tau_{\theta\phi} \\
        0 & \frac{1}{a(t)^2 r^2\sin^2\theta} \tau_{r\phi} & \frac{1}{a(t)^2 r^2\sin^2\theta} \tau_{\theta\phi} &  \frac{1}{a(t)^2 r^2 \sin^2\theta} (\pd_\phi\Phi)^2 -E
    \end{pmatrix}\,.
    \endgroup
\end{align}
To remove the remaining off-diagonal components, two of the three spatial derivatives $\pd_r\Phi$, $\pd_\theta\Phi$, and $\pd_\phi\Phi$ must vanish. However, equality of the diagonal spatial components then forces the remaining derivative to vanish as well. Hence, all derivatives of $\Phi$ must be zero, leaving
\begin{align}
    \Phi = \text{const.}
\end{align}
and reducing the energy-momentum tensor to
\begin{align}
    T\ud{\mu}{\nu} &= 
    \begin{pmatrix}
    -V(\Phi) & 0 & 0 & 0 \\
    0 & -V(\Phi) & 0 & 0 \\
    0 & 0 & -V(\Phi) & 0 \\
    0 & 0 & 0 & -V(\Phi)
    \end{pmatrix}\,.
\end{align}
We now relax the assumption $\pd_t\Phi = 0$ and instead require that two of the three spatial derivatives vanish. The condition $T\ud{r}{r} = T\ud{\theta}{\theta} = T\ud{\phi}{\phi}$ again enforces that \emph{all} spatial derivatives must vanish. This automatically removes all off-diagonal components. The resulting energy-momentum tensor becomes
\begin{align}
    T\ud{\mu}{\nu} &=
    \begin{pmatrix}
    -\frac12 (\pd_t\Phi)^2 -V(\Phi) & 0 & 0 & 0 \\
    0 & \frac12 (\pd_t\Phi)^2-V(\Phi) & 0 & 0 \\
    0 & 0 & \frac12 (\pd_t\Phi)^2-V(\Phi) & 0 \\
    0 & 0 & 0 & \frac12 (\pd_t\Phi)^2-V(\Phi)
    \end{pmatrix}\,,
\end{align}
where $\Phi = \Phi(t)$. The previous result is recovered as a special case when $\Phi$ is constant. No other configurations yield a homogeneous and isotropic energy-momentum tensor. We therefore conclude that, for the Klein-Gordon field, the symmetry-reduction of the energy-momentum tensor enforces precisely the same constraints on $\Phi$ as the direct symmetry-reduction of the scalar field itself, which we performed in subsection~\ref{ssec:ScalarFields}, if we assume a simple Lagrangian for the scalar field.

\subsection{Energy-Momentum Tensor for the Maxwell Field}\label{ssec:MaxwellEMT}
In this subsection we consider electromagnetism in curved spacetime, described by the field strength tensor
\begin{align}
    F_{\mu\nu} \ce \pd_\mu A_\nu - \pd_\nu A_\mu
\end{align}
and the Lagrangian density
\begin{align}
    L = -\frac14\sqrt{-g} g^{\alpha\mu} g^{\beta\nu} F_{\alpha\beta}F_{\mu\nu}\,.
\end{align}
From equation~\eqref{eq:HilbertSEM}, this gives rise to the well-known Hilbert energy-momentum tensor
\begin{align}
    T\ud{\mu}{\nu} &= g^{\mu\rho}g^{\lambda\sigma}F_{\rho\sigma} F_{\nu\lambda} - \frac14 \delta\ud{\mu}{\nu} g^{\alpha\rho} g^{\beta\sigma} F_{\alpha\beta} F_{\rho\sigma}\,.
\end{align}
As in the case of the Klein-Gordon field, we now substitute the FLRW metric for $g_{\mu\nu}$. The components of the energy-momentum tensor then take the form
\begin{align}\label{eq:EMSEMinComp}
    T\ud{\mu}{\nu} &=
    \begingroup
    \setlength\arraycolsep{7pt}
    \begin{pmatrix}
        -E & -\tau_{tr} & -\tau_{t\theta} & -\tau_{t\phi} \\
        \left(\frac{\chi(r)}{a(t)}\right)^2 \tau_{tr} & p_r & \left(\frac{\chi(r)}{a(t)}\right)^2 \tau_{r\theta} & \left(\frac{\chi(r)}{a(t)}\right)^2 \tau_{r\phi} \\
        \frac{1}{a(t)^2 r^2} \tau_{t\theta} & \frac{1}{a(t)^2 r^2} \tau_{r\theta} &  p_\theta & \frac{1}{a(t)^2 r^2} \tau_{\theta\phi} \\
        \frac{1}{a(t)^2 r^2\sin^2\theta} \tau_{t\phi} & \frac{1}{a(t)^2 r^2\sin^2\theta} \tau_{r\phi} & \frac{1}{a(t)^2 r^2\sin^2\theta} \tau_{\theta\phi} &  p_\phi
    \end{pmatrix}\,,
    \endgroup
\end{align}
where the following abbreviations have been introduced:
\begin{align}
    E &\ce \frac{1}{2a(t)^2 r^2}\left[(F_{tr})^2 r^2 \chi(r)^2 + (F_{t\theta})^2 + \frac{1}{\sin^2\theta}(F_{t\phi})^2 + \left(\frac{\chi(r)}{a(t)}\right)^2\left\{ (F_{r\theta})^2 + \frac{1}{\sin^2\theta} (F_{r\phi})^2\right\}\right] \notag\\
    p_r &\ce \frac{1}{a(t)^2r^2}\left[(F_{t\theta})^2+\frac{1}{\sin^2\theta}(F_{t\phi})^2 + \left(\frac{\chi(r)}{a(t)}\right)^2\left\{(F_{r\theta})^2 + \frac{1}{\sin^2\theta}(F_{r\phi})^2\right\}\right]-E \notag\\
    p_\theta &\ce E - \frac{1}{a(t)^2 r^2}\left[\frac{1}{\sin^2\theta}\left(\frac{\chi(r)}{a(t)}\right)^2 (F_{r\phi})^2 + (F_{t\theta})^2\right] \notag\\
    p_\phi &\ce E - \frac{1}{a(t)^2 r^2}\left[\frac{1}{\sin^2\theta}(F_{t\phi})^2 + \left(\frac{\chi(r)}{a(t)}\right)^2 (F_{r\theta})^2\right]
\end{align}
and
\begin{align}\label{eq:DefTauEM}
    \tau_{tr} &\ce \frac{1}{a(t)^2 r^2}\left(F_{t\theta}F_{r\theta} + \frac{1}{\sin^2\theta} F_{t\phi} F_{r\phi}\right) & \tau_{r\theta} &\ce \frac{1}{\left(a(t) r \sin\theta\right)^2}F_{r\phi} F_{\theta\phi} - F_{tr}F_{t\theta} \notag\\
    \tau_{t\theta} &\ce \frac{1}{a(t)^2 r^2}\left(\frac{1}{\sin^2\theta} F_{t\phi} F_{\theta\phi} - F_{tr} F_{r\theta}\,r^2 \chi(r)^2\right) & \tau_{r\phi} &\ce -\left(\frac{1}{a(t)^2 r^2}F_{r\theta} F_{\theta\phi} + F_{tr} F_{t\phi}\right)  \notag\\
    \tau_{t\phi} &\ce -\frac{1}{a(t)^2 r^2}\left(F_{t\theta} F_{\theta\phi} + F_{tr} F_{r\phi}\, r^2 \chi(r)^2\right) & \tau_{\theta\phi} &\ce \left(\frac{\chi(r)}{a(t)}\right)^2 F_{r\theta}F_{r\phi} - F_{t\theta}F_{t\phi}\,.
\end{align}

\paragraph{Homogeneity and isotropy conditions.}
To ensure that $T\ud{\mu}{\nu}$ is compatible with spatial homogeneity and isotropy, we must first require that all off-diagonal components vanish. This gives the conditions
\begin{align}
    \tau_{tr} &= 0\,, & \tau_{t\theta} &= 0\,, & \tau_{t\phi} &= 0\,,\notag\\
    \tau_{r\theta} &= 0\,, & \tau_{r\phi} &= 0\,,\notag\\
    \tau_{\theta\phi} &= 0\,.
\end{align}
Inspection of~\eqref{eq:DefTauEM} shows that each $\tau$ consists of a sum of two quadratic terms in the components of $F_{\mu\nu}$. The quadratic structure of these equations necessitates case distinctions, which lead to six inequivalent sets of solutions:
\begin{align}
    &\text{Set 1:} & F_{tr} &= 0\,, & F_{t\phi} &= 0\,, & F_{r\theta} &= 0\,, & F_{\theta\phi} &= 0 \notag\\
    &\text{Set 2:} & F_{tr} &= 0\,, & F_{t\theta} &= 0\,, & F_{r\phi} &= 0\,, & F_{\theta\phi} &= 0  \notag\\
    &\text{Set 3:} &  F_{t\theta} &= 0\,, & F_{t\phi} &= 0\,, & F_{r\theta} &= 0\,, & F_{r\phi} &= 0 \notag\\
    &\text{Set 4:} &  F_{tr} &= 0\,, & F_{t\theta} &= 0\,, & F_{t\phi} &= 0\,, & F_{r\theta} &= 0\,, & F_{\theta\phi} &= 0 \notag\\
    &\text{Set 5:} &  F_{tr} &= 0\,, & F_{t\theta} &= 0\,, & F_{t\phi} &= 0\,, & F_{r\theta} &= 0\,, & F_{r\phi} &= 0 \notag\\
    &\text{Set 6:} & F_{tr} &= 0\,, & F_{t\theta} &= 0\,, & F_{t\phi} &= 0\,, & F_{r\phi} &= 0\,, & F_{\theta\phi} &= 0 \,.
\end{align}

\paragraph{Diagonal forms.}
Substituting these sets into~\eqref{eq:EMSEMinComp} yields six distinct diagonal tensors. For the first three sets we find:
\begin{align}
    &\text{Set 1:} & T\ud{\mu}{\nu} &= \frac{1}{2 a(t)^2 r^2}\left\{(F_{t\theta})^2 + \left(\frac{\chi(r)}{a(t) \sin\theta}\right)^2 (F_{r\phi})^2\right\} \diag\left(-1,1,-1,1\right) \notag\\
    &\text{Set 2:} & T\ud{\mu}{\nu} &= \frac{1}{2 a(t)^2 r^2}\left\{\frac{1}{\sin^2\theta} (F_{t\phi})^2 + \left(\frac{\chi(r)}{a(t)}\right)^2 (F_{r\theta})^2\right\}\diag\left(-1,1,1,-1\right) \notag\\
    &\text{Set 3:} & T\ud{\mu}{\nu} &= \frac{1}{2 a(t)^2 r^2}\left\{r^2\chi(r)^2\,(F_{tr})^2 + \frac{1}{a(t)^2 r^2 \sin^2\theta}(F_{\theta\phi})^2\right\} \diag\left(-1,-1,1,1\right)\,.
\end{align}
The expressions in curly brackets are non-negative, while each diagonal matrix contains two positive and two negative entries.
Imposing the isotropy condition $T\ud{r}{r} = T\ud{\theta}{\theta} = T\ud{\phi}{\phi}$ then leads to equations of the form
\begin{align}
    (F_{ab})^2 + (F_{cd})^2 = -\left((F_{ab})^2 + (F_{cd})^2\right)\,
\end{align}
which implies $(F_{ab})^2 + (F_{cd})^2 = 0$. Consequently, each quadratic term must vanish separately, forcing all $F$-components to be zero. Thus, for sets 1--3,
\begin{align}
    F_{\mu\nu} &= 0 &\text{and} && T_{\mu\nu} &= 0\,.
\end{align}
For the remaining sets (4--6), where five of the six $F$-components vanish, the tensors reduce to
\begin{align}
    &\text{Set 4:} & T\ud{\mu}{\nu} &= \frac{\chi(r)^2}{2 a(t)^4 r^2 \sin^2\theta}(F_{r\phi})^2 \diag\left(-1,1,-1,1\right) \notag\\
    &\text{Set 5:} & T\ud{\mu}{\nu} &= \frac{1}{2 a(t)^4 r^2 \sin^2\theta}(F_{\theta\phi})^2 \diag\left(-1,-1,1,1\right) \notag\\
    &\text{Set 6:} & T\ud{\mu}{\nu} &= \frac{\chi(r)^2}{2 a(t)^4 r^2}(F_{r\theta})^2 \diag\left(-1,1,1,-1\right)\,.
\end{align}
Again, isotropy demands that the prefactor vanish, implying $F_{\mu\nu}=0$.

\paragraph{Interpretation.}
We therefore conclude that imposing homogeneity and isotropy on the electromagnetic energy-momentum tensor in an FLRW universe yields
\begin{align}
    T_{\mu\nu} &= 0 &\text{and} &&  F_{\mu\nu} &= 0\,.
\end{align}
This is entirely consistent with the symmetry-reduction of the vector potential discussed in subsection~\ref{ssec:VectorFields}, where we found that a homogeneous and isotropic vector field must take the form
\begin{align}
    A^\mu = (A^0(t), 0, 0, 0)\,,
\end{align}
which indeed gives $F_{\mu\nu}=0$ and hence $T_{\mu\nu}=0$. Nevertheless, additional field configurations exist for which $F_{\mu\nu}=0$ while $A^\mu$ itself is not homogeneous or isotropic. In differential-form notation, $F=\dd A$, so $F=0$ implies $\dd A=0$. By the Poincar\'{e} lemma, this condition ensures that $A$ is exact:
\begin{align}
    A_\mu = \pd_\mu \Phi\,,
\end{align}
where $\Phi$ is an arbitrary scalar field. Every vector potential of this form gives $F_{\mu\nu}=0$, though it may break homogeneity and isotropy. However, since such an $A_\mu$ is pure gauge, this does not represent a physical breaking of symmetry: homogeneity and isotropy are preserved at the level of the field strength.

In summary, we have shown that imposing cosmological symmetries at the level of the energy-momentum tensor can admit field configurations that do not themselves share these symmetries. For electromagnetism, these configurations correspond to pure-gauge potentials with vanishing field strength. In the next subsection, we will see that for the massless Kalb-Ramond field, the situation is more subtle: one obtains a non-zero, symmetry-preserving field strength and an infinite family of underlying field configurations which neither share these symmetries nor are they pure gauge.

\subsection{Energy-Momentum Tensor for the Kalb-Ramond Field}\label{ssec:T2Form}
The Kalb–Ramond field is the $2$-form $B_{\mu\nu} = -B_{\nu\mu}$, with the associated field strength $3$-form
\begin{align}
    H_{\mu\nu\rho} \ce \pd_\mu B_{\nu\rho} + \pd_\nu B_{\rho\mu} + \pd_\rho B_{\mu\nu}\,,
\end{align}
whose Lagrangian density is quadratic in first-order derivatives:
\begin{align}\label{eq:LagrangianBField}
    L = -\frac{1}{12} \sqrt{-g} g^{\mu\sigma}g^{\nu\kappa}g^{\rho\lambda} H_{\mu\nu\rho} H_{\kappa\lambda\sigma}\,.
\end{align}
This Lagrangian density is invariant under the gauge transformations
\begin{align}
    B_{\mu\nu} \quad\mapsto\quad B_{\mu\nu} + \pd_{\mu}\zeta_\nu - \pd_\nu \zeta_\mu\,,
\end{align}
where $\zeta_\mu$ is an arbitrary $1$-form. Furthermore, the field strength satisfies the Bianchi identity
\begin{align}\label{eq:BiacnhiIdH}
    \pd_{[\beta} H_{\alpha\mu\nu]} = 0\,.
\end{align}
As will become evident, this identity plays a crucial role in identifying all possible field configurations $B_{\mu\nu}$ that yield a homogeneous and isotropic energy-momentum tensor. The latter can be derived directly from~\eqref{eq:LagrangianBField} using the definition of the Hilbert energy-momentum tensor~\eqref{eq:HilbertSEM}:
\begin{align}\label{eq:SEMTensorB}
    T\ud{\mu}{\nu} = \frac{1}{12}\left(6 g^{\alpha\beta}g^{\gamma\delta}g^{\mu\lambda} - \delta\ud{\mu}{\nu} g^{\beta\gamma} g^{\delta\epsilon} g^{\lambda\alpha} H_{\alpha\gamma\epsilon}\right) H_{\lambda\beta\delta}\,.
\end{align}
Using the FLRW metric, the components of $T\ud{\mu}{\nu}$ can be written as
\begin{align}
    T\ud{\mu}{\nu} &=
    \begingroup
    \setlength\arraycolsep{7pt}
    \begin{pmatrix}
        -E & -\tau_{tr} & -\tau_{t\theta} & -\tau_{t\phi} \\
        \left(\frac{\chi(r)}{a(t)}\right)^2 \tau_{tr} & p_r & \tau_{r\theta} & \tau_{r\phi} \\
        \frac{1}{a(t)^2 r^2}\tau_{t\theta} & \frac{1}{r^2 \chi(r)^2}\tau_{r\theta} & p_\theta & \tau_{\theta\phi} \\
        \frac{1}{a(t)^2 r^2 \sin^2\theta} \tau_{t\phi} & \frac{1}{r^2 \chi(r)^2 \sin^2\theta} \tau_{r\phi} & \frac{1}{\sin^2\theta}\tau_{\theta\phi} & p_\phi
    \end{pmatrix}\,,
    \endgroup
\end{align}
where we introduce the shorthand notations
\begin{align}
    E &= \frac12\frac{1}{a(t)^4 r^4\sin^2\theta}\left[(H_{tr\theta})^2 r^2 \chi(r)^2 \sin^2\theta + (H_{tr\phi})^2\chi(r)^2 r^2 + (H_{t\theta\phi})^2 + (H_{r\theta\phi})^2\left(\frac{\chi(r)}{a(t)}\right)^2\right]\notag\\
    p_r &= \frac{1}{a(t)^4 r^4 \sin^2\theta}\left[(H_{t\theta\phi})^2 + \left(\frac{\chi(r)}{a(t)}\right)^2 (H_{r\theta\phi})^2\right] - E \notag\\
    p_\theta &= \frac{1}{a(t)^2 r^2 \sin^2\theta} \left(\frac{\chi(r)}{a(t)}\right)^2\left[(H_{tr\phi})^2 + \frac{1}{a(t)^2 r^2}(H_{r\theta\phi})^2\right] - E \notag\\
    p_\phi &= E - \frac{1}{a(t)^4 r^4 \sin^2\theta}\left[(H_{t\theta\phi})^2 + (H_{tr\phi})^2r^2 \chi(r)^2\right]
\end{align}
and
\begin{align}
    \tau_{tr} &\ce \frac{1}{a(t)^4 r^4 \sin^2\theta} H_{t\theta\phi} H_{r\theta\phi}& \tau_{r\theta} &\ce \frac{1}{a(t)^2 \sin^2\theta} \left(\frac{\chi(r)}{a(t)}\right)^2H_{tr\phi} H_{t\theta\phi} \notag\\
    \tau_{t\theta} &\ce -\frac{1}{a(t)^2 \sin^2\theta} \left(\frac{\chi(r)}{a(t)}\right)^2 H_{tr\phi} H_{r\theta\phi} & \tau_{r\phi} &\ce \frac{1}{a(t)^2 r^2} \left(\frac{\chi(r)}{a(t)}\right)^2 H_{tr\theta} H_{t\theta\phi} \notag\\
    \tau_{t\phi} &\ce \frac{1}{a(t)^2 r^2} \left(\frac{\chi(r)}{a(t)}\right)^2 H_{tr\theta} H_{r\theta\phi} & \tau_{\theta\phi} &\ce -\frac{1}{a(t)^2 r^2} \left(\frac{\chi(r)}{a(t)}\right)^2 H_{tr\theta} H_{tr\phi} \,.
\end{align}
For $T\ud{\mu}{\nu}$ to be homogeneous and isotropic, it must satisfy three conditions:
\begin{enumerate}
    \item All off-diagonal components must vanish;
    \item The pressures must be equal, i.e. $p_r = p_\theta = p_\phi$;
    \item The diagonal components must depend only on time.
\end{enumerate}
Since all components of the energy-momentum tensor are quadratic in the field strength, multiple configurations can ensure that the off-diagonal components vanish. This is a phenomenon we encountered earlier in subsection~\ref{ssec:MaxwellEMT}. Here, we identify four distinct solution sets:
\begin{align}
    &\text{Set 1:} & H_{tr\theta} &= 0\,, & H_{tr\phi} &= 0\,, & H_{t\theta\phi} &= 0 \notag\\
    &\text{Set 2:} & H_{tr\theta} &= 0\,, & H_{tr\phi} &= 0\,, & H_{r\theta\phi} &= 0 \notag\\
    &\text{Set 3:} & H_{tr\theta} &= 0\,, & H_{t\theta\phi} &= 0\,, & H_{r\theta\phi} &= 0 \notag\\
    &\text{Set 4:} & H_{tr\phi} &= 0\,, & H_{t\theta\phi} &= 0\,, & H_{r\theta\phi} &= 0\,.
\end{align}
Substituting each set into~\eqref{eq:SEMTensorB} yields four distinct diagonal forms of the energy–momentum tensor:
\begin{align}\label{eq:FourFormsOfTmunu}
    &\text{Set 1:} & T\ud{\mu}{\nu} &= \frac12\frac{(H_{r\theta\phi})^2}{a(t)^4 r^4 \sin^2\theta}\left(\frac{\chi(r)}{a(t)}\right)^2\diag\left(-1, 1, 1, 1\right) \notag\\
    &\text{Set 2:} & T\ud{\mu}{\nu} &= \frac12\frac{(H_{t\theta\phi})^2}{a(t)^4 r^4 \sin^2\theta}\diag\left(-1, 1, -1, -1\right) \notag\\
    &\text{Set 3:} & T\ud{\mu}{\nu} &= \frac12\frac{(H_{tr\phi})^2}{a(t)^2 \sin^2\theta} \left(\frac{\chi(r)}{a(t)}\right)^2\diag\left(-1, -1, 1, -1\right) \notag\\
    &\text{Set 4:} & T\ud{\mu}{\nu} &= \frac12\frac{(H_{tr\theta})^2}{a(t)^2 r^2} \left(\frac{\chi(r)}{a(t)}\right)^2\diag\left(-1, -1, -1, 1\right)
\end{align}
Only the tensor derived from Set~1 satisfies the condition $p_r = p_\theta = p_\phi$. For Sets~2--4 this equality holds only if the remaining field-strength component vanishes, which implies $T\ud{\mu}{\nu}=0$. In those cases, $H_{\alpha\mu\nu}=0$, leading via the Poincar\'{e} lemma to $B_{\mu\nu} = \pd_\mu\zeta_\nu - \pd_\nu\zeta_\mu$, i.e. the $2$-form is pure gauge. Such configurations do not contribute to the Einstein field equations.

By contrast, Set~1 yields a non-vanishing $T\ud{\mu}{\nu}$ consistent with homogeneity and isotropy, corresponding to non-trivial field configurations $B_{\mu\nu}$. The Bianchi identity~\eqref{eq:BiacnhiIdH} implies
\begin{align}\label{eq:BianchiIdComponent}
    \pd_t H_{r\theta\phi} = 0\,.
\end{align}
One can easily check that this is the only equation that results from~\eqref{eq:BiacnhiIdH}, since in solution Set 1 three of the four components of $H_{\alpha\mu\nu}$ are zero. The only non-vanishing component is $H_{r\theta\phi}$ and because of the anti-symmetrization, the only non-vanishing contribution comes from $\beta = t$. Hence, we can conclude from the Bianchi identity that $H_{r\theta\phi}$ is independent of time.  To ensure that the energy density and pressure depend only on time, $H_{r\theta\phi}$ must be proportional to $\frac{r^2 \sin\theta}{\chi(r)}$:
\begin{align}\label{eq:FormOfHrthetaphi}
    H_{r\theta\phi} = h \frac{r^2 \sin\theta}{\chi(r)}\,,
\end{align}
where $h$ is a constant\footnote{The factor $h$ cannot depend on $r$, $\theta$, $\phi$, since the energy density and the pressure terms of $T\ud{\mu}{\nu}$ would inherit these dependencies. Moreover, it cannot depend on $t$ since $H_{r\theta\phi}$ is time-independent, according to the Bianchi identity.}.  We can now determine the implications for the $B_{\mu\nu}$ components that contribute to $H_{r\theta\phi}$. From the definition of the field strength, we have
\begin{align}
    H_{r\theta\phi} = \pd_r B_{\theta\phi} - \pd_\theta B_{r\phi} + \pd_\phi B_{r\theta}\,.
\end{align}
It follows that $B_{r\theta}$, $B_{r\phi}$, and $B_{\theta\phi}$ must be of the form
\begin{align}
    h f(r, \theta, \phi) + c(t, ...)\,,
\end{align}
with $h \neq 0$, where $c$ depends on time and on the two spatial coordinates that do not appear in the derivative operator acting on the given component. For instance, for $B_{\theta\phi}$, the function $c$ depends on $t$, $\theta$, and $\phi$, but not on $r$.

We can systematically construct $f(r,\theta,\phi)$ and the corresponding functions $c(r, ...)$ by integrating $h, \frac{r^2 \sin\theta}{\chi(r)}$ with respect to $r$, $\theta$, and $\phi$, respectively, and equating the results to $B_{\theta\phi}$, $B_{r\phi}$, and $B_{r\theta}$. This yields
\begin{align}\label{eq:Bsol}
    B_{\theta\phi} &= h \sin\theta\int \frac{r^2 \dd r}{\chi(r)} = h \epsilon_1 \sin\theta \left(\frac{1}{k^{3/2}}\arctan\left(\frac{\sqrt{k} r}{\chi(r) - 1}\right) - \frac{1}{2k} r\,\chi(r)\right) + c_1(t, \theta, \phi) \notag\\
    B_{r\phi} &= h \frac{r^2}{\chi(r)} \int \sin\theta\dd \theta = h \epsilon_2 \frac{r^2 \cos\theta}{\chi(r)} - c_2(t, r, \phi) \notag\\
    B_{r\theta} &= h \frac{r^2 \sin\theta}{\chi(r)} \int \dd \phi = h \epsilon_3 \frac{r^2 \sin\theta}{\chi(r)} \phi + c_3(t, r, \theta)\,,
\end{align}
where the $c_i$ are arbitrary integration functions\footnote{The sign in front of $c_2$ has been chosen to be $-1$ for later convenience (cf.~equation~\eqref{eq:CurlEq}).}, and the real constants $\epsilon_i$ satisfy
\begin{align}
    \epsilon_1 + \epsilon_2 + \epsilon_3 = 1\,.
\end{align}
Equations~\eqref{eq:Bsol} give the most general solution for $B_{\theta\phi}$, $B_{r\phi}$, and $B_{r\theta}$ that guarantees, by construction, that
\begin{align}
    H_{r\theta\phi} &= \pd_r B_{\theta\phi} - \pd_{\theta} B_{r\phi} + \pd_{\phi} B_{r\theta} \notag\\
    &= \left(\epsilon_1 + \epsilon_2 + \epsilon_3\right) h \frac{r^2 \sin\theta}{\chi(r)} = h \frac{r^2 \sin\theta}{\chi(r)}\,.
\end{align}
Next, recall that we previously imposed conditions to eliminate the off-diagonal components of $T\ud{\mu}{\nu}$. These translate into the following constraints on the $B_{\mu\nu}$ components:
\begin{align}
    H_{tr\theta} &= \pd_t B_{r\theta} - \pd_r B_{t\theta} + \pd_\theta B_{tr} = 0 \notag\\
    H_{tr\phi} &= \pd_t B_{r\phi} - \pd_r B_{t\phi} + \pd_\phi B_{tr} = 0\notag\\
    H_{t\theta\phi} &= \pd_t B_{\theta\phi} - \pd_\theta B_{t\phi} + \pd_\phi B_{t\theta} = 0\,.
\end{align}
Substituting the solutions~\eqref{eq:Bsol} into these relations and simplifying, we obtain a system of first-order partial differential equations for $B_{tr}$, $B_{t\theta}$, and $B_{t\phi}$ of the form
\begin{align}\label{eq:CurlEq}
    \pd_\theta B_{t\phi} - \pd_\phi B_{t\theta} &= \dot{c}_1(t, \theta, \phi) \notag\\
    \pd_\phi B_{tr} - \pd_r B_{t\phi} &= \dot{c}_2(t, r, \phi) \notag\\
    \pd_r B_{t\theta} - \pd_\theta B_{tr} &= \dot{c}_3(t, r, \theta)\,,
\end{align}
where $\dot{c}_i$ denotes the time derivative of $c_i(t, ...)$. This system can be expressed more compactly by treating $B_{tr}$, $B_{t\theta}$, $B_{t\phi}$ and $c_1$, $c_2$, $c_3$ as components of two three-dimensional vectors. Specifically, we define
\begin{align}
    \boldsymbol{B} &\ce (B_{tr}, B_{t\theta}, B_{t\phi})^\text{T} & \text{and} && \boldsymbol{C} &\ce (c_1, c_2, c_3)^\text{T}\,,
\end{align}
where $^\text{T}$ denotes the transpose of the row vector. In terms of these vectors, equation~\eqref{eq:CurlEq} can be rewritten as
\begin{align}\label{eq:CompactEq}
    \nabla\times\boldsymbol{B} = \dot{\boldsymbol{C}}\,.
\end{align}
To find the most general solution, one could integrate both sides. However, before proceeding, it is essential to ensure integrability. From vector calculus, we know that the identity
\begin{align}
    \nabla\cdot\left(\nabla\times\boldsymbol{B}\right) = 0\,
\end{align}
holds for any vector field $\boldsymbol{B}$. Taking the divergence of~\eqref{eq:CompactEq} therefore yields the integrability condition
\begin{align}
    \nabla\cdot\dot{\boldsymbol{C}} = \pd_t \left(\nabla\cdot\boldsymbol{C}\right) = 0\,,
\end{align}
where we used the commutativity of partial derivatives to exchange $\pd_t$ and $\nabla$. This condition is automatically satisfied, since $\boldsymbol{C}$ is divergence-free:
\begin{align}
    \nabla\cdot{\boldsymbol{C}} &= \pd_r c_1(t, \theta, \phi) + \pd_\theta c_2(t,r,\phi) + \pd_\phi c_3(t, r, \theta) = 0\,.
\end{align}
This is no coincidence: it is precisely this property of $\boldsymbol{C}$ that ensures the solution~\eqref{eq:Bsol} produces a time-independent $H_{r\theta\phi}$.

Because $\boldsymbol{C}$ is divergence-free, we know that locally there exists a vector potential $\boldsymbol{P}$ such that\footnote{This is analogous to electromagnetism, where the vanishing divergence of the magnetic field ensures the (local) existence of a vector potential.}
\begin{align}\label{eq:DefP}
    \boldsymbol{C} = \nabla\times\boldsymbol{P}\,,
\end{align}
in terms of which equation~\eqref{eq:CompactEq} can be written as
\begin{align}
    \nabla\times\left(\boldsymbol{B} - \dot{\boldsymbol{P}}\right) = 0\,.
\end{align}
This implies that the vector $\boldsymbol{B} - \dot{\boldsymbol{P}}$ is curl-free and therefore locally expressible as the gradient of a scalar field $\Psi(t,r,\theta,\phi)$:
\begin{align}
    \boldsymbol{B} - \dot{\boldsymbol{P}} = \nabla\Psi(t,r,\theta,\phi)\,.
\end{align}
Hence, the most general solution to~\eqref{eq:CompactEq} is\footnote{Recall from electromagnetism that the electric field can be written as $\boldsymbol{E} = -\nabla\Phi - \pd_t \boldsymbol{A}$, where $\Phi$ is the electric potential and $\boldsymbol{A}$ the magnetic vector potential. This closely resembles~\eqref{eq:MostGeneralB}. Indeed, our vector field $\boldsymbol{B}$ satisfies an equation analogous to one of the source-free Maxwell equations, namely $\nabla\times\boldsymbol{E} = -\pd_t\boldsymbol{B}_\text{mag}$, where $\boldsymbol{B}_\text{mag}$ is the magnetic field. Moreover, our vector field $\boldsymbol{C}$ plays a similar role to $\boldsymbol{B}_\text{mag}$, satisfying $\nabla\cdot\boldsymbol{C}=0$, the second source-free Maxwell equation.}
\begin{align}\label{eq:MostGeneralB}
    \boldsymbol{B}(t,r,\theta, \phi) = \nabla\Psi(t,r,\theta, \phi) + \dot{\boldsymbol{P}}(t,r,\theta, \phi)\,,
\end{align}
for an arbitrary scalar field $\Psi$ and any vector field $\boldsymbol{P}$ satisfying~\eqref{eq:DefP}. Because of the close resemblance between the scalar and vector potential of electromagnetism and the $\Psi$ and $\boldsymbol{P}$ fields encountered here, it should be of no surprise that $\boldsymbol{B}$ is invariant under the simultaneous gauge transformations
\begin{align}\label{eq:GaugeFreedom}
    \boldsymbol{P}\quad &\mapsto\quad  \boldsymbol{P} + \nabla\chi &\text{and}&& \Psi \quad& \mapsto\quad  \Psi + \pd_t \chi\,,
\end{align}
where $\chi$ is an arbitrary scalar field. This gauge freedom is crucial when constructing an explicit (local) vector potential $\boldsymbol{P}$, since it allows one component of $\boldsymbol{P}$ to be chosen freely.

Let us, for instance, fix $P_r(t,r,\theta,\phi)$ to be an arbitrary smooth function. The remaining components $P_\theta$ and $P_\phi$ can then be obtained by integrating equation~\eqref{eq:DefP}. Considering first its $\phi$-component, we find
\begin{align}
    \pd_r P_\theta = \pd_\theta P_r + c_3(t, r, \theta)\,.
\end{align}
Integrating over $r$ yields
\begin{align}\label{eq:Ptheta}
    P_\theta(t, r, \theta, \phi) = \int_{r_0}^r\left(\pd_\theta P_r(t,s,\theta,\phi) + c_3(t, s, \theta)\right)\dd s + \alpha(t,\theta,\phi)\,,
\end{align}
where $r_0$ is an arbitrary reference radius, $\alpha(t,\theta,\phi)$ is an $r$-independent integration function, and $\pd_\theta P_r$ is known since $P_r$ was freely specified.

Analogously, from the $\theta$-component of~\eqref{eq:DefP} we obtain
\begin{align}\label{eq:Pphi}
    P_\phi(t,r,\theta,\phi) = \int_{r_0}^r \left(\pd_\phi P_r(t,s,\theta, \phi) - c_2(t,s,\phi)\right)\dd s + \beta(t,\theta,\phi)\,,
\end{align}
where $\beta(t,\theta,\phi)$ is another integration function independent of $r$. We should not forget that the $r$-component of equation~\eqref{eq:DefP} contains derivatives of $P_\theta$ and $P_\phi$, which might lead to integrability conditions. Explicitly, the $r$-component of that equation reads
\begin{align}
    \pd_\theta P_\phi - \pd_\phi P_\theta = c_1(t,\theta, \phi)\,. 
\end{align}
Substituting the expressions for $P_\theta$ and $P_\phi$ into this equation yields the integrability condition
\begin{align}
    \pd_\theta \beta(t,\theta,\phi) - \pd_\phi \alpha(t,\theta, \phi) = c_1(t,\theta, \phi)\,.
\end{align}
Since the $r$-dependent integrals cancel, this relation constrains the functions $\alpha$ and $\beta$. A convenient choice is to set $\alpha(t,\theta,\phi)=0$, after which integration over $\theta$ gives
\begin{align}\label{eq:IntCond}
    \alpha(t, \theta, \phi) &= 0 & \text{and} && \beta(t, \theta, \phi) &= \int_{\theta_0}^\theta c_1(t,s, \phi) \dd s + \gamma(t, \phi)\,,
\end{align}
where $\gamma(t,\phi)$ is an arbitrary function independent of $r$ and $\theta$.

Let us summarize what we did and what we found up to this point: Starting from the energy–momentum tensor~\eqref{eq:SEMTensorB}, we found four distinct ways of reducing $T\ud{\mu}{\nu}$ to a diagonal form, but only the first yields a non-trivial tensor potentially compatible with homogeneity and isotropy. Enforcing these symmetries, together with the Bianchi identity~\eqref{eq:BianchiIdComponent}, required $H_{r\theta\phi}$ to be time-independent and of a specific form (see~\eqref{eq:FormOfHrthetaphi}).

Using the definition of $H_{r\theta\phi}$ in terms of derivatives of $B_{\theta\phi}$, $B_{r\phi}$, and $B_{r\theta}$, we obtained the general structure of these components (equation~\eqref{eq:Bsol}). The remaining equations, $H_{tr\theta} = H_{tr\phi} = H_{t\theta\phi} = 0$, were shown to translate into a compact vectorial relation involving the curl of $\boldsymbol{B}$ and the time derivative of $\boldsymbol{C}$.

We then introduced $\boldsymbol{P}$ via~\eqref{eq:DefP} and integrated to find the general solution~\eqref{eq:MostGeneralB}, which combines a scalar potential $\Psi$ and a time-varying vector potential $\boldsymbol{P}$. The gauge freedom~\eqref{eq:GaugeFreedom} allowed us to fix $P_r$ arbitrarily and determine $P_\theta$ and $P_\phi$ through the integrals~\eqref{eq:Ptheta} and~\eqref{eq:Pphi}, subject to the integrability condition~\eqref{eq:IntCond}.

Putting everything together, the most general $2$-form $B_{\mu\nu}$ that yields a homogeneous and isotropic energy–momentum tensor is
\begin{align}
    B_{\mu\nu} =
    \begingroup
    \setlength\arraycolsep{7pt}
    \begin{pmatrix}
        0 & \pd_r\Psi + P_r & \pd_\theta\Psi + P_\theta & \pd_\phi\Psi + P_\phi \\
        -(\pd_r\Psi + P_r) & 0 & h \epsilon_3 \frac{r^2 \sin\theta}{\chi(r)} \phi + c_3 & h \epsilon_2 \frac{r^2 \cos\theta}{\chi(r)} - c_2 \\
        -(\pd_\theta\Psi + P_\theta) & -\left(h \epsilon_3 \frac{r^2 \sin\theta}{\chi(r)} \phi + c_3\right) & 0 & h \epsilon_1  \int^r\frac{s^2 \sin\theta}{\chi(s)}\dd s \\
        -(\pd_\phi\Psi + P_\phi) & -\left(h \epsilon_2 \frac{r^2 \cos\theta}{\chi(r)} - c_2\right) & -h \epsilon_1  \int^r\frac{s^2 \sin\theta}{\chi(s)}\dd s & 0
    \end{pmatrix}\,,
    \endgroup
\end{align}
where
\begin{align}
    \int^r\frac{s^2 \sin\theta}{\chi(s)}\dd s = \sin\theta\left(\frac{1}{k^{3/2}}\arctan\left(\frac{\sqrt{k} r}{\chi(r) - 1}\right) - \frac{1}{2k} r\,\chi(r)\right) + c_1(t,\theta,\phi)\,.
\end{align}
From subsection~\ref{ssec:2Forms} we know that a homogeneous and isotropic $2$-form must vanish. Since the $B_{\mu\nu}$ obtained above is manifestly nonzero, it cannot be both homogeneous and isotropic. At most, it could satisfy one of these symmetries individually; however, direct inspection shows that it is neither. In particular, isotropy would require $\pd_\phi B_{\mu\nu}=0$, which is violated because $\Psi$, $P_r$, $P_\theta$, $P_\phi$, and the functions $c_1$ and $c_2$ may all depend explicitly on~$\phi$.

Similarly, one can verify by direct computation that $B_{\mu\nu}$ is not homogeneous, since
\begin{align}
    \lie_{\T_3} B_{\mu\nu} \neq 0\,.
\end{align}
For simplicity, let us set $k=0$ and consider the $tr$-component of this Lie derivative. We then find
\begin{align}
    \left(\lie_{\T_3}B\right)_{tr} &= \frac{\sin\theta}{r^2} B_{t\theta} + \cos\theta \pd_r B_{tr} - \frac{\sin\theta}{r} \pd_\theta B_{tr} \notag\\
    &= \frac{\sin\theta}{r^2} \left(\pd_\theta \Psi + P_\theta\right) + \cos\theta \pd_r\left(\pd_r \Psi + P_r\right) - \frac{\sin\theta}{r} \pd_\theta\left(\pd_r \Psi + P_r\right)\,.
\end{align}
Evidently, there is no reason for this expression to vanish. We can therefore conclude that our $B_{\mu\nu}$ is \emph{neither homogeneous nor isotropic}. Moreover, it is not pure gauge, i.e., it cannot be written in terms of a $1$-form $\zeta_\mu$ as $B_{\mu\nu} = \pd_\mu \zeta_\nu - \pd_\nu \zeta_\mu$, since such a form would imply a vanishing field strength. In contrast, we have explicitly found that one component of the field strength is nonzero, yielding a non-vanishing energy–momentum tensor.

Interestingly, even though $B_{\mu\nu}$ is neither homogeneous nor isotropic, the corresponding field strength $H_{\alpha\mu\nu}$ constructed from it takes precisely the form of the most general homogeneous and isotropic $3$-form. As shown in subsection~\ref{ssec:3Forms}, the most general homogeneous and isotropic $3$-form is given by
\begin{align}
    K_{tr\theta} &= 0\,, & K_{tr\phi} &= 0\,, & K_{t\theta\phi} &= 0\,, & K_{r\theta\phi} &= h(t)\frac{r^2 \sin\theta}{\chi(r)}\,.
\end{align}
This coincides exactly with the structure of the field strength $H_{\alpha\mu\nu}$ we obtained earlier, with the only difference being that in our case $h$ is a constant, whereas for a generic $3$-form it may depend on time. The constancy of $h$ arises because $H$ is not an arbitrary $3$-form, but rather the exterior derivative of a $2$-form, $H = \dd B$, and therefore must satisfy the Bianchi identity~\eqref{eq:BiacnhiIdH}.

In summary, we have identified an infinite family of nontrivial $2$-forms $B_{\mu\nu}$ that are neither homogeneous, nor isotropic, nor pure gauge, yet which yield a homogeneous and isotropic field strength and energy–momentum tensor. Each choice of the functions $\Psi(t,r,\theta,\phi)$, $P_r(t,r,\theta,\phi)$, $c_1(t,\theta,\phi)$, $c_2(t,r,\phi)$, $c_3(t,r,\theta)$, $\alpha(t,\theta,\phi)$, $\beta(t,\theta,\phi)$, and the constants $h$, $\epsilon_1$, $\epsilon_2$, and $\epsilon_3$, subject to appropriate constraints, defines a distinct $B_{\mu\nu}$ with these remarkable properties.

\section{The Scalar-Vector-Tensor Decomposition of Cosmological Perturbations}\label{sec:SVT}
The Cosmological Principle posits that the Universe is homogeneous and isotropic \emph{on sufficiently large scales}. On smaller scales, however, we observe significant inhomogeneities and anisotropies. Galaxies, stars, and planets are examples of matter clumps that are distributed neither homogeneously nor isotropically. Consequently, while the metric of the Universe is well approximated by the FLRW form, it is not exact. Understanding the formation of these symmetry-breaking structures requires us to move beyond the idealized FLRW model. In this section, we provide a pedagogical discussion of the main concepts of cosmological perturbation theory, culminating in a transparent derivation of the scalar–vector–tensor decomposition. Throughout, we adopt the following conventions:
\begin{enumerate}
    \item We focus exclusively on the flat case, $k=0$. Modifications necessary for $k=\pm 1$ will be mentioned only briefly.
    \item We work in Cartesian coordinates, rather than spherical coordinates. Specifically, the spatial coordinates are denoted $x$, $y$, and $z$, or collectively by $\x \ce (x,y,z)$.
    \item Instead of cosmological time $t$, we employ conformal time $\eta$, related to $t$ via
    \begin{align}
        \dd \eta = \frac{\dd t}{a(t)}\,.
    \end{align}
    In these coordinates, the FLRW line element for $k=0$ reads
    \begin{align}\label{eq:FLRWConfTime}
        \dd s^2_\text{FLRW} = a(\eta)^2 \left(-\dd \eta^2 + \dd x^2 + \dd y^2 + \dd z^2\right) = a(\eta)^2 \left(-\dd \eta^2 + \dd \x^2\right)\,,
    \end{align}
    showing that, in $(\eta, \x)$ coordinates, the FLRW metric is simply $a(\eta)^2$ times the Minkowski metric.
    \item The generators of rotations and translations in Cartesian coordinates for $k=0$ take the simple form
    \begin{align}\label{eq:GeneratorsOfRotationsCartesian}
        \R_1 &= y\,\pd_z - z\, \pd_y\,, & \R_2 &= z\,\pd_x - x\,\pd_z\,, & \R_3 &= x\,\pd_y - y\,\pd_x
    \end{align}
    and
    \begin{align}\label{eq:GeneratorsOfTranslationsCartesian}
        \T_1 &= \pd_x\,, & \T_2 &= \pd_y\,, & \T_3 &= \pd_z\,.
    \end{align}
\end{enumerate}
A key tool for systematically studying deviations from the FLRW Universe is the introduction of small perturbations in the metric and matter content. We achieve this by introducing a small bookkeeping parameter $\epsilon$, with $|\epsilon|\ll 1$, to track the perturbative order. The metric $g_{\mu\nu}$ and the energy–momentum tensor $T_{\mu\nu}$ can then be written as
\begin{align}\label{eq:PerturbativeAnsatz}
    g_{\mu\nu} &= \bar{g}_{\mu\nu} + \epsilon\,  h_{\mu\nu} &\text{and}&& T_{\mu\nu} &= \bar{T}_{\mu\nu} + \epsilon\, \tau_{\mu\nu}\,,
\end{align}
where a bar denotes a background quantity or a quantity constructed solely from background fields. Since we aim to study deviations from FLRW, the background metric $\bar{g}_{\mu\nu}$ is simply the $k=0$ FLRW metric, sourced by the homogeneous and isotropic energy–momentum tensor $\bar{T}_{\mu\nu}$. We call the triple  $(\M, \bar{g}_{\mu\nu}, \bar{T}_{\mu\nu})$ the \emph{background} and we demand that it obeys the Einstein field equations
\begin{align}\label{eq:BackgroundEFE}
    \bar{G}_{\mu\nu}(\bar{g}) = 8\pi\, \bar{T}_{\mu\nu}
\end{align}
together with the energy–momentum conservation law
\begin{align}\label{eq:BackgroundConservation}
    \bar{\nabla}_\mu \bar{T}\ud{\mu}{\nu} = 0\,.
\end{align}
where $\bar{\nabla}_\mu$ is the unique torsion-free covariant derivative compatible with $\bar{g}_{\mu\nu}$, expressed in terms of the background Christoffel symbols. Because $\bar{T}_{\mu\nu}$ is homogeneous and isotropic, it must take the perfect fluid form~\eqref{eq:PerfectFluidEMT}, with energy density $\bar{\rho}(\eta)$ and pressure $\bar{p}(\eta)$.

In the perturbative ansatz~\eqref{eq:PerturbativeAnsatz}, we introduced two new symmetric rank–$2$ tensor fields: the metric perturbation $h_{\mu\nu}$ and the matter perturbation $\tau_{\mu\nu}$. These fields encode small deviations from homogeneity and isotropy, and their Lie derivatives with respect to the generators of translations and rotations \emph{do not vanish}.
In order for our perturbative ansatz to describe the real Universe, we require that the full metric $g_{\mu\nu}$ and energy–momentum tensor $T_{\mu\nu}$ satisfy Einstein’s field equations:
\begin{align}
    G_{\mu\nu}(\bar{g} + \epsilon\, h) = 8\pi\, \left(\bar{T}_{\mu\nu} + \epsilon\, \tau_{\mu\nu}\right)\,.
\end{align}
By virtue of the contracted Bianchi identities, $\nabla_\mu G\ud{\mu}{\nu} = 0$, we must also demand that $T_{\mu\nu}$ obeys the energy–momentum conservation law\footnote{There is no bar on the covariant derivative here. This indicates that $\nabla_\mu$ is the torsion-free covariant derivative compatible with the perturbed metric $g_{\mu\nu} = \bar{g}_{\mu\nu} + \epsilon\, h_{\mu\nu}$.}:
\begin{align}\label{eq:EnergyMomentumConservation}
    \nabla_\mu T\ud{\mu}{\nu} = 0\,.
\end{align}
Assuming small perturbations around the fixed background $(\M, \bar{g}_{\mu\nu}, \bar{T}_{\mu\nu})$, we expand the Einstein equations to first order in $\epsilon$:
\begin{align}
    \bar{G}_{\mu\nu}(\bar{g}) + \epsilon\, \left[\frac{\dd}{\dd \epsilon} G_{\mu\nu}(\bar{g}+\epsilon\, h)\right]_{\epsilon=0} &= 8\pi\left(\bar{T}_{\mu\nu} + \epsilon\, \tau_{\mu\nu}\right)\,.
\end{align}
Using the background field equations~\eqref{eq:BackgroundEFE}, this simplifies to the linearized Einstein equations:
\begin{align}\label{eq:LinearizedEquations}
    H_{\mu\nu}(\bar{g}, h) = 8\pi\, \tau_{\mu\nu}\,,
\end{align}
where we define the perturbed Einstein tensor as
\begin{align}
    H_{\mu\nu}(\bar{g}, h) &\ce \left[\frac{\dd}{\dd \epsilon} G_{\mu\nu}(\bar{g}+\epsilon\, h)\right]_{\epsilon=0}\,.
\end{align}
The explicit expression for $H_{\mu\nu}$, linear in $h_{\mu\nu}$, reads
\begin{align}
    H_{\mu\nu} &= -\frac12\left( \bar{g}^{\alpha\beta}\bar{\nabla}_{(\mu}\bar{\nabla}_{\nu)}-2\bar{g}^{\gamma(\alpha}\bar{\nabla}_{\gamma}\delta\ud{\beta)}{(\mu}\bar{\nabla}_{\nu)} + \bar{g}^{\gamma\delta}\delta\ud{\alpha}{(\mu}\delta\ud{\beta}{\nu)} \bar{\nabla}_\gamma\bar{\nabla}_\delta \right. \notag\\
    &\phantom{=- \frac12...} \left.+\bar{g}^{\alpha(\delta}\bar{g}^{\gamma)\beta} \bar{g}_{\mu\nu}\bar{\nabla}_\gamma\bar{\nabla}_\delta - \bar{g}^{\alpha\beta}\bar{g}^{\gamma\delta} \bar{g}_{\mu\nu} \bar{\nabla}_\gamma \bar{\nabla}_\delta - \bar{g}_{\mu\nu} \bar{R}^{\alpha\beta} + \bar{R}\,\delta\ud{\alpha}{(\mu} \delta\ud{\beta}{\nu)}\right)h_{\alpha\beta}\,.
\end{align}
All coefficients depend solely on the background metric $\bar{g}_{\mu\nu}$ and its curvature. Similarly, the energy–momentum conservation law~\eqref{eq:EnergyMomentumConservation} can be expanded to first order in $\epsilon$:
\begin{align}
    \bar{\nabla}_\mu \bar{T}\ud{\mu}{\nu} + \epsilon\, \left[\frac{\dd}{\dd \epsilon} \nabla_\mu T\ud{\mu}{\nu}(\bar{g} + \epsilon\, h)\right]_{\epsilon=0} &= 0\,.
\end{align}
Observe that both covariant derivative operators, $\nabla_\mu$ and $\bar{\nabla}_\mu$, appear in this expansion. The zeroth-order term vanishes due to the background conservation law~\eqref{eq:BackgroundConservation}. The first-order term is then
\begin{align}
    \bar{\nabla}_\mu \tau\ud{\mu}{\nu} + \bar{T}\ud{\beta}{\nu} \delta \Gamma\ud{\alpha}{\alpha\beta} - \bar{T}\ud{\beta}{\alpha}\delta \Gamma\ud{\alpha}{\beta\nu} &= 0\,,
\end{align}
where the perturbation of the Christoffel symbols is
\begin{align}
    \delta\Gamma\ud{\alpha}{\mu\nu} \ce \left[\frac{\dd}{\dd \epsilon}\Gamma\ud{\alpha}{\mu\nu}(\bar{g} + \epsilon\,h)\right]_{\epsilon=0} = \frac12 \bar{g}^{\alpha\lambda}\left(\bar{\nabla}_\mu h_{\nu\lambda} + \bar{\nabla}_\nu h_{\mu\lambda} - \bar{\nabla}_\lambda h_{\mu\nu}\right)\,.
\end{align}
Using this, the linearized energy–momentum conservation law simplifies to
\begin{align}\label{eq:ExpandedConservationLaw}
    \bar{\nabla}_\mu \tau\ud{\mu}{\nu} + \frac12 \left(\bar{T}\ud{\alpha}{\nu} \bar{\nabla}_\alpha h\ud{\beta}{\beta} - \bar{T}^{\alpha\beta} \bar{\nabla}_\nu h_{\alpha\beta}\right) = 0\,.
\end{align}

Both the linearized Einstein equations~\eqref{eq:LinearizedEquations} and the linearized conservation law~\eqref{eq:ExpandedConservationLaw} are linear in the perturbations $h_{\mu\nu}$ and $\tau_{\mu\nu}$. 

Cosmological perturbation theory and its goal can be summarized as follows:
The background spacetime $(\M, \bar{g}_{\mu\nu}, \bar{T}_{\mu\nu})$ is fixed, with $\bar{g}_{\mu\nu}$ being the flat FLRW metric sourced by a perfect fluid $\bar{T}_{\mu\nu}$. All perturbative fields evolve on this background. The metric perturbations $h_{\mu\nu}$ satisfy the linearized Einstein equations~\eqref{eq:LinearizedEquations}, while the matter perturbations obey the linearized conservation law~\eqref{eq:ExpandedConservationLaw}. The goal is to study these equations and their solution space, in order to understand the evolution of small deviations from homogeneity and isotropy in our Universe.

\subsection{Analyzing the Equations for Matter and Metric Perturbations}\label{ssec:AnalyzingLinearizedEqs}
The perturbative approach of the previous subsection produced a set of coupled partial differential equations. We rewrite these equations here in a slightly different form, as they represent the starting point for a  more in-depth analysis:
\begin{align}\label{eq:CoupledSystem}
    \bar{\mathcal{E}}\du{\mu\nu}{\alpha\beta}h_{\alpha\beta} &= 8\pi\, \tau_{\mu\nu} \notag\\
    \bar{\nabla}_\mu \tau\ud{\mu}{\nu} &= \bar{\mathcal{F}}\du{\nu}{\alpha\beta} h_{\alpha\beta}
\end{align}
For later convenience, we introduced the Lichnerowicz operator $\bar{\mathcal{E}}\du{\mu\nu}{\alpha\beta}$ and the operator $\bar{\mathcal{F}}\du{\nu}{\alpha\beta}$, defined by
\begin{align}
    \bar{\mathcal{E}}\du{\mu\nu}{\alpha\beta} &\ce -\frac12\left( \bar{g}^{\alpha\beta}\bar{\nabla}_{(\mu}\bar{\nabla}_{\nu)}-2\bar{g}^{\gamma(\alpha}\bar{\nabla}_{\gamma}\delta\ud{\beta)}{(\mu}\bar{\nabla}_{\nu)} + \bar{g}^{\gamma\delta}\delta\ud{\alpha}{(\mu}\delta\ud{\beta}{\nu)} \bar{\nabla}_\gamma\bar{\nabla}_\delta \right. \notag\\
    &\phantom{=- \frac12...} \left.+\bar{g}^{\alpha(\delta}\bar{g}^{\gamma)\beta} \bar{g}_{\mu\nu}\bar{\nabla}_\gamma\bar{\nabla}_\delta - \bar{g}^{\alpha\beta}\bar{g}^{\gamma\delta} \bar{g}_{\mu\nu} \bar{\nabla}_\gamma \bar{\nabla}_\delta - \bar{g}_{\mu\nu} \bar{R}^{\alpha\beta} + \bar{R}\,\delta\ud{\alpha}{(\mu} \delta\ud{\beta}{\nu)}\right) \notag\\
    \bar{\mathcal{F}}\du{\nu}{\alpha\beta} &\ce \frac12 \left(\bar{T}^{\alpha\beta} \bar{\nabla}_\nu - \bar{g}^{\alpha\beta} \bar{T}\ud{\lambda}{\nu}\bar{\nabla}_\lambda\right)\,.
\end{align}

Our task now is to better understand these differential equations and the solution space to these equations. The key observation which allows us to make progress is to realize that the operators $\bar{\mathcal{E}}\du{\mu\nu}{\alpha\beta}$ and $\bar{\mathcal{F}}\du{\nu}{\alpha\beta}$ are exclusively constructed from background quantities. It therefore seems plausible that these operators commute with the symmetries of the background in the sense that
\begin{align}\label{eq:CommutingOperators}
    [\lie_\xi, \bar{\mathcal{E}}\du{\mu\nu}{\alpha\beta}]h_{\alpha\beta} &= 0 &\text{and} && [\lie_\xi, \bar{\mathcal{F}}\du{\nu}{\alpha\beta}]h_{\alpha\beta} &= 0\,,
\end{align}
where $\xi$ stands for a generator of translations or rotations. Indeed, we can easily prove that these commutation relations are true. Clearly, the background metric satisfies
\begin{align}
    \lie_\xi \bar{g}_{\mu\nu} = 0\,.
\end{align}
As we saw in subsection~\ref{ssec:TProposition}, this implies that the Ricci tensor and the Ricci scalar are also invariant under rotations and translations:
\begin{align}
    \lie_\xi \bar{R}_{\mu\nu} &= 0 &\text{and} && \lie_\xi \bar{R} = 0\,.
\end{align}
Finally, the only remaining quantity appearing in the Lichnerowicz operator is the covariant derivative with respect to the Christoffel symbols of the background metric. As we know from the identity~\eqref{eq:CommutatorLieNabla}, the commutator of the Lie derivative with any torsion-free covariant derivative operator is proportional to the Lie derivative of the Christoffel symbols of that operator. In our case, we obtain
\begin{align}\label{eq:CommutatorBackgroundNabla}
    [\lie_\xi, \bar{\nabla}_\mu]h_{\alpha\beta} &= \PD{\left(\bar{\nabla}_\mu h_{\alpha\beta}\right)}{\bar{\Gamma}\ud{\lambda}{\rho\sigma}} \lie_\xi \bar{\Gamma}\ud{\lambda}{\rho\sigma} = 0\,,
\end{align}
because the Lie derivative of the background Christoffel symbols vanishes. Notice that identity~\eqref{eq:CommutatorLieNabla} is true for any tensor $\Psi\ud{\bullet}{\circ}$, including a tensor of the form $\Psi_{\nu\alpha\beta} \ce \bar{\nabla}_\nu h_{\alpha\beta}$. It follows that
\begin{align}
    [\lie_\xi, \bar{\nabla}_\mu]\Psi_{\nu\alpha\beta} = \PD{\left(\nabla_\mu \Psi_{\nu\alpha\beta}\right)}{\bar{\Gamma}\ud{\lambda}{\rho\sigma}}\lie_\xi \bar{\Gamma}\ud{\lambda}{\rho\sigma} = 0\,.
\end{align}
This equation together with~\eqref{eq:CommutatorBackgroundNabla} now implies
\begin{align}
    \lie_\xi\left(\nabla_\mu\nabla_\nu h_{\alpha\beta}\right) = \lie_\xi\left(\nabla_\mu \Psi_{\nu\alpha\beta}\right) &= \nabla_\mu\left(\lie_\xi \Psi_{\nu\alpha\beta}\right) \notag\\
    &= \nabla_\mu \nabla_\nu \left(\lie_\xi h_{\alpha\beta}\right)\,.
\end{align}
We have therefore successfully shown that every operator appearing in the Lichnerowicz operator commutes with the Lie derivative $\lie_\xi$. A similar reasoning applies to the operator $\bar{F}\du{\nu}{\alpha\beta}$, where we need to use the fact that the Lie derivative of the background energy-momentum tensor vanishes. Thus, we have shown that the equations~\eqref{eq:CommutingOperators} are indeed correct.

The importance of this result lies in the observation that if $h_{\mu\nu}$ and $\tau_{\mu\nu}$ are solutions to the linearized Einstein and conservation equations, then $\lie_\xi h_{\mu\nu}$ and $\lie_\xi \tau_{\mu\nu}$ are also solutions. To see this, we substitute $\lie_\xi h_{\mu\nu}$ and $\lie_\xi \tau_{\mu\nu}$ into the equations and we use the commutators~\eqref{eq:CommutingOperators} \emph{en reverse}:
\begin{align}
    \bar{\mathcal{E}}\du{\mu\nu}{\alpha\beta}\left(\lie_\xi h_{\alpha\beta}\right) - 8\pi \left(\lie_\xi \tau_{\mu\nu}\right) &= \lie_\xi\left(\bar{\mathcal{E}}\du{\mu\nu}{\alpha\beta} h_{\alpha\beta}\right) - 8\pi \left(\lie_\xi \tau_{\mu\nu}\right) \notag\\
    &= \lie_\xi\underbrace{\left(\bar{\mathcal{E}}\du{\mu\nu}{\alpha\beta} h_{\alpha\beta} - 8\pi \tau_{\mu\nu} \right)}_{=0} = 0 \notag\\
    \bar{\nabla}_\mu \left(\lie_\xi \tau\ud{\mu}{\nu}\right) - \bar{\mathcal{F}}\du{\nu}{\alpha\beta}\left(\lie_\xi h_{\alpha\beta}\right)& = \lie_\xi\left(\bar{\nabla}_\mu \tau\ud{\mu}{\nu}\right) - \lie_\xi\left(\bar{\mathcal{F}}\du{\nu}{\alpha\beta} h_{\alpha\beta}\right) \notag\\
    &= \lie_\xi\underbrace{\left(\bar{\nabla}_\mu \tau\ud{\mu}{\nu} - \bar{\mathcal{F}}\du{\nu}{\alpha\beta} h_{\alpha\beta}\right)}_{=0} = 0\,.
\end{align}
We used that the terms in the round brackets vanish, since $h_{\alpha\beta}$ and $\tau_{\alpha\beta}$ are by assumption solutions of~\eqref{eq:CoupledSystem}. 

The fact that the operators $\bar{\mathcal{E}}\du{\mu\nu}{\alpha\beta}$ and $\bar{\mathcal{F}}\du{\nu}{\alpha\beta}$ commute with $\lie_\xi$ can be used to greatly simplify the equations we wish to solve. In linear algebra, whenever two matrices commute, we know that it is possible to find a vector space basis which simultaneously diagonalizes both matrices. In the present case, we have commuting linear operators acting on a function space. However, this function space can also be interpreted as an infinite dimensional vector space and our task is to find basis functions which diagonalize both operators. 

\paragraph{The Laplace equation as example:} To illustrate the general ideas and techniques, we consider the Laplace equation restricted to a cylindrical surface of unit radius as simple warm-up example. Given a scalar field $h$ in coordinates $(\phi,z)$, this equation reads
\begin{align}
    \Delta h = \left(\pd^2_\phi + \pd^2_z\right)h = 0\,.
\end{align}
The Laplace operator $\Delta$ commutes with the Lie derivative along the generator of rotations in the $\phi$-direction, i.e.,
\begin{align}
    [\lie_{\pd_\phi}, \Delta]h = 0\,.
\end{align}
These are two commuting linear operators, and we seek a set of basis functions $\{f_i\}$ such that the action of $\lie_{\pd_\phi}$ and $\Delta$ on $h$ becomes diagonal. This is achieved by imposing that the function $\{f_i\}$ are eigenfunctions of one of the operators. Concretely, we impose
\begin{align}\label{eq:EigenfunctionCondition}
    \lie_{\pd_\phi} f_i = \lambda\,f_i\,
\end{align}
for some real or complex constant $\lambda$. Solving this condition will give us a set of eigenfunctions which in turn allows us to expand $h$ in terms of an eigenbasis. The effect of this operation is that we organize the solution space of the Laplace equation in a way which respects the symmetries of the operator $\Delta$. Having a well-organized solution space helps in actually solving the equation and pinpointing a specific solution. Moreover, this condition has the great advantage that it simplifies the Laplace equation by effectively eliminating $\phi$ from $\Delta$. Intuitively, $\Delta$ is invariant under rotations along the $\phi$-direction, and thus there should be a way of rendering it independent of $\phi$. The above eigenfunction condition achieves precisely this.

To see this, we solve the eigenfunction condition by integration, which yields
\begin{align}
    f_\lambda(\phi, z) = \e^{\lambda\, \phi} H_\lambda(z)
\end{align}
for some arbitrary real or complex valued function $H_\lambda(z)$ and where we chose to label the functions $f_i$ in the set $\{f_i\}$ by their eigenvalue $\lambda$. This is possible since every eigenvalue gives rise to a distinct function $f_\lambda\propto \e^{\lambda\,\phi}$. Thus, our basis is described by the uncountable set $\{f_\lambda\}_{\lambda\in\mathbb{K}}$ with $\mathbb{K} = \bbR$ or $\mathbb{K} = \mathbb{C}$. 

Because we restricted the Laplace equation to a cylinder, we need to impose that the function $h(\phi, z)$, which lives on that cylinder, is $2\pi$-periodic in $\phi$. That is, we need to impose the condition $h(0, z) \overset{!}{=} h(2\pi, z)$ for all values of $z$. This is only possible if the basis functions $f_\lambda$ satisfy this condition, which leads to
\begin{align}
    \e^{0} \overset{!}{=} \e^{2\pi\,\lambda} \quad\Longrightarrow\quad \lambda = i\,m\quad\text{with } m\in\bbZ\,.
\end{align}
This means that our uncountable set of basis functions actually becomes countable and it can be relabeled by the integer $m$, i.e., the basis is given by
\begin{align}
    \{f_m(\phi,z)\}_{m\in\bbZ} = \{\e^{i\,m\,\phi} H_m(z)\}_{m\in\bbZ}\,.
\end{align}
Now that we have found an eigenbasis, we can express $h$ as an infinite linear combination of these basis functions:
\begin{align}
    h(\phi, z) = \sum_{m\in\bbZ} \e^{i\, m\,\phi} H_m(z)\,.
\end{align}
By plugging this expansion into the Laplace equation, we obtain
\begin{align}\label{eq:SimplifiedLaplaceEq}
    \left(\pd^2_z - m^2\right) \sum_{m\in\bbZ}\e^{i\,m\,\phi}H_m(z) = 0\,.
\end{align}
As anticipated, the $\phi$-dependence of the Laplace operator was eliminated, leaving us with a simpler differential equation to solve. The $\phi$-dependence can even be completely eliminated by noting that $\{\e^{i\,m\,\phi}\}_{m\in\bbZ}$ are linearly independent. Therefore, the only way the sum over $\e^{i\,m\,\phi} \left(\pd^2_z - m^2\right)H_m(z)$ can be zero, is if every term $\left(\pd^2_z - m^2\right)H_m(z)$ is zero individually. In other words, we end up with the infinite set of second order ordinary differential equations
\begin{align}
    \left(\pd^2_{z} - m^2\right)H_{m}(z) = 0 \quad \text{for } m\in\bbZ\,.
\end{align}
What we have achieved is that $\phi$ completely dropped out of the equations. Since there is a symmetry in the $\phi$-direction, we would intuitively expect that this variable drops out. Moreover, and perhaps more importantly, we have not only diagonalized $\lie_{\pd_\phi}$, but also $\Delta$, which led to a simpler differential equation. When we say that $\Delta$ is diagonalized, we mean that it does not mix functions with different values of $m$. Rather, the operator $(\pd^2_z-m^2)$ acts on each term in the expansion of $h$ individually.

\subsection{Diagonalization of the operators \texorpdfstring{$\bar{\mathcal{E}}\du{\mu\nu}{\alpha\beta}$}{E} and \texorpdfstring{$\bar{\mathcal{F}}\du{\nu}{\alpha\beta}$}{F}}
These ideas and techniques are directly applicable in our current context. Just as in the simple example of the Laplace equation, we begin by seeking a basis of eigenfunctions $f_{\alpha\beta}$ in which to expand $h_{\alpha\beta}$ and~$\tau_{\alpha\beta}$. We impose
\begin{align}\label{eq:EigenfunctionConditionf}
    \lie_{\xi_i} f^{(\lambda_i)}_{\alpha\beta} \overset{!}{=} \lambda_i\, f^{(\lambda_i)}_{\alpha\beta}\,,
\end{align}
where $\lambda_i$ is a label, rather than an index, and $\xi_i$ is either one of the generators of rotations or of translations. Given that we work in Cartesian coordinates and with $k=0$, the generators of translations possess a very simple form (see equation~\eqref{eq:GeneratorsOfTranslationsCartesian}). Thus, if we choose the generators of translations for $\xi$, the eigenfunction condition~\eqref{eq:EigenfunctionConditionf} is straightforward to integrate, giving us
\begin{align}
    f^{(\boldsymbol{\lambda})}_{\alpha\beta}(\eta, \x) = \e^{\boldsymbol{\lambda}\cdot\x} e^{(\boldsymbol{\lambda})}_{\alpha\beta}(\eta)\,,
\end{align}
where $\boldsymbol{\lambda} = (\lambda_1, \lambda_2, \lambda_3)$ and $e^{(\boldsymbol{\lambda})}_{\alpha\beta}(\eta)$ is an arbitrary tensor which is symmetric in $\alpha$ and $\beta$, and which depends on conformal time $\eta$. This tensor is also labeled by $\boldsymbol{\lambda}$ to emphasize that for each eigenvalue $\lambda_i$ we can have a different $e_{\alpha\beta}$. Moreover, the vector $\boldsymbol{\lambda}$ can be either real or complex. Unlike the example of the Laplace equation, where we considered functions $h(\phi, z)$ restricted to a cylinder, in cosmological perturbation theory we have no such restriction. This restriction forced $\lambda$ to be purely imaginary. In perturbation theory, however, we need to make this choice based on other arguments.

If we choose $\boldsymbol{\lambda}$ to be real, our basis would consist of functions which grow or decay exponentially. We can choose such a basis, but this makes the physical interpretation more opaque, since exponential growth is rather unphysical and exponential decay does not occur in most realistic situations. Similarly, if we chose a basis where the $\lambda_i$ has a complex \emph{and} and real part, we end up describing perturbations by oscillatory contributions which are exponentially growing or exponentially decaying. Again, this makes the physical interpretation more difficult. However, we can chose the $\lambda_i$ to be purely imaginary. This allows us to expand the perturbations in terms of oscillatory functions, which is the same as saying that we end up describing them in terms of a Fourier transform.  

Therefore, let us choose $\boldsymbol{\lambda} = i\, \k$, where $\k$ is a real-valued vector living in three-dimensional Euclidean space. This turns the eigenbasis into
\begin{align}\label{eq:Eigenbasisf}
    \{f^{(\k)}_{\alpha\beta}(\eta, \x)\}_{\k\in\bbR^3} = \{\e^{i\,\k\cdot \x} e^{(\k)}_{\alpha\beta}(\eta)\}_{\k\in\bbR^3}\,.
\end{align}
Given a symmetric tensor field $f_{\alpha\beta}(\eta, \x)$, we can now expand it in this basis. Since $\k$ takes on continuous rather than discrete values, the sum over $\k$ is really an integral:
\begin{align}\label{eq:ExpansionInEigenbasis}
    f_{\alpha\beta}(\eta, \x) = \int_{\bbR^3} \e^{i\,\k\cdot \x} e^{(\k)}_{\alpha\beta}(\eta)\,\dd^3 \k\,.
\end{align}
By expanding the tensor $f_{\alpha\beta}(\eta, \x)$ in this way, we make sure that the action of $\lie_{\T_i}$ for $i=1,2,3$, where $\T_i$ generates translations in the direction $i$, is diagonalized. The key factor to ensure this property is the exponential $\e^{i\, \k\cdot\x}$, since it satisfies the eigenvector condition for any choice of $e^{(\k)}_{\alpha\beta}(\eta)$. We have therefore the liberty to choose $e^{(\k)}_{\alpha\beta}(\eta)$ in a way which is convenient to us. By setting $e^{(\k)}_{\alpha\beta}(\eta) = \hat{f}_{\alpha\beta}(\eta, \k)$, where $\hat{f}_{\alpha\beta}$ are the Fourier coefficients of $f_{\alpha\beta}$, we obtain
\begin{align}\label{eq:Expansionf}
    f_{\alpha\beta}(\eta, \x) = \int_{\bbR^3} \e^{i\,\k\cdot \x} \hat{f}_{\alpha\beta}(\eta,\k)\, \dd^3\k\,.
\end{align}
The right-hand side of this equation is simply the Fourier transform of $\hat{f}_{\alpha\beta}(\eta, \k)$. Moreover, because all $\x$-dependence is contained in the exponential, every spatial derivative in the Lichnerowicz operator or in $\bar{\mathcal{F}}\du{\nu}{\alpha\beta}$ becomes a multiplication by a component of $\k$ once we plug in the expansion~\eqref{eq:Expansionf} for $h_{\alpha\beta}$ and $\tau_{\alpha\beta}$. Consequently, just as in the case of the Laplace equation, the operators $\bar{\mathcal{E}}\du{\mu\nu}{\alpha\beta}$ and $\bar{\mathcal{F}}\du{\nu}{\alpha\beta}$ become diagonal in $\k$-space: modes with different $\k$ do not mix.

In addition, all spatial derivatives are converted into algebraic factors involving $\k$. The linearized Einstein and conservation equations therefore reduce to ordinary differential equations in conformal time $\eta$, rather than partial differential equations in both space and time.

It is worth pausing to reflect on the conceptual meaning of this step. We started from the linearized Einstein and conservation equations, seeking to understand their solution space. This led us to study the operators $\bar{\mathcal{E}}\du{\mu\nu}{\alpha\beta}$ and $\bar{\mathcal{F}}\du{\nu}{\alpha\beta}$, which depend only on the background quantities and are therefore invariant under spatial translations and rotations. Drawing on linear algebra, we know that commuting linear operators can be simultaneously diagonalized. To make this explicit, we imposed the eigenfunction condition~\eqref{eq:EigenfunctionConditionf} and constructed a basis~\eqref{eq:Eigenbasisf}, giving the expansion~\eqref{eq:ExpansionInEigenbasis}.

Of course, one could skip this detour and directly plug the Fourier transform of $h_{\alpha\beta}$ and $\tau_{\alpha\beta}$ into the linearized equations, as is often done in the literature. While valid, that approach may seem arbitrary or unmotivated to someone studying cosmological perturbation theory for the first time. Our discussion shows that the Fourier transform emerges naturally as a basis that diagonalizes the linear operators due to the translational symmetry of the background.

In fact, using the Fourier transform is not strictly necessary. Any expansion in terms of exponential functions $\e^{\boldsymbol{\lambda}\cdot \x}$---with $\boldsymbol{\lambda}$ real, imaginary, or complex---would achieve the same diagonalization. The coefficients $e^{(\k)}_{\alpha\beta}(\eta)$ can be kept arbitrary; the key point is that the operators $\bar{\mathcal{E}}\du{\mu\nu}{\alpha\beta}$ and $\bar{\mathcal{F}}\du{\nu}{\alpha\beta}$ do not mix modes and this can be seen in any basis of the form~\eqref{eq:Eigenbasisf}. The Fourier transform is simply a convenient, well-known choice that allows us to exploit familiar results from Fourier analysis.

An alternative, more formal perspective uses representation theory and Schur's lemma. A linear operator commuting with the action of a symmetry group (here, translations and rotations) must act as a multiple of the identity on each irreducible representation\footnote{The formal statement is as follows: Let $G$ be a group, $V$ a vector space, and $\rho:G\to GL(V,\mathbb{K})$ an irreducible representation over the field $\mathbb{K}$. Then every endomorphism $f:V\to V$ commuting with $\rho$ must be a multiple of the identity, i.e., $f \rho(g) = \rho(g) f$ for all $g\in G$ implies $f=\lambda\mathbb{1}$ for some $\lambda\in\mathbb{K}$.}. This immediately implies that the operators $\bar{\mathcal{E}}\du{\mu\nu}{\alpha\beta}$ and $\bar{\mathcal{F}}\du{\nu}{\alpha\beta}$ do not mix modes, and each $\k$-mode can be treated independently as an ordinary differential equation in $\eta$. While elegant, this approach is less intuitive for those without a background in representation theory. By spelling out the steps in terms of commuting operators and eigenfunctions, we gain a more transparent and concrete understanding of what is happening, as we did above. Of course our approach is linked to representation theory: The background is invariant with respect to the group of rotations and the group of translations in $\bbR^3$. We can expand any tensor in terms of irreducible representations of the three-dimensional translation group, which are simply given by $\e^{i\k\cdot\x}$. Notice that this means the translation group is abelian and only possesses one-dimensional irreducible representations. Finally, because our linearized operators commute with the irreducible representations, different modes do not mix.

\subsection{The Transformation-Behavior of Modes under Rotations}
The basis~\eqref{eq:Eigenbasisf} introduced in the previous subsection diagonalizes the Lie derivative with respect to translations as well as the operators $\bar{\mathcal{E}}\du{\mu\nu}{\alpha\beta}$ and $\bar{\mathcal{F}}\du{\nu}{\alpha\beta}$ entering the linearized Einstein and conservation equations. However, it fails to simultaneously diagonalize the Lie derivative with respect to rotations. First of all, the commutator of Lie derivatives with respect to rotations and translations is in general not zero:
\begin{align}
    [\lie_{\R_i}, \lie_{\T_j}] = \epsilon_{ijk}\lie_{\T_k}\,.
\end{align}
Thus, we should not expect these operators to be simultaneously diagonalizable. Secondly, we can verify by a direct computation that $\lie_{\R_i}$ is not diagonal in the basis~\eqref{eq:Eigenbasisf}. To that end, we determine the action of $\lie_{\R_x}$ on a single basis element:
\begin{align}
    \lie_{\R_i}\left(\e^{i\,\k\cdot\x} e^{(\k)}_{\alpha\beta}\right) &= i\lie_{\R_i}\left(\k\cdot\x\right)\e^{i\,\k\cdot \x} e^{(\k)}_{\alpha\beta} + \e^{i\,\k\cdot\x}\lie_{\R_i}e^{(\k)}_{\alpha\beta}\notag\\
    &= \left[i\,(\x\times\k)_i\, \delta\du{\alpha}{\sigma} \delta\du{\beta}{\rho} +2 \delta\du{(\alpha}{\sigma}\pd_{\beta)} \R^{\rho}_i \right] \e^{i\,\k\cdot\x}e^{(\k)}_{\sigma\rho}\,,
\end{align}
where $(\x\times\k)_i$ denotes the $i$-th element of the vector
\begin{align}
    \x\times\k = 
    \begin{pmatrix}
        y\,k_z - z\,k_y \\
        z\, k_x - x\, k_z \\
        x\,k_y - y\, k_x
    \end{pmatrix}\,.
\end{align}
Observe that the right hand side of this equation fails to be a \emph{constant} multiple of the basis element $\e^{i\,\k\cdot\x} e^{(\k)}_{\alpha\beta}$ because the square bracket is a complex function of $\x$ and because the Kronecker deltas imply a sum over different components of $e^{(\k)}_{\alpha\beta}$. The term $\lie_{\R_i}\left(\k\cdot\x\right)$ explicitly mixes modes with different $\k$, reflecting the fact that $\lie_{\R_i}$ generates infinitesimal rotations that transform $\k$ into a rotated vector $\k'$. 

A clearer and more geometric view of this mode mixing arises by considering how the Fourier transform of $h_{\alpha\beta}$ behaves under a finite rotation $R_{\n}(\theta)$ around an axis $\n$ by an angle $\theta$. In four dimensions, this rotation acts on the coordinates as
\begin{align}
    (\eta, \x) \quad \mapsto\quad (\eta, R_{\n}(\theta)\x) = (\eta, \hat{\n} (\hat{\n}\cdot \x) + \cos\theta (\hat{\n}\times\x)\times\hat{\n} + \sin\theta (\hat{\n}\times\x))\,,
\end{align}
where $\hat{\n}$ is the vector $\n$ renormalized such that it has unit length, $\|\hat{\n}\| = 1$.
As expected, rotations leave the time coordinate unaffected and any multiple of the vector $\n$ is left invariant:
\begin{align}\label{eq:AxisInvariance}
    R_{\n}(\theta)\n = \hat{\n} \underbrace{(\hat{\n}\cdot \n)}_{=\|\n\|} + \cos\theta \underbrace{(\hat{\n}\times\n)}_{=\boldsymbol{0}}\times\hat{\n} + \sin\theta \underbrace{(\hat{\n}\times\n)}_{=\boldsymbol{0}} = \n\,.
\end{align}
Under a generic diffeomorphism $x^\mu  \mapsto x'^\mu$, a rank-$2$ tensor field $h_{\alpha\beta}$ transforms as
\begin{align}
    h_{\alpha\beta}(x) \quad\mapsto\quad h'_{\alpha\beta}(x) = \PD{x'^\mu}{x^\alpha}\PD{x'^\nu}{x^\beta} h_{\mu\nu}(x')\,.
\end{align}
In our specific case, the Jacobian matrix $\PD{x'^\mu}{x^\alpha}$ has components
\begin{align}
    \PD{x'^0}{x^0} &= \PD{\eta'}{\eta} = 1 & \PD{x'^0}{x^{i}} &= \PD{\eta'}{x^{i}} = 0 \notag\\
    \PD{x'^{i}}{x^0} &= \PD{x'^{i}}{\eta} = 0 & \PD{x'^{i}}{x^{j}} &= \PD{\left(R_{\n}(\theta)\x\right)^{i}}{x^{j}} = [R_{\n}(\theta)]\ud{i}{j} \,,
\end{align}
where $[R_{\n}(\theta)]\ud{i}{j}$ denotes, by a slight abuse of notation, the matrix obtained by differentiating the $i$-th component of the vector $R_{\n}(\theta)\x$ with respect to the $j$-th component of $\x$. We then find that the components of $h_{\alpha\beta}$ transform as
\begin{align}
    h'_{00}(\eta, \x) &= h_{00}(\eta, R_{\n}\x) & h'_{0i}(\eta, \x) = [R_{\n}]\ud{j}{i}\,h_{0j}(\eta, R_{\n}\x) \notag\\
    h'_{ij}(\eta, \x) &= [R_{\n}]\ud{k}{i}\,[R_{\n}]\ud{\ell}{j}\, h_{k\ell}(\eta, R_{\n}\x)\,,
\end{align}
where we suppressed the argument $\theta$ for better readability. One observes that different components of $h_{\alpha\beta}$ transform differently under rotations: The $00$-component transforms like a scalar field, the $0i$-components like a three-dimensional vector field, and the $ij$-components like a three-dimensional rank-$2$ tensor field. We will return to this behavior further below. 

What is more important for the time being, is the observation that the argument of each transformed component is $R_{\n}\x$, not $\x$. Substituting this into the Fourier transform gives
\begin{align}
    h_{\alpha\beta}(\eta, R_{\n}\x) &= \int_{\bbR^3} \e^{i\, \k\cdot(R_{\n}\x)} \hat{h}_{\alpha\beta}(\eta, \k)\dd^3 \k \notag\\
    &= \int_{\bbR^3} \e^{i\, \k\cdot\x} \hat{h}_{\alpha\beta}(\eta, R_{\n}\k)\dd^3 \k\,.
\end{align}
To get from the first to the second line, we performed a change of integration variables of the form $\k\mapsto\k' = R_{\n}\k$. Since $R_{\n}$ is an orthogonal matrix, one obtains
\begin{align}
    (R_{\n}\k)\cdot(R_{\n}\x) &= \k\cdot\x &&\text{and} & \dd^3\k' = |\det(R_{\n})| \dd^3\k = \dd^3 \k\,.
\end{align}
Hence, a rotation in real space induces a transformation of the Fourier modes according to
\begin{align}
    \hat{h}_{\alpha\beta}(\eta, \k) \quad\mapsto\quad \hat{h}_{\alpha\beta}(\eta, R_{\n}\k)\,,
\end{align}
showing explicitly that rotations mix different $\k$-modes.

Recall that our goal is to find a simple way to organize the solution space and that the basis we selected allows us to study the equations mode by mode. Even tough we saw it is impossible to find a basis which simultaneously diagonalized the action of $\lie_{\T_i}$ and $\lie_{\R_i}$, we can at least find a way of characterizing how a given mode transforms under rotations.

To that end, we will temporarily work exclusively in Fourier space. A given mode $\hat{h}_{\alpha\beta}(\eta,\k)$ transforms under a rotation as
\begin{align}
    \hat{h}'_{00}(\eta, \k) &= \hat{h}_{00}(\eta, R_{\n}\k) & \hat{h}'_{0i}(\eta, \k) = [R_{\n}]\ud{j}{i}\,\hat{h}_{0j}(\eta, R_{\n}\k) \notag\\
    \hat{h}'_{ij}(\eta, \k) &= [R_{\n}]\ud{k}{i}\,[R_{\n}]\ud{\ell}{j}\, \hat{h}_{k\ell}(\eta, R_{\n}\k)\,.
\end{align}
Notice that if we choose $\n$ to align with $\k$, then these transformation laws simplify to
\begin{align}
    \hat{h}'_{00}(\eta, \k) &= \hat{h}_{00}(\eta, \k) & \hat{h}'_{0i}(\eta, \k) = [R_{\k}]\ud{j}{i}\,\hat{h}_{0j}(\eta, \k) \notag\\
    \hat{h}'_{ij}(\eta, \k) &= [R_{\k}]\ud{k}{i}\,[R_{\k}]\ud{\ell}{j}\, \hat{h}_{k\ell}(\eta, \k)\,.
\end{align}
because $R_{\k}$ leaves $\k$ invariant, as we saw in~\eqref{eq:AxisInvariance}. Evidently, if we choose a rotation axis $\n\propto \k$, the mode mixing disappears and we conclude that under rotations the $00$-component transforms as a scalar, the $0i$-components as a vector, and the $ij$-components as a rank-$2$ tensor. Notice that unlike before, here we are talking about actual scalars, vectors, and tensors, rather than scalar fields, vector fields, and tensor fields.

This suggests the following approach to organize our solution space: At the level of a fixed mode $\k$, and working in Fourier space, we can classify how Fourier coefficients transform under rotations about the axis $\k$. In other words, instead of studying how Fourier modes transform under generic $SO(3)$ transformations, which introduce mode-mixing, we only consider the so-called little group consisting of rotations around a given axis $\k$. This group is the one which avoids mode mixing. 

Given a generic vector $\v$, unless it is parallel or antiparallel to $\k$, the rotation $R_{\k}(\theta)$ will not leave it invariant. Rather, $R_{\k}(\theta)$ will rotate the components of $\v$ which lie in the plane orthogonal to $\k$ by an angle $\theta$, while leaving the component of $\v$ parallel to $\k$ invariant. This intuitive fact suggests to decompose $\v$ as
\begin{align}\label{eq:VectorDecomposition}
    \v &= \v_\parallel + \v_\perp &\text{with} && \v_\parallel &\ce (\v\cdot\k)\,\k  && \text{ and } & \v_\perp &\ce \v - \frac{1}{\|k\|^2} (\v\cdot\k)\,\k\,.
\end{align}
This decomposition guarantees that $\v_\parallel$ is parallel or antiparallel to $\k$, depending on the sign of $(\v\cdot\k)$, while $\v_\perp$ lives in the plane perpendicular to $\k$, which is the same as saying that its scalar product with $\k$ vanishes, $\v_\perp\cdot\k = 0$.

It follows that any vector $\v$ can be decomposed into a component $\v_\parallel$ which transforms as a scalar under $R_{\k}(\theta)$ and a piece $\v_\perp$ which transforms as a vector:
\begin{align}
    R_{\k}(\theta)\v_\parallel &= (\v\cdot\k)\,\underbrace{R_{\k}(\theta)\k}_{=\k} = (\v\cdot\k)\,\k = \v_\parallel \notag\\
    R_{\k}(\theta)\v_\perp &= \underbrace{R_{\k}(\theta)\v}_{\neq \v} - \frac{1}{\|\k\|^2}(\v\cdot\k) \underbrace{R_{\k}(\theta)\k}_{=\k} \neq \v_\perp\,.
\end{align}
Notice that, as expected and intuitively clear, the rotated vector $\v_\perp$ still lies in the plane orthogonal to $\k$. That is, its scalar product with $\k$ vanishes:
\begin{align}
    (R_{\k}\v_\perp)\cdot \k &= (R_{\k}\v)\cdot \k - \frac{1}{\|\k\|^2} (\v\cdot\k)\, (\k\cdot\k) \notag\\
    &= (R_{\k}\v)\cdot \k - (R_{\k}\v)\cdot(R_{\k}\k) \notag\\
    &= (R_{\k}\v)\cdot \k - (R_{\k}\v) \cdot\k = 0\,,
\end{align}
where we used $\k\cdot\k = \|\k\|^2$ on the first line, $(R_{\k}\v)\cdot(R_{\k}\k) = (\v\cdot\k)$ to obtain the second one, and $R_{\k}\k = \k$ for the third line.

In a similar fashion, it is possible to decompose a spatial tensor $M_{ij}$ into components invariant under $R_{\k}(\theta)$, components which transform as vectors, and components which transform as tensors. We begin by observing that any tensor can be decomposed into a trace part and a trace-free part. In three dimensional Euclidean space, this decomposition reads 
\begin{align}
    M_{ij} &= \frac13 D\, \delta_{ij} + E_{ij} &&\text{with} & D &\ce M\ud{i}{i} &&\text{and} & E_{ij}&\ce M_{ij}-\frac13 D\, \delta_{ij}\,,
\end{align}
where $D$ is the trace of $M$, $E$ its trace-free part, and $\delta$ the Euclidean metric. Under a rotation $R_{\k}(\theta)$, the trace part remains invariant, while the trace-free part changes. To see which component of the trace-free part transform in which manner, we decompose $E$ with respect to the orthonormal vector space basis $\{\khat, \ee_1, \ee_2\}$. In words: The vectors $\khat$, $\ee_1$ and $\ee_2$ are normalized and mutually orthogonal. With respect to this basis, and assuming that $E$ is symmetric ($E_{ij} = E_{ji}$), its six components are given by
\begin{align}\label{eq:ComponentsOfE}
    F_{00} &\ce E_{ij} \hat{k}^{i} \hat{k}^{j} \notag\\
    F_{01} &\ce E_{ij} \hat{k}^{i} e^{j}_1\,, & F_{02} &\ce E_{ij} \hat{k}^{i} e^{j}_2 \notag\\
    F_{11} &\ce E_{ij}e^{i}_1e^{j}_1\,, & F_{12} &\ce E_{ij} e^{i}_1 e^{j}_2\,, & F_{22}&\ce E_{ij}e^{i}_2 e^{j}_2\,.
\end{align}
The indices $0$, $1$, and $2$ should not be confused with spacetime indices. Rather, the index $0$ refers to $\hat{\k}$, while the indices $1$ and $2$ are associated with $\ee_1$ and $\ee_2$, respectively. The tensor $E$ can now be written as
\begin{align}
    E_{ij} = F_{00}\, \hat{k}_i \hat{k}_j + 2 F_{01}\,\hat{k}_{(i}e^1_{j)} + 2 F_{02}\, \hat{k}_{(i}e^{2}_{j)} + F_{11}\, e^1_ie^1_j + 2F_{12} e^1_{(i}e^2_{j)} + F_{22}\, e^2_i e^2_j\,.
\end{align}
This expression reproduces the relations~\eqref{eq:ComponentsOfE} when the respective contractions with $\khat$, $\ee_1$, and $\ee_2$ are formed. This expansion now allows us to read off how the individual components transform under a rotation $R_{\k}(\theta)$:
\begin{align}
    E'_{ij} &= [R_{\k}(\theta)]\ud{k}{i} [R_{\k}(\theta)]\ud{\ell}{j} E_{k\ell} \notag\\
    &= F_{00}\, \hat{k}_i \hat{k}_j + 2 F_{01}\,\hat{k}_{(i}\tilde{e}^1_{j)} + 2 F_{02}\, \hat{k}_{(i}\tilde{e}^{2}_{j)} + F_{11}\, \tilde{e}^1_i \tilde{e}^1_j + 2F_{12} \tilde{e}^1_{(i} \tilde{e}^2_{j)} + F_{22}\, \tilde{e}^2_i \tilde{e}^2_j\,,
\end{align}
where we introduced
\begin{align}
    \tilde{e}^1_i &\ce [R_{\k}(\theta)]\ud{\ell}{i}\,e^1_\ell & \text{and} && \tilde{e}^2_i &\ce [R_{\k}(\theta)]\ud{\ell}{i}\,e^2_\ell
\end{align}
as shorthand notations. We can now read off the transformation behavior and we find
\begin{align}
    F_{00} &:  \quad\text{Scalar component} \notag\\
    F_{01}, F_{02} &: \quad\text{Vector components} \notag\\
    F_{11}, F_{12}, F_{22} &: \quad \text{Tensor components}
\end{align}
With this, we have completely decomposed the Fourier mode $\hat{h}_{\alpha\beta}(\eta, \k)$ into components which transform as scalars, vectors, and tensors under rotations which leave $\k$ fixed. To summarize, we found that $\hat{h}_{00}(\eta, \k)$ transforms as a scalar, $\hat{h}_{0i}(\eta, \k)$ contains one scalar and two vector components, while $\hat{h}_{ij}(\eta, \k)$ can be decomposed into a trace part, containing one scalar component, and a trace-free part which contains one scalar, two vectors and seemingly three tensor components. Together, this gives us four scalar, four vector, and three tensor components. Hence, eleven components in total for a tensor containing only ten components. The counting error stems from neglecting the fact that $E_{ij}$ is trace-less by definition, which translates into
\begin{align}\label{eq:TracelessCondition}
    \delta^{ij}E_{ij} = F_{00} + F_{11} + F_{22} = 0\,.
\end{align}
Hence, one of the tensor components is eliminated by the trace-freeness of $E$, leaving us with only two tensors and thus ten components in total.

In summary, when studying the linearized Einstein and conservation equations in Fourier space, each mode $\k$ can be further classified according to its transformation behavior under rotations about $\k$. Because rotations about $\k$ do not mix scalar, vector, and tensor components, we can decompose the equations---and their solution space---into three independent sectors: scalar, vector, and tensor. This decomposition considerably simplifies both the analysis and the physical interpretation of cosmological perturbations.

In the next subsection, we will translate this classification from Fourier space to real space, arriving at the well-known scalar–vector–tensor (SVT) decomposition.

\subsection{The Scalar-Vector-Tensor Decomposition}
Given a symmetric tensor field $h_{\alpha\beta}(\eta, \x)$ in real space, we know from the previous subsection that its Fourier modes contain four scalar, four vector, and two tensor components. In particular, we know that the $\hat{h}_{00}(\eta, \k)$ component transforms as a scalar, $\hat{h}_{0i}(\eta,\k)$ transforms as a vector, and $\hat{h}_{ij}(\eta,\k)$ transforms as a tensor. 

It is customary to introduce a scalar potential $\hat{\Phi}(\eta, \k)$, defined via
\begin{align}
    \hat{h}_{00}(\eta,\k) = -2\hat{\Phi}(\eta, \k)\,.
\end{align}
The prefactor of $-2$ is just a convention. Furthermore, as we know from the previous subsection, the vector $\hat{h}_{0i}(\eta,\k)$ can be further decomposed into a component transforming as a scalar, and two components transforming as vectors. In analogy with equation~\eqref{eq:VectorDecomposition}, we define
\begin{align}
    \hat{B}(\eta,\k) &\ce i\,\hat{h}_{0i}(\eta,\k)k^{i} &\text{and} && \hat{B}^\text{t}_i(\eta,\k) \ce \hat{h}_{0i}(\eta, \k) - \frac{1}{\|k\|^2} (\hat{h}_{0\ell}(\eta, \k)k^{\ell}) k_i\,,
\end{align}
where the superscript $^\text{t}$ stands for ``transverse''. This is a reminder that $\hat{B}^\text{t}_i$ is orthogonal to $\k$, i.e., $\hat{B}^\text{t}_i k^{i} = 0$. Notice the factor of $i$ in front of $\hat{h}_{0i}$. This arises because, rather than expanding our vector in a basis where $\k$ is a real basis vector, we use the imaginary basis vector  $i\, \k$. The motivation for this choice comes from the Fourier transform: every factor of $i\, k_i$ in Fourier space corresponds to a partial derivative $\pd_i$ in real space. Furthermore, replacing $k_i$ by $i\,k_i$ converts the norm $\|\k\|^2$  on the right hand side of $\hat{B}^\text{t}_i$ into $-\|\k\|^2$, which cancels the sign arising from $(\hat{h}_{0\ell}\,i\,k^\ell)\,i\,k^\ell$. Consequently, the expression for $\hat{B}^\text{t}_i$ in the complex basis is the same as in the real basis considered earlier.

For the tensorial part $\hat{h}_{ij}(\eta, \k)$ we introduce the scalar $\hat{\Psi}$, defined as
\begin{align}
    \hat{\Psi}(\eta, \k) \ce \frac16 \delta^{ij}\hat{h}_{ij}(\eta, \k) = \frac16 D(\eta, \k)\,,
\end{align}
in order to capture the trace part. Here, $D$ is the scalar introduced in the previous subsection. For the trace-free part, we write
\begin{align}
    E_{ij}(\eta, \k) = \hat{E}^\text{S}_{ij}(\eta, \k) +  \hat{E}^\text{V}_{ij}(\eta, \k) + \hat{E}^\text{T}_{ij}(\eta, \k)\,,
\end{align}
where $E_{ij}$ is the symmetric, trace-free tensor we introduced in the previous subsection and $\hat{E}^\text{S}_{ij}$, $\hat{E}^\text{V}_{ij}$, and $\hat{E}^\text{T}_{ij}$ represent its components which transform as scalars, vectors, and tensors, respectively. 

Taking everything together, we can write the modes $\hat{h}_{\alpha\beta}(\eta, \k)$ as
\begin{align}
    \hat{h}_{\alpha\beta}(\eta, \k) =
    \begingroup
    \setlength\arraycolsep{7pt}
    \def\arraystretch{1.5}
    \begin{pmatrix}
        -2\hat{\Phi} & \hat{B}\, k_i + \hat{B}^\text{t}_i \\
        \hat{B}\, k_i + \hat{B}^\text{t}_i & 2 \hat{\Psi}\delta_{ij} + \hat{E}^\text{S}_{ij} +  \hat{E}^\text{V}_{ij} + \hat{E}^\text{T}_{ij}
    \end{pmatrix}\,.
    \endgroup
\end{align}
In order to find the corresponding decomposition in real space, we only need to compute 
\begin{align}
    h_{\alpha\beta}(\eta, \x) = \int_{\bbR^3} \e^{i\, \k\cdot\x}\hat{h}_{\alpha\beta}(\eta, \k)\dd^3\k\,.
\end{align}
For the $00$- and $0i$-components this is straightforward. Using the fact that the Fourier transform of $i\,k_i$ times a Fourier mode is equal to the partial derivative $\pd_i$ of a function in real space, we obtain
\begin{align}
    h_{00}(\eta, \x) &= -2\int_{\bbR^3} \e^{i\, \k\cdot\x} \hat{\Phi}(\eta, \k)\dd^3\k = -2\Phi(\eta, \x) \notag\\
    h_{0i}(\eta, \x) &= \int_{\bbR^3}\e^{i\, \k\cdot\x} i k_i \hat{B}(\eta, \k)\dd^3 \k + \int_{\bbR^3}\e^{i\, \k\cdot\x} \hat{B}^\text{t}_i(\eta, \k) \dd^3 \k \notag\\
    &= \pd_i B(\eta, \x) + B^\text{t}_i(\eta, \x)
\end{align}
where the vector $B^\text{t}_i$ is divergence-free, i.e., it satisfies the equation
\begin{align}
    \nabla\cdot\boldsymbol{B}^\text{t} = \delta^{ij} \pd_{j} B^\text{t}_i = 0\,.
\end{align}
This property is a direct consequence of $\hat{B}^\text{t}$ being transverse to $\k$, as can be seen by Fourier transforming $k^{i} \hat{B}^\text{t}_i=0$ to real space. 

Next, we turn our attention to $2\hat{\Psi}\delta_{ij} + \hat{E}^\text{S}_{ij} +  \hat{E}^\text{V}_{ij} + \hat{E}^\text{T}_{ij}$. In order to perform the Fourier transform, we need to make two changes compared to the previous subsection:
\begin{enumerate}
    \item Rather than expanding $E_{ij}$ in the orthonormal basis $\{\khat, \ee_1, \ee_2\}$, we have to expand in the basis $\{i\, \k, \ee_1, \ee_2\}$. The basis vectors are still mutually orthogonal and $\ee_1$ and $\ee_2$ are still normalized. However, $\k$ no longer has unit length. These changes are necessary so that every occurrence of $i\,k_i$ can be translated into a spatial derivative~$\pd_i$ and because in the integral we ``sum'' over every possible $\k$, not just the ones which are normalized.
    \item The tensor $E_{ij}$ is traceless, which implies a certain relation between some of its components (cf. equation~\eqref{eq:TracelessCondition}). However, in practice it is more convenient to use the definition
    \begin{align}
        E_{ij} \ce \hat{h}_{ij} - \frac13 \delta^{k\ell}\hat{h}_{k\ell}\delta_{ij}\,,
    \end{align}
    which guarantees that $E_{ij}$ is traceless, rather than imposing~\eqref{eq:TracelessCondition} on the components.
\end{enumerate}
With these changes, the expansion of $E_{ij}$ in the basis $\{i\,\k, \ee_1, \ee_2\}$ gives rise to the following three pieces:
\begin{align}
    \hat{E}^\text{S}_{ij} &\ce -\left(k_i k_j - \frac13 k_\ell k^\ell \delta_{ij}\right)\hat{S} \notag\\
    \hat{E}^\text{V}_{ij} &\ce i \left(k_i \hat{V}_j + k_j \hat{V}_i\right) \notag\\
    \hat{E}^\text{T}_{ij} &\ce \hat{T}_{11} e^1_i e^1_j + 2 \hat{T}_{12} e^1_{(i}e^2_{j)} + \hat{T}_{22}e^2_i e^2_j - \frac13 \left(\hat{T}_{11}e^1_\ell e^\ell_1 + \hat{T}_{22} e^2_\ell e^\ell_2\right)\,,
\end{align}
which under the rotation $R_{\k}(\theta)$ transform as scalar, vector, and tensor, respectively, and where we defined
\begin{align}
    \hat{S} &\ce -\frac{E_{ij}k^{i} k^{j}}{\|\k\|^4}\,, & \hat{V}_i &\ce \hat{V}_1 e^1_i + \hat{V}_2 e^2_i &\text{with} && \hat{V}_{1,2} &\ce -i \frac{E_{ij}k^{i} e^{j}_{1,2}}{\|\k\|^2} \notag\\
    \hat{T}_{11} &\ce E_{ij}e^{i}_1 e^{j}_1\,, & \hat{T}_{12} &\ce E_{ij}e^{i}_1 e^{j}_2\,, &\text{and} && \hat{T} &\ce E_{ij}e^{i}_2 e^{j}_2\,.
\end{align}
Observe that each piece individually is trace-less, i.e.,
\begin{align}
    \delta^{ij}\hat{E}^\text{S}_{ij} &= 0\,, & \delta^{ij}\hat{E}^\text{V}_{ij} &= 0\,, &&\text{and} & \delta^{ij}\hat{E}^\text{T}_{ij} &= 0\,.
\end{align}
Moreover, the piece $\hat{E}^\text{T}_{ij}$ which transforms as a tensor under rotations around the axis $\k$ is transverse to that vector:
\begin{align}
    k^{i} \hat{E}^\text{T}_{ij} = 0\,.
\end{align}
The same is not true for the other two pieces, even tough the vector $\hat{V}_i$ is orthogonal to $\k$:
\begin{align}
    k^{i} \hat{V}_i = 0\,.
\end{align}
We can now finally perform a Fourier transform to obtain the corresponding decomposition of $E_{ij}$ in real space. For the scalar piece we find
\begin{align}
    E^\text{S}_{ij}(\eta, \x) &= -\int_{\bbR^3} \e^{i\,\k\cdot\x} \left(k_i k_j - \frac13 k_\ell k^\ell \delta_{ij}\right)\hat{S} \dd^3\k \notag\\
    &= \left(\pd_i \pd_j - \frac13 \delta_{ij}\pd_\ell \pd^\ell \right)\int_{\bbR^3} \e^{i\,\k\cdot\x} \hat{S}(\eta, \k)\dd^3 \k \notag\\
    &= \left(\pd_i \pd_j - \frac13 \delta_{ij}\pd_\ell \pd^\ell \right) S(\eta, \x)\,.
\end{align}
Similarly, we compute for the vectorial piece
\begin{align}
    E^\text{V}_{ij} &= i \int_{\bbR^3}\e^{i\,\k\cdot\x} \left(k_i \hat{V}(\eta, \k) + k_j \hat{V}(\eta, \k)\right)\dd^3 \k \notag\\
    &= \pd_i \int_{\bbR^3} \e^{i\,\k\cdot\x} \hat{V}_j(\eta, \k)\dd^3\k + \pd_j \int_{\bbR^3} \e^{i\,\k\cdot\x} \hat{V}_i(\eta, \k)\dd^3\k \notag\\
    &= \pd_i V_j(\eta, \x) + \pd_j V_i(\eta, \x)\,.
\end{align}
Observe that the orthogonality of $\hat{V}_i$ with respect to $\k$ is translated into the divergenclessness of the vector $V_i$:
\begin{align}
    \nabla\cdot\boldsymbol{V} = \pd^{i}V_i &= 0\,.
\end{align}
Finally, for the tensorial piece we simply define
\begin{align}
    h^\text{tt}_{ij}(\eta, \x) &\ce \int_{\bbR^3} \e^{i\, \k\cdot\x} \hat{E}^\text{T}_{ij}(\eta, \k)\dd^3 \k\,.
\end{align}
The superscript $^\text{tt}$ stands for ``transverse-traceless''. That $h^\text{tt}_{ij}$ is traceless is clear, since it inherits this property from $\hat{E}^\text{T}_{ij}$. Transversality in this context means that
\begin{align}
    \pd^{i} h^\text{tt}_{ij}(\eta,\x) = 0\,,
\end{align}
which it inherits from the orthogonality of $\hat{E}^\text{T}_{ij}$ with respect to $\k$. 

We can finally express $h_{\alpha\beta}$ through its scalar-vector-tensor decomposition:
\begin{align}\label{eq:SVT}
    h_{\alpha\beta} =
    \begingroup
    \setlength\arraycolsep{7pt}
    \def\arraystretch{1.5}
    \begin{pmatrix}
        -2\Phi & & \pd_i B + B^\text{t}_i \\[5pt]
        \pd_i B + B^\text{t}_i & & 2\Psi \delta_{ij} + \left(\pd_i \pd_j -\frac13 \delta_{ij}\pd_\ell \pd^\ell\right) S + 2 \pd_{(i}V_{j)} +  h^\text{tt}_{ij}
    \end{pmatrix}\,.
    \endgroup   
\end{align}
This decomposition expresses $h_{\alpha\beta}$ in terms of four scalar fields $\Phi$, $\Psi$, $B$, $S$; two divergenceless vector fields $B^\text{t}_i$ and $V_j$ (contributing four components in total); and a symmetric, transverse, and traceless tensor $h^\text{tt}_{ij}$, which contributes two additional components corresponding to gravitational waves.

With the scalar-vector-tensor (SVT) decomposition~\eqref{eq:SVT} in real space we achieve a separation of the linearized Einstein and conservation equations~\eqref{eq:LinearizedEquations} into different sectors. In Fourier space we argued that fields which transform differently under rotations about $\k$ cannot be mixed, which leads to a separation of the equations into a scalar sector, a vector sector, and a tensor sector. The decomposition~\eqref{eq:SVT} achieves the same separation in real space: four equations arise in the scalar sector, four in the vector sector, and two in the tensor sector.

In equations \ref{eq:CoupledSystem} there is also $\tau_{\mu\nu}$ from the matter fields apart from the metric perturbations $h_{\mu\nu}$.
At this point it is convenient to express $\tau_{\mu\nu}$ in terms of variables which facilitate a physical interpretation. Give that $\bar{T}_{\mu\nu}$ must have the form of an energy-momentum tensor of a perfect fluid described by an energy density $\bar{\rho}$ and a pressure $\bar{p}$, it is only natural to seek a description of $\tau$ in terms of fluid variables (for a derivation, see appendix~\ref{app:B}):
\begin{align}\label{eq:TauParametrization}
    \tau\ud{0}{0} &= -\delta \rho(\eta, \x) && \text{(energy density perturbation)} \notag\\
    \tau\ud{0}{j} &= (\bar{\rho}+\bar{p})\,v_j(\eta, \x) && \text{(momentum density perturbations)} \notag\\
    \tau\ud{i}{0} &= -(\bar{\rho}+\bar{p})\, v^{i}(\eta, \x) && \text{(momentum flux perturbations)} \notag\\
    \tau\ud{i}{j} &= \delta p(\eta, \x)\, \delta\ud{i}{j} + \pi\ud{i}{j}(\eta, \x) && \text{(stress tensor perturbations)}\,.
\end{align}
We used the background metric $\bar{g}_{\mu\nu}$ to raise the first index and we introduced the energy density perturbation $\delta \rho(\eta, \x)$, the pressure perturbation $\delta p(\eta, \x)$, the peculiar velocity field $v^{i}(\eta, \x)$, and the anisotropic stress tensor $\pi_{ij}(\eta, \x)$. The latter is a spatial tensor encoding the perturbations in off-diagonal components of the energy-momentum tensor. Formally, we can define it by subtracting the diagonal elements $T\ud{k}{k}\,\delta\ud{i}{j}$ from $T\ud{i}{j}$:
\begin{align}
    \pi\ud{i}{j} &\ce T\ud{i}{j} - T\ud{k}{k}\,\delta\ud{i}{j} = T\ud{i}{j} - \left(\bar{p} + \delta p\right)\, \delta\ud{i}{j}\,.
\end{align}
It follows that the anisotropic stress is symmetric ($\pi_{ij} = \pi_{ji}$) and trace-less by definition ($\pi\ud{k}{k} = 0$). Counting degrees of freedom, $\tau_{\mu\nu}$ introduces ten new functions to parametrize the energy-momentum perturbations:
\begin{itemize}
    \item Two scalar fields, $\delta \rho(\eta, \x)$ and $\delta p(\eta, \x)$;
    \item One three-dimensional vector field $v^{i}(\eta, \x)$;
    \item One symmetric, trace-free spatial tensor $\pi_{ij}$, which due to these properties only has five independent components. 
\end{itemize}
The number of functions needed to parametrize the matter perturbations matches the number introduced to model the metric perturbations, since $h_{\mu\nu}$ possesses ten independent components. 

Evidently, both equations \ref{eq:CoupledSystem} are linear in $h_{\mu\nu}$ and $\tau_{\mu\nu}$, which is an improvement compared to the full set of non-linear Einstein equations. However, we need to address a mismatch between the number of functions we are trying to solve for and the number of equations we have at our disposal: The linearized Einstein equations constitute ten second order partial differential equations, while the linearized conservation law provides only an additional four first order equations. Thus, there are $10+4$ equations for the $10+10$ fields $h_{\mu\nu}$ and $\tau_{\mu\nu}$.

We can understand the nature of the problem in physical terms: The equation $\nabla_\mu T\ud{\mu}{\nu} = 0$ only tells us something about the conservation of energy and momentum of the matter system, but nothing about its internal thermodynamics. The two quantities which are most closely related with notions of energy and momentum are the energy density fluctuation $\delta\rho(\eta,\x)$ and the velocity perturbation $v^{i}(\eta, \x)$. We shall therefore interpret the second equation in~\eqref{eq:CoupledSystem} as evolution equations for these quantities. The pressure fluctuations $\delta p(\eta, \x)$ and the anisotropic stress $\pi\ud{i}{j}(\eta, \x)$ remain undetermined.

To remedy this situation, we need to introduce constitutive relations which tell us how pressure and stress respond to density and velocity perturbations. In the case of pressure, what is needed is an equation of state $p = p(\rho)$, while for the anisotropic stress we need to resort to kinetic theory and the Boltzmann equations. Conceptually, we have been describing the matter content of the Universe as a fluid in terms of coarse grained observables---energy density, pressure, (fluid) velocity, and anisotropic stress. However, matter is composed of individual particles moving in space and interacting with each other. Kinetic theory provides a statistical description of these particles, which is governed by the Boltzmann equation \cite{Ma:1995ey}.

Thus, we assume that an equation of state and an expression for the anisotropic stress has been provided. Concerning the concrete matter fields and their perturbations,
in the next few subsections, for completeness, we also discuss the SVT decomposition of scalar fields, vector fields, and $2$-forms.

\subsection{Scalar Fields}
The SVT decomposition of scalar fields is trivial, as one might intuitively expect. A scalar field can of course be represented by its Fourier transform:
\begin{align}
    \Phi(\eta, \x) = \int_{\bbR^3} \e^{i\, \k\cdot\x} \hat{\Phi}(\eta, \k) \dd^3\k\,.
\end{align}
Under rotations about the axis $\k$, each Fourier mode $\hat{\Phi}(\eta, \k)$ transforms as a scalar. Thus, as expected, the SVT decomposition of $\Phi$ is trivial.

\subsection{Vector Fields}
Following the derivation of the SVT decomposition for the metric, we anticipate that a generic vector field can be decomposed into two scalar functions and a transverse vector component. This can be seen by first expressing $A_\mu$ as\footnote{Using the $1$-form $A_\mu$ rather than the vector field $A^\mu$ is merely a matter of convenience. All arguments apply equally to $A^\mu$.}
\begin{align}\label{eq:FTA}
    A_\mu(\eta, \x) = \int_{\bbR^3}\e^{i\,\k\cdot\x} \hat{A}_\mu(\eta, \k) \dd^3\k\,.
\end{align}
Under a rotation about the axis $\k$, the modes $\hat{A}_\mu(\eta, \k)$ transform as
\begin{align}
    \hat{A}'_0(\eta, \k) &= \hat{A}_0(\eta, \k)\,, 
    &\text{and} && 
    \hat{A}_{i}(\eta, \k) &= [R_{\k}]\ud{j}{i}\hat{A}_{j}(\eta, \k)\,.
\end{align}
As expected, the temporal component is invariant, while the spatial part $\hat{A}_i(\eta, \k)$ transforms as a vector.  

Introducing the orthogonal basis $\{i\,\k, \ee_1, \ee_2\}$, we can decompose the spatial part as
\begin{align}
    \hat{A}_{i}(\eta, \k) &= i\,\hat{B}(\eta, \k)\, \k_i + \hat{B}^\text{t}_i(\eta, \k)\,, 
    &\text{with} && 
    \hat{B}^\text{t}_i \ce \hat{B}^\text{t}_1 \ee_1 + \hat{B}^\text{t}_2 \ee_2\,.
\end{align}
The vector $\hat{B}^\text{t}$ is orthogonal to $\k$, which justifies the superscript $^\text{t}$ for ``transverse''. Substituting this decomposition into~\eqref{eq:FTA}, we obtain the corresponding expression in real space:
\begin{align}
    A_i(\eta, \x) &= \int_{\bbR^3} \e^{i\,\k\cdot\x}\!\left(i\,\hat{B}(\eta, \k)\, \k_i + \hat{B}^\text{t}_i(\eta, \k)\right) \dd^3\k = \pd_i B(\eta, \x) + B^\text{t}_i(\eta, \x)\,,
\end{align}
where
\begin{align}
    B(\eta, \x) &\ce \int_{\bbR^3} \e^{i\,\k\cdot\x} \hat{B}(\eta, \k)\,\dd^3\k\,, 
    &\text{and} && 
    B^\text{t}_i(\eta, \x) &\ce \int_{\bbR^3} \e^{i\,\k\cdot\x}\hat{B}^\text{t}_i(\eta, \k)\,\dd^3\k\,.
\end{align}
The transverse vector field $B^\text{t}_i(\eta, \x)$ satisfies
\begin{align}
    \pd^{i} B^\text{t}_i(\eta, \x) = 0\,,
\end{align}
a property it inherits from the condition $k^{i} \hat{B}^\text{t}_i(\eta, \k) = 0$.  

Finally, if we relabel the temporal component $A_0(\eta, \x)$ as $\Phi(\eta, \x)$---to emphasize that it transforms as a scalar---the SVT decomposition of $A_\mu$ can be written as
\begin{align}
    A_\mu(\eta, \x) &= \big(\Phi(\eta, \x),\, \pd_i B(\eta, \x) + B^\text{t}_i(\eta, \x)\big)\,.
\end{align}
As expected, this decomposition involves two scalar functions, $\Phi$ and $B$, together with a transverse vector field $B^\text{t}$. Altogether, they represent four degrees of freedom, equal to the number of components of~$A_\mu$. How many of those would propagate at the end will depend on the Lagrangian. A massless vector field should propagate two and a massive vector field three.

\subsection{\texorpdfstring{$2$-Forms}{2-Forms}}
To derive the SVT decomposition of a $2$-form $B_{\mu\nu}$, we start with its Fourier representation:
\begin{align}\label{eq:FTB}
    B_{\mu\nu}(\eta, \x) = \int_{\bbR^3} \e^{i\, \k\cdot\x} \hat{B}_{\mu\nu}(\eta, \k)\,\dd^3\k\,.
\end{align}
Next, we examine how the modes $\hat{B}_{\mu\nu}(\eta, \k)$ transform under rotations about the axis $\k$. Since rotations act only on spatial indices, we might naively expect the $00$-component to transform as a scalar, the $0i$-components as vectors, and the $ij$-components as spatial tensors.

However, because $\hat{B}_{\mu\nu}$ is anti-symmetric, there is no $00$-component. The remaining components transform as
\begin{align}
    \hat{B}'_{0i}(\eta, \k) &= [R_{\k}]\ud{j}{i}\hat{B}_{0j}(\eta, \k) &\text{and} && \hat{B}'_{ij}(\eta, \k) &= [R_{\k}]\ud{k}{i}[R_{\k}]\ud{\ell}{j} \hat{B}_{k\ell}(\eta, \k)\,.
\end{align}
Since $\hat{B}_{0i}$ transforms as a spatial vector, we can immediately write its real-space decomposition as
\begin{align}
    B_{0i}(\eta, \x) &= \pd_i C(\eta, \x) + C^\text{t}_i(\eta, \x) &\text{with} && \pd^{i}C^\text{t}_i &= 0\,.
\end{align}

To decompose $\hat{B}_{ij}$, we follow the procedure used for the spatial part of the metric. Normally, one would split a tensor into trace and trace-free parts, but $\hat{B}_{ij}$ is automatically trace-free due to anti-symmetry. We therefore directly expand it in the orthogonal basis $\{i\,\k, \ee_1, \ee_2\}$. Its non-zero components are
\begin{align}
    \hat{D}_{01} &\ce i\, \hat{B}_{ij} k^{i} \ee^{j}_1\,, & \hat{D}_{02} &= i\, \hat{B}_{ij}k^{i}\ee^{j}_2\,, & \hat{D}_{12} &\ce \hat{B}_{ij}\ee^{i}_1 \ee^{j}_2\,.
\end{align}
so that
\begin{align}\label{eq:BijDecomp}
    \hat{B}_{ij} &= 2 i\, \hat{D}_{01} k_{[i}\ee^1_{j]} + 2i\, \hat{D}_{02} k_{[i}\ee^2_{j]} + 2\,\hat{D}_{12}\ee^1_{[i}\ee^2_{j]}\,,
\end{align}
where the factors of $2$ compensate for the $\frac12$ in the anti-symmetrization of $i$ and $j$. Fourier-transforming the first two terms in~\eqref{eq:BijDecomp} gives
\begin{align}
    2i&\int_{\bbR^3}\e^{i\,\k\cdot\x}\left(\hat{D}_{01}(\eta, \k) k_{[i}\ee^1_{j]} + \hat{D}_{02}(\eta, \k)k_{[i}\ee^2_{j]}\right)\dd^3\k \notag\\
    &= \pd_i D_{01}(\eta, \x) \ee^1_j - \pd_j D_{01}(\eta, \x) \ee^1_i + \pd_i D_{02}(\eta, \x) \ee^2_j - \pd_j D_{02}(\eta, \x) \ee^2_i \notag\\
    &= \pd_i\left[D_{01}(\eta, \x) \ee^1_j + D_{02}(\eta, \x) \ee^2_j\right] - \pd_j\left[D_{01}(\eta, \x) \ee^1_i + D_{02}(\eta, \x) \ee^2_i\right]\,.
\end{align}
The last line suggests the introduction of the transverse vector 
\begin{align}
    S^\text{t}_i(\eta, \x) \ce D_{01}(\eta, \x) \ee^1_i + D_{02}(\eta, \x) \ee^2_i\,.
\end{align}
In compact form:
\begin{align}
    2i\int_{\bbR^3}\e^{i\,\k\cdot\x}\left(\hat{D}_{01}(\eta, \k) k_{[i}\ee^1_{j]} + \hat{D}_{02}(\eta, \k)k_{[i}\ee^2_{j]}\right)\dd^3\k &= \pd_i S^\text{t}_j(\eta, \x) - \pd_j S^\text{t}_i(\eta, \x)\,,
\end{align}
with $\pd^i S^\text{t}_i = 0$ because the Fourier modes are orthogonal to $\k$. Thus, this vector encodes two degrees of freedom.

The third term in~\eqref{eq:BijDecomp} can be written as
\begin{align}\label{eq:Option1}
    2&\int_{\bbR^3}\e^{i\,\k\cdot\x} \hat{D}_{12}(\eta, \k) \ee^1_{[i}\ee^2_{j]}\dd^3\k = \epsilon_{ij} \Phi(\eta, \x)\,,
\end{align}
where
\begin{align}
    \Phi(\eta, \x) &\ce \int_{\bbR^3}\e^{i\,\k\cdot\x} \hat{D}_{12}(\eta, \k)\dd^3\k &\text{and} && \epsilon_{ij} &\ce 2 \ee^1_{[i}\ee^2_{j]}\,.
\end{align}
Equivalently, one can parametrize this term as
\begin{align}\label{eq:Option2}
    2&\int_{\bbR^3}\e^{i\,\k\cdot\x} \hat{D}_{12}(\eta, \k) \ee^1_{[i}\ee^2_{j]}\dd^3\k = \epsilon_{ijk}\pd^k \beta(\eta,\x)\,.
\end{align}
The key point is that the decomposition~\eqref{eq:BijDecomp} contains one anti-symmetric tensor component, which requires a scalar field and anti-symmetrization. Both~\eqref{eq:Option1} and~\eqref{eq:Option2} achieve this.

Thus, the SVT decomposition of $B_{\mu\nu}$ can be expressed in terms of two scalars, $C$ and $\Phi$ (or $\beta$), and two transverse vectors, $C^\text{t}_i$ and $S^\text{t}_i$, yielding six degrees of freedom---exactly the number of independent, non-zero components of $B_{\mu\nu}$.

Choosing the parametrization~\eqref{eq:Option2} for concreteness, we can write the SVT of $B_{\mu\nu}$ as
\begin{align}
    B_{0i}(\eta, \x) &= \pd_i C(\eta, \x) + B^\text{t}_i(\eta, \x) \notag\\
    B_{ij}(\eta, \x) &= \pd_i S^\text{t}_j(\eta, \x) - \pd_j S^\text{t}_i(\eta, \x) + \epsilon_{ijk}\pd^k \beta(\eta, \x)\,.
\end{align}
In the literature one also often encounters an alternative, but equivalent SVT decomposition for $2$-forms. The decomposition for $B_{0i}$ remains unchanged. However, to simplify the discussion of the $B_{ij}$ part, one can use the fact that these components can be written as
\begin{align}
    B_{ij}(\eta, \x) &= \epsilon_{ijk} b^k(\eta, \x)\,,
\end{align}
for some vector $b^k(\eta, \x)$. Thus, the problem of finding the SVT decomposition of $B_{ij}$ is reduced to finding the decomposition for $b^k$. This is a problem we already solved:
\begin{align}
    b^k(\eta, \x) = \pd^k \beta(\eta, \x) + b^k_\text{t}(\eta, \x)\,,
\end{align}
where $\pd_k b^k_\text{t}(\eta, \x) = 0$. Hence, the SVT decomposition becomes
\begin{align}
    B_{0i}(\eta, \x) &= \pd_i C(\eta, \x) + B^\text{t}_i(\eta, \x) \notag\\
    B_{ij}(\eta, \x) &= \epsilon_{ijk}\left(\pd^k \beta(\eta, \x) + b^k_\text{t}(\eta, \x)\right)\,,
\end{align}
featuring two scalar fields, $C$ and $\beta$, and two transverse vectors, $C^\text{t}_i$ and $b^i_\text{t}$.

\section{Conclusion}\label{sec:Conclusion}
\begin{displayquote}
\emph{``Every theoretical physicist who is any good knows six or seven different theoretical representations for exactly the same physics.''}\\
\phantom{}\hfill-- Richard P. Feynman
\end{displayquote}

This paper has been largely conceptual in nature, re-examining familiar subjects from a new perspective. Its three main themes are closely connected. The first, developed in sections~\ref{sec:Symmetries} and~\ref{sec:Metric}, concerns the role of homogeneity and isotropy as spacetime symmetries. We explored how these symmetries can be formulated in terms of $1$-parameter families of diffeomorphisms and their generating vector fields, and we clarified the conceptual distinction between rotations and translations. In section~\ref{sec:MatterFields} we then demonstrated how well-known results follow from a systematic symmetry reduction of various fields and of a generic energy-momentum tensor.

These considerations culminated in a new derivation of the FLRW metric which, to the best of our knowledge, has not appeared previously in the literature. Unlike the standard derivations found in standard textbooks~\cite{EinsteinBook, WeinbergBook, MisnerThorneWheelerBook, HawkingEllisBook, WaldBook, SchutzBook, WeinbergCosmologyBook, CarrollBook, MalamentBook, RovelliBook, BaumannBook, Blau}, our treatment does not rely on properties of maximally symmetric spaces. Instead, we impose the rotational and translational Killing equations for the metric.

The rotational sector poses no difficulties and is well documented~(see, e.g., \cite{CarrollBook, DAmbrosio:2021}). The real challenge lies in enforcing translational symmetry, because the corresponding Killing vector fields are not known a priori. As argued in section~\ref{sec:Symmetries}, postulating translational Killing vectors without first determining the metric is essentially equivalent to postulating the metric itself.

We addressed this difficulty in subsection~\ref{ssec:GenVecFRotAndTransl} by analyzing the conceptual meaning of translations in a general curved spacetime $(\M, g_{\mu\nu})$. This led to an algebraic characterization of translations which, together with spherical symmetry of $g_{\mu\nu}$, allowed us to constrain the form of the translational generators. In subsection~\ref{ssec:ImposingTranslInv}, using a general spherically symmetric metric as a starting point, we solved the Killing equations for translations; both for certain metric components and for the functions parametrizing the generators. In this way we obtained simultaneously the FLRW metric and its translational Killing vectors for $k \in {-1,0,1}$.

The second theme concerns the symmetries of the Einstein tensor and, more generally, of the Hilbert energy-momentum tensor. This was the focus of section~\ref{sec:SymmetryReducedT}. We argued that, rather than imposing symmetries on matter fields, it is equally natural to impose them directly on the energy-momentum tensor. There are two main reasons: first, many matter fields are gauge fields, making symmetry requirements at the level of the fields themselves subtle; and second, the metric couples to matter only via the energy-momentum tensor.

Subsection~\ref{ssec:TProposition} presents and proves two intuitive propositions. The first states that if a metric admits a given set of continuous symmetries, then the Einstein tensor inherits those symmetries. The second states that if the metric and matter fields share a set of continuous symmetries, then the Hilbert energy-momentum tensor inherits them as well.

One might expect the converse to hold: that if $G_{\mu\nu}$ and $T_{\mu\nu}$ possess a given set of continuous symmetries, then the metric and matter fields themselves must share them. However, as we demonstrated in subsections~\ref{ssec:MaxwellEMT} and~\ref{ssec:T2Form} by means of explicit counterexamples, this expectation fails for the energy-momentum tensor. In both the Maxwell and Kalb–Ramond cases, the homogeneous and isotropic forms of $T_{\mu\nu}$ admit field configurations that break these symmetries. The Kalb–Ramond example is particularly striking: the resulting field configuration is significantly more general than those appearing in the literature and does not reduce to a pure gauge.

The potential cosmological implications of these findings remain open to future work. Another question we believe to be important concerns the Einstein tensor itself: given $G_{\mu\nu}$ with continuous symmetries generated by vector fields $\xi$, which metric configurations are compatible with it? In particular, can one construct a metric that violates one or more of the symmetries of $G_{\mu\nu}$, in the sense that $\lie_\xi g_{\mu\nu} \neq 0$? Imposing homogeneity and isotropy directly on the Einstein tensor may, in principle, admit metric configurations that are themselves neither homogeneous nor isotropic. If such configurations exist, they would carry significant implications for cosmology, potentially revealing hidden geometric structures compatible with symmetric gravitational dynamics. Exploring this possibility is technically demanding, and we leave this intriguing and potentially far-reaching question to future work.

The third theme of this paper concerns cosmological perturbation theory and the well-known scalar–vector–tensor (SVT) decomposition. Section~\ref{sec:SVT} begins with a general and conceptually oriented discussion of metric and matter perturbations, the linearized Einstein equations, and the linearized conservation laws.
Our aim was to give an accessible and pedagogical treatment of perturbation theory, with particular emphasis on the logic underlying the SVT decomposition. Our exposition departs from standard presentations and aims to provide a fresh perspective. We analyzed the structure of the linearized Einstein equations and conservation equations, their solution space, and the most transparent way to organize perturbations. Along the way, analogies with simpler equations were used to clarify the main technical tools.

These efforts culminate in an elementary and transparent derivation of the SVT decomposition, in which each step is conceptually motivated and clearly explained. Our derivations of both the FLRW metric and the SVT decomposition reflects Feynman's point, as we provide new theoretical lenses through which the same physics can be illuminated from different perspectives.

\newpage
\appendix
\section{Important Results on Lie Derivatives}\label{app:A}
The proofs of Propositions~\ref{prop:Gmunu} and~\ref{prop:Tmunu} rely on three identities, which we repeat here for convenience and which we prove for the sake if a self-contained exposition. The first identity concerns the Lie derivative of the connection:
\begin{align}
    \lie_\xi \Gamma\ud{\alpha}{\mu\nu} = \frac12 g^{\alpha\lambda} \left(\nabla_\mu \left(\lie_\xi g_{\nu\lambda}\right) + \nabla_\nu \left(\lie_\xi g_{\mu\lambda}\right) - \nabla_\lambda\left(\lie_\xi g_{\mu\nu}\right)\right)\,.
\end{align}
Here, $\Gamma\ud{\alpha}{\mu\nu}$ is the unique torsion-free and metric compatible connection. As is well-known, it is given by
\begin{align}\label{eq:Gammag}
    \Gamma\ud{\alpha}{\mu\nu} = \frac12 g^{\alpha\lambda}\left(\pd_\mu g_{\nu\lambda} + \pd_\nu g_{\mu\lambda} - \pd_\lambda g_{\mu\nu}\right)\,.
\end{align}
The second identity is the Palatini identity, which relates the Lie derivative of the Ricci tensor to the Lie derivative of a torsion-free connection $\Gamma\ud{\alpha}{\mu\nu}$. Finally, we also derive a formula for the commutator of the Lie derivative with an arbitrary torsion-free connection. This formula plays a role in the proof of Proposition~\ref{prop:Tmunu}, as well as in our analysis of the linearized Einstein and conservation equations in subsection~\ref{ssec:AnalyzingLinearizedEqs}.

\subsection{The Lie Derivative of the Connection}\label{app:A_LieDChristoffel}
In order to prove the above identity, it is most convenient to exploit the metric-compatibility of the Levi-Civita connection:
\begin{align}\label{eq:MetComp}
    \nabla_\alpha g_{\mu\nu} = 0\,.
\end{align}
Our strategy is to apply a $1$-parameter family of diffeomorphisms $\phi_s:I\times\M\to \M$ with $\phi_{s=0}= \text{id}$ to the metric and to study how the metric-compatibility condition behaves in a neighborhood of $s=0$. 

By applying such a diffeomorphism to the metric, we can generate a new metric which depends on the parameter $s$,
\begin{align}
    g^{(s)}_{\mu\nu} \ce (\phi^*_s g)_{\mu\nu}\,,
\end{align}
where $^*$ denotes the pull-back operation. Notice that for $s=0$ we recover the original metric. Moreover, any change we induce in the metric will propagate to the connection, since it depends on $g_{\mu\nu}$ and its derivatives (see~\eqref{eq:Gammag}). Thus, $\Gamma\ud{\alpha}{\mu\nu}$ also depends on $s$ and $\nabla_\alpha$ inherits this dependence as well. 

However, diffeomorphisms cannot change geometric facts. This means that $\nabla^{(s)}_\alpha$ and $g^{(s)}_{\mu\nu}$ still need to satisfy the metric-compatibility condition:
\begin{align}
    \nabla^{(s)}_\alpha g^{(s)}_{\mu\nu} = 0\,.
\end{align}
This condition needs to hold for all values of $s$ for which the $1$-parameter family of diffeomorphisms is defined. In particular, if $\phi_s$ is smooth in $s$, we can Taylor-expand the above condition up to first order in $s$. Since the zeroth order term is simply~\eqref{eq:MetComp}, thanks to $\phi_{s=0} = \text{id}$, the Taylor expansion becomes
\begin{align}
    \left.\frac{\dd}{\dd s}\nabla^{(s)}_\alpha g^{(s)}_{\mu\nu}\right|_{s=0} = 0\,.
\end{align}
Writing out the covariant derivative, we obtain
\begin{align}\label{eq:MetCompWrittenOut}
    \left.\frac{\dd}{\dd s}\left(\pd_\alpha g^{(s)}_{\mu\nu} - \Gamma\ud{(s)\lambda}{\alpha\mu}g^{(s)}_{\lambda\nu} - \Gamma\ud{(s)\lambda}{\alpha\nu}g^{(s)}_{\lambda\mu}\right)\right|_{s=0} &=0\,.
\end{align}
Next, we expand $g^{(s)}_{\mu\nu}$ and $\Gamma\ud{(s)\alpha}{\mu\nu}$ up to first order in $s$ around $s=0$:
\begin{align}\label{eq:ExpgandGamma}
    g^{(s)}_{\mu\nu} &= g_{\mu\nu} + s\,\lie_\xi g_{\mu\nu} + + \mathcal{O}(s^2) &\text{and} && \Gamma\ud{(s)\lambda}{\alpha\beta} &= \Gamma\ud{\lambda}{\alpha\beta} + s\, \lie_\xi \Gamma\ud{\lambda}{\alpha\beta} + \mathcal{O}(s^2)\,,
\end{align}
where we introduced the generating vector field
\begin{align}
    \xi \ce \left.\frac{\dd\phi_s}{\dd s}\right|_{s=0}
\end{align}
and the Lie derivatives
\begin{align}
    \lie_\xi g_{\mu\nu} &\ce \left.\frac{\dd}{\dd s} g^{(s)}_{\mu\nu}\right|_{s=0} & \text{and} && \lie_\xi \Gamma\ud{\alpha}{\mu\nu} &\ce \left.\frac{\dd}{\dd s} \Gamma\ud{(s)\alpha}{\mu\nu}\right|_{s=0}\,.
\end{align}
If we substitute~\eqref{eq:ExpgandGamma} into~\eqref{eq:MetCompWrittenOut}, we see that terms of second order or higher drop out, since we evaluate~\eqref{eq:MetCompWrittenOut} at $s=0$. Only the zeroth and first order terms of~\eqref{eq:ExpgandGamma} survive and we obtain
\begin{align}\label{eq:FirstOrderMetComp}
    \left.\frac{\dd}{\dd s}\nabla^{(s)}_\alpha g^{(s)}_{\mu\nu}\right|_{s=0} &= \left.\frac{\dd}{\dd s}\left(\pd_\alpha g^{(s)}_{\mu\nu} - \Gamma\ud{(s)\lambda}{\alpha\mu}g^{(s)}_{\lambda\nu} - \Gamma\ud{(s)\lambda}{\alpha\nu}g^{(s)}_{\lambda\mu}\right)\right|_{s=0} \notag\\
    &= \pd_\alpha \left(\lie_\xi g_{\mu\nu}\right) - \Gamma\ud{\lambda}{\alpha\mu} \left(\lie_\xi g_{\lambda\nu}\right) - \Gamma\ud{\lambda}{\alpha\nu} \left(\lie_\xi g_{\lambda\mu}\right) \notag\\
    &\phantom{=} - \left(\lie_\xi\Gamma\ud{\lambda}{\alpha\mu}\right) g_{\lambda\nu} - \left(\lie_\xi\Gamma\ud{\lambda}{\alpha\nu}\right) g_{\lambda\mu} \notag\\
    &=\quad \nabla_\alpha\left(\lie_\xi g_{\mu\nu}\right) - \left(\lie_\xi\Gamma\ud{\lambda}{\alpha\mu}\right) g_{\lambda\nu} - \left(\lie_\xi\Gamma\ud{\lambda}{\alpha\nu}\right) g_{\lambda\mu} = 0\,.
\end{align}
In order to isolate $\lie_\xi\Gamma\ud{\alpha}{\mu\nu}$ from the last line of~\eqref{eq:FirstOrderMetComp}, we perform two cyclic permutation of its indices. This results in two equations of equal structure. The equation obtained from the first cyclic permutation is \emph{added} to the original equation, while the second cyclic permutation is \emph{subtracted}. This results in the following equation:
\begin{align}
    \phantom{+}\, \nabla_\alpha\left(\lie_\xi g_{\mu\nu}\right) - \left(\lie_\xi\Gamma\ud{\lambda}{\alpha\mu}\right) g_{\lambda\nu} - \left(\lie_\xi\Gamma\ud{\lambda}{\alpha\nu}\right) g_{\lambda\mu} = 0 \notag\\
    +\,\nabla_\mu\left(\lie_\xi g_{\nu\alpha}\right) - \left(\lie_\xi\Gamma\ud{\lambda}{\mu\nu}\right) g_{\lambda\alpha} - \left(\lie_\xi\Gamma\ud{\lambda}{\mu\alpha}\right) g_{\lambda\nu} = 0 \notag\\
    -\left(\,\nabla_\nu\left(\lie_\xi g_{\alpha\mu}\right) - \left(\lie_\xi\Gamma\ud{\lambda}{\nu\alpha}\right) g_{\lambda\mu} - \left(\lie_\xi\Gamma\ud{\lambda}{\nu\mu}\right) g_{\lambda\alpha}\right) = 0\notag \\
    \cline{1-2}
    \nabla_\alpha\left(\lie_\xi g_{\mu\nu}\right) + \nabla_\mu\left(\lie_\xi g_{\nu\alpha}\right) - \nabla_\nu\left(\lie_\xi g_{\alpha\mu}\right) - 2 \left(\lie_\xi\Gamma\ud{\lambda}{\alpha\mu}\right) g_{\lambda\nu} = 0\,.
\end{align}
Notice that only one Lie derivative of the connection appears on the last line. In order to isolate that Lie derivative, we contract the whole equation by $g^{\nu\rho}$ and divide by $2$. After relabeling $\mu$ as $\beta$, we obtain
\begin{align}
    \lie_\xi\Gamma\ud{\rho}{\alpha\beta} = \frac12 g^{\rho\nu}\left(\nabla_\alpha\left(\lie_\xi g_{\beta\nu}\right) + \nabla_\beta\left(\lie_\xi g_{\nu\alpha}\right) - \nabla_\nu\left(\lie_\xi g_{\alpha\beta}\right)\right)\,,
\end{align}
which is the identity we wanted to prove.

\subsection{The Lie Derivative of a Generic Torsion-Free Connection}\label{app:A_LieDGenericGamma}
In this subsection, $\bar{\Gamma}\ud{\alpha}{\mu\nu}$ is still assumed to be torsion-free, but it is no longer assumed to be metric-compatible. Thus,
\begin{align}
    \bar{\nabla}_\alpha g_{\mu\nu} \neq 0\,.
\end{align}
Our goal is to derive a compact expression for the Lie derivative of such a connection, but now we can no longer apply the strategy of the previous subsection.

From the definition of the Lie derivative of a connection (see for instance~\cite{Heisenberg:2023}), which is valid for \emph{any} affine connection, one obtains
\begin{align}\label{eq:DefLieDGamma}
    \lie_\xi \bar{\Gamma}\ud{\alpha}{\mu\nu} = \xi^\beta\pd_\beta\bar{\Gamma}\ud{\alpha}{\mu\nu} + \bar{\Gamma}\ud{\alpha}{\beta\nu} \pd_\mu\xi^\beta + \bar{\Gamma}\ud{\alpha}{\mu\beta} \pd_\nu\xi^\beta - \bar{\Gamma}\ud{\beta}{\mu\nu} \pd_\beta\xi^\alpha + \pd_\mu\pd_\nu \xi^\alpha\,.
\end{align}
It should be noted that even tough the connection is \emph{not} a tensor, its Lie derivative is a proper $(1,2)$-tensor field. To make this fact more manifest, our strategy is to cast every term in a manifestly covariant form, starting with $\pd_\mu\pd_\nu\xi^\alpha$. Using
\begin{align}\label{eq:deldelxi}
    \bar{\nabla}_\mu \bar{\nabla}_\nu \xi^\alpha &= \pd_\mu\pd_\nu \xi^\alpha + \bar{\Gamma}\ud{\alpha}{\mu\gamma}\left(\bar{\Gamma}\ud{\gamma}{\nu\beta}\xi^\beta + \pd_\nu\xi^\gamma\right) - \bar{\Gamma}\ud{\gamma}{\mu\nu}\left(\bar{\Gamma}\ud{\alpha}{\gamma\beta}\xi^\beta + \pd_\gamma\xi^\alpha\right)\notag\\
    &\phantom{=} + \bar{\Gamma}\ud{\alpha}{\nu\beta}\pd_\mu\xi^\beta + \xi^\beta \pd_\mu\bar{\Gamma}\ud{\alpha}{\nu\beta}
\end{align}
we can substitute $\pd_\mu\pd_\nu \xi^\alpha$ in~\eqref{eq:DefLieDGamma}. This results in
\begin{align}\label{eq:LieDSubstituted}
     \lie_\xi \bar{\Gamma}\ud{\alpha}{\mu\nu} &= \bar{\nabla}_\mu \bar{\nabla}_\nu \xi^\alpha + \left(\pd_\beta \bar{\Gamma}\ud{\alpha}{\mu\nu} - \pd_\mu \bar{\Gamma}\ud{\alpha}{\nu\beta} + \bar{\Gamma}\ud{\alpha}{\beta\gamma}\bar{\Gamma}\ud{\gamma}{\mu\nu} - \bar{\Gamma}\ud{\alpha}{\mu\gamma} \bar{\Gamma}\ud{\gamma}{\mu\gamma}\right)\xi^\beta
\end{align}
Observe that the terms in the bracket are of the form $\pd\bar{\Gamma} - \pd\bar{\Gamma}$ and $\bar{\Gamma}\bar{\Gamma} - \bar{\Gamma}\bar{\Gamma}$. This is the same structure we know from the curvature tensor, which is given by
\begin{align}
    \bar{R}\du{\mu\beta\nu}{\alpha} = \pd_\beta\bar{\Gamma}\ud{\alpha}{\mu\nu} - \pd_\mu\bar{\Gamma}\ud{\alpha}{\beta\nu} + \bar{\Gamma}\ud{\alpha}{\beta\gamma}\bar{\Gamma}\ud{\gamma}{\mu\nu} - \bar{\Gamma}\ud{\alpha}{\mu\gamma}\bar{\Gamma}\ud{\gamma}{\beta\nu}\,.
\end{align}
Indeed, the terms in the bracket are \emph{exactly} equal to the curvature tensor. Thus, we obtain
\begin{align}
    \lie_\xi \bar{\Gamma}\ud{\alpha}{\mu\nu} &= \bar{\nabla}_{\mu} \bar{\nabla}_{\nu}\xi^\alpha + \bar{R}\du{\mu\beta\nu}{\alpha}\xi^\beta\,.
\end{align}
This result will be useful in subsection~\ref{app:A_Commutator}.

\subsection{The Palatini Identity}\label{app:A_PalatiniIdentity}
Let $\bar{\Gamma}\ud{\alpha}{\mu\nu}$ be an affine connection which is torsion-free but not necessarily metric-compatible. The Ricci tensor associated with such a connection reads
\begin{align}\label{eq:Ricci}
    \bar{R}_{\mu\nu} = \pd_\alpha\bar{\Gamma}\ud{\alpha}{\mu\nu} - \pd_\mu \bar{\Gamma}\ud{\alpha}{\alpha\nu} + \bar{\Gamma}\ud{\alpha}{\alpha\beta}\bar{\Gamma}\ud{\beta}{\mu\nu} - \bar{\Gamma}\ud{\alpha}{\mu\beta}\bar{\Gamma}\ud{\beta}{\alpha\nu}\,.
\end{align}
Just like in~\ref{app:A_LieDChristoffel}, if we apply a $1$-parameter family of diffeomorphisms $\phi_s:I\times\M\to\M$ to the connection, it changes to first order in $s$ as
\begin{align}\label{eq:ChangeInGamma}
    \bar{\Gamma}\ud{\alpha}{\mu\nu}\quad \overset{\phi_s}{\mapsto}\quad \bar{\Gamma}\ud{\alpha}{\mu\nu} + s\, \lie_\xi\bar{\Gamma}\ud{\alpha}{\mu\nu}\,.
\end{align}
This change in the connection induces a change in the Ricci tensor. If we substitute~\eqref{eq:ChangeInGamma} into~\eqref{eq:Ricci}, the Ricci tensor changes as
\begin{align}
    \bar{R}_{\mu\nu} \overset{\phi_s}{\mapsto}\quad \bar{R}_{\mu\nu} + s &\left[
    \pd_\alpha\left(\lie_\xi \bar{\Gamma}\ud{\alpha}{\mu\nu}\right)
    - \pd_\mu\left(\lie_\xi\bar{\Gamma}\ud{\alpha}{\alpha\nu}\right)
    + \left(\lie_\xi\bar{\Gamma}\ud{\alpha}{\alpha\beta}\right) \bar{\Gamma}\ud{\beta}{\mu\nu}
    - \left(\lie_\xi\bar{\Gamma}\ud{\alpha}{\mu\beta}\right)\bar{\Gamma}\ud{\beta}{\alpha\nu} \right. \notag\\
    &\left.
    + \bar{\Gamma}\ud{\alpha}{\alpha\beta}\left(\lie_\xi\bar{\Gamma}\ud{\beta}{\mu\nu}\right)
    - \bar{\Gamma}\ud{\alpha}{\mu\beta}\left(\lie_\xi\bar{\Gamma}\ud{\beta}{\alpha\nu}\right)
    \right]\,,
\end{align}
where we only retained terms up to first order in $s$. Thus, the Lie derivative of the Ricci tensor can be written as
\begin{align}\label{eq:DefLieDRicci}
    \lie_\xi \bar{R}_{\mu\nu} &\equiv \lim_{s\to0}\frac{1}{s}\left((\phi^*_s\bar{R})_{\mu\nu} - \bar{R}_{\mu\nu}\right) \notag\\
    &= \pd_\alpha\left(\lie_\xi \bar{\Gamma}\ud{\alpha}{\mu\nu}\right)
    - \pd_\mu\left(\lie_\xi\bar{\Gamma}\ud{\alpha}{\alpha\nu}\right)
    + \left(\lie_\xi\bar{\Gamma}\ud{\alpha}{\alpha\beta}\right) \bar{\Gamma}\ud{\beta}{\mu\nu}
    - \left(\lie_\xi\bar{\Gamma}\ud{\alpha}{\mu\beta}\right)\bar{\Gamma}\ud{\beta}{\alpha\nu} \notag\\
    &\phantom{=} + \bar{\Gamma}\ud{\alpha}{\alpha\beta}\left(\lie_\xi\bar{\Gamma}\ud{\beta}{\mu\nu}\right)
    - \bar{\Gamma}\ud{\alpha}{\mu\beta}\left(\lie_\xi\bar{\Gamma}\ud{\beta}{\alpha\nu}\right)\,.
\end{align}
Next, we make use of the fact that even tough $\bar{\Gamma}\ud{\alpha}{\mu\nu}$ is \emph{not} a tensor, its Lie derivative $\lie_\xi\bar{\Gamma}\ud{\alpha}{\mu\nu}$ is a proper $(1,2)$-tensor field. Thus, the covariant derivative is well-defined and given by
\begin{align}
    \bar{\nabla}_\beta\left(\lie_\xi\bar{\Gamma}\ud{\alpha}{\mu\nu}\right) = \pd_\beta\left(\lie_\xi \bar{\Gamma}\ud{\alpha}{\mu\nu}\right) + \bar{\Gamma}\ud{\alpha}{\beta\gamma}\left(\lie_\xi \bar{\Gamma}\ud{\gamma}{\mu\nu}\right) - \bar{\Gamma}\ud{\gamma}{\beta\mu}\left(\lie_\xi \bar{\Gamma}\ud{\alpha}{\gamma\nu}\right) - \bar{\Gamma}\ud{\gamma}{\beta\nu} \left(\lie_\xi \bar{\Gamma}\ud{\alpha}{\mu\gamma}\right)\,.
\end{align}
This equation can be used to rewrite $\pd_\alpha\left(\lie_\xi \bar{\Gamma}\ud{\alpha}{\mu\nu}\right)$ and $\pd_\mu\left(\lie_\xi\bar{\Gamma}\ud{\alpha}{\alpha\nu}\right)$ in terms of $\bar{\nabla}_\alpha\left(\lie_\xi\bar{\Gamma}\ud{\alpha}{\mu\nu}\right)$ and $\bar{\nabla}_\mu\left(\lie_\xi\bar{\Gamma}\ud{\alpha}{\alpha\nu}\right)$ in~\eqref{eq:DefLieDRicci}. One also finds that the remaining terms in~\eqref{eq:DefLieDRicci} cancel, such that one is left with
\begin{align}
    \lie_\xi \bar{R}_{\mu\nu} = \nabla_\alpha\left(\lie_\xi\bar{\Gamma}\ud{\alpha}{\mu\nu}\right) - \nabla_\mu\left(\lie_\xi\bar{\Gamma}\ud{\alpha}{\alpha\nu}\right)\,.
\end{align}
This is the Palatini identity, which we wanted to prove. 

\subsection{The Commutator of a Lie Derivative and a Covariant Derivative}\label{app:A_Commutator}
Finally, we turn to the commutator of the Lie derivative and covariant derivative. We assume a torsion-free covariant derivative $\bar{\nabla}_\alpha$, but one that is not necessarily metric-compatible.

Let us recall that both Lie derivative and covariant derivative take a $(p,q)$-tensor field\footnote{Scalar fields are regarded as $(0,0)$-tensor fields.} as input and they produce a new tensor field. In the case of the Lie derivative, the new tensor field is still of type $(p,q)$. The covariant derivative, on the other hand, produces a $(p,q+1)$-tensor field. Therefore, the commutator $[\lie_\xi, \bar{\nabla}_\alpha]$, when acting in a $(p,q)$-tensor $\Psi\ud{\bullet}{\circ} \equiv \Psi\ud{\mu_1\cdots \mu_p}{\nu_1\cdots\nu_q}$, always produces a $(p,q+1)$-tensor field.

With this in mind, let us first examine the action of the commutator on a scalar field $\Psi$. Any covariant derivative acting on a scalar field simply produces
\begin{align}
    \bar{\nabla}_\alpha\Phi = \pd_\alpha\Phi\,,
\end{align}
which despite looking non-covariant, is actually a $1$-form\footnote{In differential form notation, the right hand side of this equation reads $\dd\Phi$.}. Similarly, the Lie derivative acting on a scalar field always yields
\begin{align}
    \lie_\xi \Phi = \xi^\lambda \pd_\lambda\Phi\,,
\end{align}
which is still a $0$-form, i.e., a scalar field, because the $1$-form $\pd_\lambda\Phi$ is contarcted with the vector field $\xi^\lambda$. To evaluate the commutator, we only need to know how the Lie derivative acts on $1$-forms:
\begin{align}
    \lie_\xi \omega_\alpha &= \xi^\lambda \pd_\lambda \omega_\alpha + \omega_\lambda \pd_\alpha \xi^\lambda\,.
\end{align}
Substituting $\omega_\alpha = \pd_\alpha\Phi$ into this equation gives
\begin{align}
    \lie_\xi \omega_\alpha &= \xi^\lambda\pd_\lambda \pd_\alpha\Phi + \pd_\lambda \Phi \pd_\alpha\xi^\lambda\notag\\
    &= \pd_\alpha\left(\xi^\lambda\pd_\lambda\Phi\right)\,.
\end{align}
Thus, we found that the first term of the commutator evaluates to
\begin{align}
    \lie_\xi\left(\bar{\nabla}_\alpha \Phi\right) &= \pd_\alpha\left(\xi^\lambda\pd_\lambda\Phi\right)\,.
\end{align}
To evaluate the second term, we need to apply the covariant derivative to $\xi^\lambda\pd_\lambda\Phi$. As we have note above, this is a scalar field and thus the covariant derivative $\bar{\nabla}_\alpha$ is simply converted to a partial derivative $\pd_\alpha$. In other words, the second term of the commutator evaluates to
\begin{align}
    \bar{\nabla}_\alpha\left(\lie_\xi\Phi\right) = \pd_\alpha\left(\xi^\lambda\pd_\lambda\Phi\right)\,.
\end{align}
This is the same expression we obtained for the first term, which thus means the commutator acting on a scalar field vanishes:
\begin{align}
    [\lie_\xi, \bar{\nabla}_\alpha]\Phi = 0\,.
\end{align}
This result will come in handy further below, when we determine the action of the commutator on a $1$-form. Now, however, we turn our attention to computing the commutator
\begin{align}
    [\lie_\xi, \bar{\nabla}_\alpha]V^\mu = \lie_\xi\left(\bar{\nabla}_\alpha V^\mu\right) - \bar{\nabla}_\alpha\left(\lie_\xi V^\mu\right)\,,
\end{align}
where $V^\mu$ is a vector field. As before, we compute both terms separately and subtract them in the end. For the first term we need to compute the Lie derivative of the $(1,1)$-tensor field $\bar{\nabla}_\alpha V^\mu$:
\begin{align}\label{eq:FirstTermV}
    \lie_\xi\left(\bar{\nabla}_\alpha V^\mu\right) &= \xi^\beta \bar{\nabla}_\beta \bar{\nabla}_\alpha V^\mu - (\bar{\nabla}_\beta\xi^\mu) \bar{\nabla}_\alpha V^\beta + (\bar{\nabla}_\alpha\xi^\beta)\bar{\nabla}_\beta V^\mu\,,
\end{align}
Notice that we expressed the partial derivatives contained in $\lie_\xi$ by the covariant derivative~$\bar{\nabla}_\alpha$. This is allowed because we assumed the covariant derivative to be torsion-free. 

For the second term we first express $\lie_\xi V^\mu$ in terms of the covariant derivative $\bar{\nabla}_\beta$, and then we take the $\bar{\nabla}_\alpha$ derivative:
\begin{align}\label{eq:SecondTermV}
    \bar{\nabla}_\alpha\left(\lie_\xi V^\mu\right) &= \bar{\nabla}_\alpha\left(\xi^\beta\bar{\nabla}_\beta V^\mu - (\bar{\nabla}_\beta\xi^\mu)V^\beta\right)\notag\\
    &= (\bar{\nabla}_\alpha \xi^\beta)\bar{\nabla}_\beta V^\mu + \xi^\beta \bar{\nabla}_\alpha \bar{\nabla}_\beta V^\mu - (\bar{\nabla}_\alpha \bar{\nabla}_\beta \xi^\mu)V^\beta - (\bar{\nabla}_\beta \xi^\mu) \bar{\nabla}_\alpha V^\beta\,.
\end{align}
To form the commutator, we subtract the second term~\eqref{eq:SecondTermV}
from the first term~\eqref{eq:FirstTermV}:
\begin{align}
    [\lie_\xi, \bar{\nabla}_\alpha]V^\mu &= \lie_\xi\left(\bar{\nabla}_\alpha V^\mu\right) - \bar{\nabla}_\alpha\left(\lie_\xi V^\mu\right) \notag\\
    &= \xi^\beta \bar{\nabla}_\beta \bar{\nabla}_\alpha V^\mu - (\bar{\nabla}_\beta\xi^\mu) \bar{\nabla}_\alpha V^\beta + (\bar{\nabla}_\alpha\xi^\beta)\bar{\nabla}_\beta V^\mu \notag\\
    &\phantom{=} -\left((\bar{\nabla}_\alpha \xi^\beta)\bar{\nabla}_\beta V^\mu + \xi^\beta \bar{\nabla}_\alpha \bar{\nabla}_\beta V^\mu - (\bar{\nabla}_\alpha \bar{\nabla}_\beta \xi^\mu)V^\beta - (\bar{\nabla}_\beta \xi^\mu) \bar{\nabla}_\alpha V^\beta\right) \notag\\
    &= \xi^\beta \bar{\nabla}_\beta \bar{\nabla}_\alpha V^\mu - \xi^\beta \bar{\nabla}_\alpha \bar{\nabla}_\beta V^\mu + (\bar{\nabla}_\alpha \bar{\nabla}_\beta \xi^\mu)V^\beta \notag\\
    &= \xi^\beta \left(\bar{\nabla}_\beta \bar{\nabla}_\alpha - \bar{\nabla}_\alpha \bar{\nabla}_\beta\right) V^\mu + (\bar{\nabla}_\alpha \bar{\nabla}_\beta \xi^\mu)V^\beta\,.
\end{align}
The first term on the last line is a commutator of covariant derivatives. Given that $\bar{\nabla}_\alpha$ is torsion-free, we can express this commutator uniquely in terms of the curvature tensor $\bar{R}\ud{\alpha}{\mu\nu\rho}$ associated with $\bar{\nabla}_\alpha$\footnote{For a derivation of this fact, see for instance~\cite{Heisenberg:2023}.}
\begin{align}
    [\bar{\nabla}_\beta, \bar{\nabla}_\alpha]V^\mu = \bar{R}\du{\alpha\beta\gamma}{\mu}V^\gamma
\end{align}
This gives us
\begin{align}
    [\lie_\xi, \bar{\nabla}_\alpha]V^\mu &= \xi^\beta \bar{R}\du{\alpha\beta\gamma}{\mu}V^\gamma + \left(\bar{\nabla}_\alpha \bar{\nabla}_\beta \xi^\mu\right) V^\beta
\end{align}
or, equivalently, after re-labeling the $\beta$ as $\gamma$ and $\gamma$ as $\beta$ in the first term:
\begin{align}
    [\lie_\xi, \bar{\nabla}_\alpha]V^\mu &= \left(\bar{\nabla}_\alpha\bar{\nabla}_\beta \xi^\mu + \bar{R}\du{\alpha\gamma\beta}{\mu}\xi^\gamma\right) V^\beta\,.
\end{align}
From subsection~\ref{app:A_LieDGenericGamma} we recognize the term in the round bracket as the Lie derivative of the connection $\bar{\Gamma}\ud{\mu}{\alpha\beta}$. Thus, we conclude that
\begin{align}
    [\lie_\xi, \bar{\nabla}_\alpha]V^\mu &= \left(\lie_\xi\bar{\Gamma}\ud{\mu}{\alpha\beta}\right) V^\beta\,.
\end{align}
This is a special case of the formula we wish to prove and it is indeed in agreement with equation~\eqref{eq:CommutatorLieNabla}. 

Next, we perform the same kind of computations for a $1$-form $\omega_\nu$. For the first term of the commutator, we obtain
\begin{align}\label{eq:Term1w}
    \lie_\xi\left(\bar{\nabla}_\alpha\omega_\nu\right) = \xi^\beta\bar{\nabla}_\beta\bar{\nabla}_\alpha\omega_\nu + \left(\bar{\nabla}_\alpha\omega_\beta\right)\bar{\nabla}_\nu \xi^\beta + \left(\bar{\nabla}_\alpha\xi^\beta\right) \bar{\nabla}_\beta\omega_\nu\,.
\end{align}
The second term, on the other hand, reads
\begin{align}\label{eq:Term2w}
    \bar{\nabla}_\alpha\left(\lie_\xi \omega_\nu\right) &= \xi^\beta \bar{\nabla}_\alpha\bar{\nabla}_\beta\omega_\nu + \omega_\beta \bar{\nabla}_\alpha\bar{\nabla}_\nu \xi^\beta + \left(\bar{\nabla}_\alpha\xi^\beta\right)\bar{\nabla}_\beta\omega_\nu + \left(\bar{\nabla}_\alpha\omega_\beta\right)\bar{\nabla}_\nu \xi^\beta\,.
\end{align}
Subtracting~\eqref{eq:Term2w} from~\eqref{eq:Term1w} results in
\begin{align}
    [\lie_\xi,\bar{\nabla}_\alpha]\omega_\nu &= \xi^\beta\left(\bar{\nabla}_\beta\bar{\nabla}_\alpha\omega_\nu - \bar{\nabla}_\alpha\bar{\nabla}_\beta \omega_\nu\right) - \omega_\beta \bar{\nabla}_\alpha\bar{\nabla}_\nu \xi^\beta\,.
\end{align}
The term in brackets gives again rise to a curvature term, via
\begin{align}
    [\bar{\nabla}_\beta, \bar{\nabla}_\alpha]\omega_\nu = -\bar{R}\du{\alpha\beta\nu}{\gamma}\omega_\gamma\,.
\end{align}
Thus, the commutator becomes
\begin{align}
    [\lie_\xi,\bar{\nabla}_\alpha]\omega_\nu &= -\xi^\beta\bar{R}\du{\alpha\beta\nu}{\gamma}\omega_\gamma - \omega_\beta \bar{\nabla}_\alpha\bar{\nabla}_\nu \xi^\beta\notag\\
    &= -\left(\xi^\gamma \bar{R}\du{\alpha\gamma\nu}{\beta} + \bar{\nabla}_\alpha\bar{\nabla}_\nu \xi^\beta \right)\omega_\beta \notag\\
    &= -\left(\lie_\xi \bar{\Gamma}\ud{\beta}{\alpha\nu}\right)\omega_\beta\,.
\end{align}
This is again a special case of~\eqref{eq:CommutatorLieNabla}. The general case now follows from the commutator action on scalars, vectors, and $1$-forms. To see this, we make use of the Leibniz of differentiation, which holds for the covariant as well as the Lie derivative: Given two tensors $S$ and $T$, the differentiation of the tensor product can be computed according to
\begin{align}
    \bar{\nabla}_\alpha\left(S\otimes T\right) &= \left(\bar{\nabla}_\alpha S\right)\otimes T + S\otimes\left(\bar{\nabla}_\alpha T\right) \notag\\
    \lie_\xi \left(S\otimes T\right) &= \left(\lie_\xi S\right) \otimes T + S\otimes\left(\lie_\xi T\right)\,.
\end{align}
Furthermore, any $(p,q)$-tensor $\Psi\ud{\mu_1\cdots\mu_p}{\nu_1\cdots\nu_q}$ can be represented as a linear combination of terms
\begin{align}\label{eq:TAsLinComb}
    \Theta\, V^{\mu_1}_1\otimes\cdots\otimes V^{\mu_p}_p\otimes\omega^1_{\nu_1}\otimes\cdots\otimes \omega^q_{\nu_q}\,,
\end{align}
where $\Theta$ is an appropriately chosen scalar field, $V^{\mu_1}_1\otimes\cdots\otimes V^{\mu_p}_p$ a basis element of $T\M^{\otimes p}$, and $\omega^1_{\nu_1}\otimes\cdots\otimes \omega^q_{\nu_q}$ a basis element of $T^*\M^{\otimes q}$. Each $V^\mu_i$ is a vector field and each $\omega^{i}_\nu$ a $1$-form.

Therefore, if we act with the commutator on~\eqref{eq:TAsLinComb}, we simply obtain 
\begin{align}\label{eq:CommutatorBasisElement}
    &[\lie_\xi, \bar{\nabla}_\alpha]\left(\Theta\, V^{\mu_1}_1\otimes\cdots\otimes V^{\mu_p}_p\otimes\omega^1_{\nu_1}\otimes\cdots\otimes \omega^q_{\nu_q}\right) = \notag\\
    &\phantom{-}\Theta\left(\left(\lie_\xi\bar{\Gamma}\ud{\mu_1}{\alpha\beta}\right) V^{\beta}_1\otimes\cdots\otimes V^{\mu_p}_p + \dots + \left(\lie_\xi\bar{\Gamma}\ud{\mu_p}{\alpha\beta}\right)V^{\mu_1}_1\otimes\cdots\otimes V^{\beta}_p\right)\otimes \Omega_{\nu_1\cdots\nu_q} \notag\\
    &-\Theta\, V^{\mu_1\cdots\mu_p}\otimes\left(\left(\lie_\xi\bar{\Gamma}\ud{\beta}{\alpha\nu_1}\right)\omega_{\beta}\otimes\cdots\otimes\omega_{\nu_q} + \dots + \left(\lie_\xi\bar{\Gamma}\ud{\beta}{\alpha\nu_q}\right)\omega_{\nu_1}\otimes\cdots\otimes\omega_{\beta}\right)\,,
\end{align}
where we defined
\begin{align}
    V^{\mu_1\cdots\mu_p} &\ce V^{\mu_1}_1\otimes\cdots\otimes V^{\mu_p}_p &\text{and} && \Omega_{\nu_1\cdots\nu_q} &\ce \omega^1_{\nu_1}\otimes\cdots\otimes\omega^q_{\nu_q}
\end{align}
for brevity. We see that every upper index $\mu_i$ produces a term $+\lie_\xi\bar{\Gamma}\ud{\mu_i}{\alpha\beta}$, which, through $\beta$, is contracted with the $\mu_i$ slot in $V^{\mu_1\cdots\mu_i\cdots\mu_p}$. Similarly, every lower index $\nu_i$ contributes a term $-\lie_\xi\bar{\Gamma}\ud{\beta}{\alpha\nu_i}$, which is contracted with the $\nu_i$ slot of $\Omega_{\nu_1\cdots\nu_i\cdots\nu_q}$.

This pattern is familiar from the covariant derivative of a $(p, q)$-tensor, which can be written as
\begin{align}
    \bar{\nabla}_\alpha\left(\Theta\, V^{\mu_1\cdots\mu_p}\otimes\Omega_{\nu_1\cdots\nu_q}\right) &= \left(\pd_\alpha\Theta\right)V^{\mu_1\cdots\mu_p}\otimes\Omega_{\nu_1\cdots\nu_q} \notag\\
    &+\Theta\left(\bar{\Gamma}\ud{\mu_1}{\alpha\beta}V^{\beta\cdots\mu_p} + \cdots + \bar{\Gamma}\ud{\mu_p}{\alpha\beta}V^{\mu_1\cdots\beta}\right)\otimes\Omega_{\nu_1\cdots\nu_q} \notag\\
    &-\Theta\, V^{\mu_1\cdots\mu_p}\otimes\left(\bar{\Gamma}\ud{\beta}{\alpha\nu_1} \Omega_{\beta\cdots\nu_q} + \dots + \bar{\Gamma}\ud{\beta}{\alpha\nu_q} \Omega_{\nu_1\cdots\beta}\right)\,.
\end{align}
The second and third line have the same structure as~\eqref{eq:CommutatorBasisElement}, except that every occurrence of a Lie derivative is replaced by a connection component. A convenient way of writing the right hand side of the commutator is therefore
\begin{align}
    [\lie_\xi, \bar{\nabla}_\alpha]\left(\Theta\, V^{\mu_1\cdots\mu_p}\otimes \Omega_{\nu_1\cdots\nu_q}\right) &= \PD{\left(\bar{\nabla}_\alpha \left[\Theta\, V^{\mu_1\cdots\mu_p}\otimes \Omega_{\nu_1\cdots\nu_q}\right]\right)}{\bar{\Gamma}\ud{\lambda}{\mu\nu}}\lie_\xi\bar{\Gamma}\ud{\lambda}{\mu\nu}\,.
\end{align}
The derivative of the covariant derivative of $\Theta\, V^{\mu_1\cdots\mu_p}\otimes \Omega_{\nu_1\cdots\nu_q}$ with respect to the affine connection automatically produces the correct contraction pattern with the correct $+$ or $-$ signs. Moreover, the term $\left(\pd_\alpha\Theta\right)V^{\mu_1\cdots\mu_p}\otimes\Omega_{\nu_1\cdots\nu_q}$ does not contribute, because it contains no connection coefficient. 

Finally, because every $(p, q)$-tensor field $\Psi\ud{\bullet}{\circ} = \Psi\ud{\mu_1\cdots\mu_p}{\nu_1\cdots\nu_q}$ can be written as a linear combination of terms of the form~\eqref{eq:TAsLinComb}, and because the commutator $[\lie_\xi, \bar{\nabla}_\alpha]$ inherits the linearity of $\lie_\xi$ and $\bar{\nabla}_\alpha$, it follows that
\begin{align}
    [\lie_\xi, \bar{\nabla}_\alpha]\Psi\ud{\bullet}{\circ} = \PD{\left(\bar{\nabla}_\alpha \Psi\ud{\bullet}{\circ}\right)}{\bar{\Gamma}\ud{\lambda}{\mu\nu}} \lie_\xi\bar{\Gamma}\ud{\lambda}{\mu\nu}\,,
\end{align}
which is what we wanted to prove.

\section{The Energy-Momentum Tensor of an Imperfect Fluid and its Relation to Cosmological Perturbation Theory}\label{app:B}
At the beginning of section~\ref{sec:SVT} we introduced the tensor $\tau_{\mu\nu}$ to describe the perturbations of the energy-momentum tensor. Given that the unperturbed tensor corresponds to a perfect fluid, it is helpful to describe the components of $\tau_{\mu\nu}$ in terms of fluid variables. However, we can no longer assume the fluid to be described only by energy density and isotropic pressure. Rather, we should expect the presence of anisotropies and flows. In other words, the perturbations $\tau_{\mu\nu}$ are described by the variables of an imperfect fluid.

It is important to keep in mind that these variables are always defined relative to some observer. Let's denote the four-velocity of the observer by $u^\mu$. As usual, it is normalized as
\begin{align}
    u^\mu u_\mu = -1\,.
\end{align}
Naturally, this also implies that $u^\mu$ is a timelike vector. We can therefore use $u^\mu$ to single out a time direction and to define a three-dimensional space orthogonal to it. To achieve the latter, we introduce the projection operator
\begin{align}
    h_{\mu\nu} \ce g_{\mu\nu} + u_\mu u_\nu\,.
\end{align}
It is straightforward to verify that $h_{\mu\nu}$ satisfies the projector property
\begin{align}\label{eq:OperatorProperty}
    h\ud{\mu}{\sigma} h_{\mu\nu} = h_{\nu\sigma}\,.
\end{align}
Furthermore, the trace of $h_{\mu\nu}$ in $n$-dimensions can either be defined using $g^{\mu\nu}$ or $h^{\mu\nu}$. In either case one obtains
\begin{align}
    g^{\mu\nu} h_{\mu\nu} = h^{\mu\nu} h_{\mu\nu} = n-1\,.
\end{align}
Observe that, by construction, $h_{\mu\nu}$ satisfies the orthogonality relation
\begin{align}
    h_{\mu\nu} u^\mu = 0\,.
\end{align}
This means that $h_{\mu\nu}$ has no time-time or space-time components. It is entirely composed of space-space components. We therefore call it a spacelike tensor. Schematically, we can think of $u^\mu$ and $h_{\mu\nu}$ in $n=4$ dimensions as being of the form
\begin{align}
    u^\mu &= (u^0, 0, 0, 0) &\text{and} && h_{\mu\nu} = 
    \begin{pmatrix}
        0 & 0 & 0 & 0 \\
        0 & h_{11} & h_{12} & h_{13} \\
        0 & h_{12} & h_{22} & h_{23} \\
        0 & h_{13} & h_{23} & h_{33}
    \end{pmatrix}\,.
\end{align}
Given the tuple $(u^\mu, h_{\mu\nu})$, we can decompose any $(p,q)$-tensor field into temporal or spatial parts. For instance, a generic vector field $V^\mu$ possesses one temporal and $n-1$ spatial components, defined as
\begin{align}
    &\text{Temporal component:} && V^\mu u_\mu \notag\\
    &\text{Temporal component:} && V^\mu h_{\mu\alpha}\,.
\end{align}
In other words, $V^\mu$ has one component pointing in the time direction, and $n-1$ components laying in the space orthogonal to $u^\mu$. Thus, $V^\mu$ can be written as
\begin{align}
    V^\mu &= \tau\,u^\mu + \sigma^\alpha h\ud{\mu}{\alpha} &\text{with} && \tau &\ce V^\mu u_\mu && \text{and} & \sigma^\alpha &\ce V^\mu h\ud{\sigma}{\mu}\,.
\end{align}
Similarly, we can decompose a symmetric tensor $S_{\mu\nu}$ into temporal-temporal, temporal-spatial, and spatial-spatial components:
\begin{align}
    &\text{Temporal-temporal component:} && S_{\mu\nu} u^\mu u^\nu \notag\\
    &\text{Temporal-spatial components:} && S_{\mu\nu} u^\mu h\ud{\nu}{\sigma} \notag\\
    &\text{Spatial-spatial components:} && S_{\mu\nu} h\ud{\mu}{\alpha} h\ud{\nu}{\beta}\,.
\end{align}
This procedure generalizes in a natural way to non-symmetric tensors and to tensors of arbitrary $(p,q)$-type. However, what is important to us, is that $(u^\mu, h_{\mu\nu})$ allows us to write the energy-momentum tensor $T_{\mu\nu}$ as
\begin{align}\label{eq:DecompT}
    T_{\mu\nu} = \rho\, u_\mu u_\nu + q_\mu u_\nu + q_\nu u_\nu + \Sigma_{\mu\nu}\,,
\end{align}
where we defined
\begin{align}\label{eq:DefsDecomp}
    \rho &\ce T_{\alpha\beta}u^\alpha u^\beta\,, & q_\mu &\ce -T_{\alpha\beta} u^\alpha h\ud{\beta}{\mu}\,, &\text{and} && \Sigma_{\mu\nu} &\ce T_{\alpha\beta} h\ud{\alpha}{\mu} h\ud{\beta}{\nu}\,.
\end{align}
Because $h_{\mu\nu}$ is orthogonal to $u^\mu$, it follows that $q_\mu$ and $\Sigma_{\mu\nu}$ inherit this property:
\begin{align}
    q^\mu u_\mu &= 0 &\text{and} && \Sigma_{\mu\nu} u^\mu &= 0\,.
\end{align}
Thus, $q^\mu$ and $\Sigma_{\mu\nu}$ are spacelike tensors. Informally, we can think of these tensors to have no temporal components.

It is convenient to further decompose the symmetric tensor $\Sigma_{\mu\nu}$ into a trace- and a trace-less part. This will facilitate the physical interpretation later on.

Taking into account that the trace of $h_{\mu\nu}$ is $n-1$ in $n$ dimensions, we obtain the following expression for the trace part, which we call $p$:
\begin{align}\label{eq:TracePart}
    p &\ce \frac13 \Sigma_{\mu\nu} h^{\mu\nu} = \frac13 T_{\alpha\beta} h\ud{\alpha}{\mu} \underbrace{h\ud{\beta}{\nu} h^{\mu\nu}}_{= h^{\beta\mu}} \notag\\
    &= \frac13 T_{\alpha\beta} \underbrace{h\ud{\alpha}{\mu} h^{\mu\beta}}_{= h^{\alpha\beta}} \notag\\
    &= \frac13 T_{\alpha\beta} h^{\alpha\beta}\,.
\end{align}
We used the operator property~\eqref{eq:OperatorProperty} twice to arrive at the final expression. The trace-free part of $\Sigma_{\mu\nu}$, which we call $\pi_{\mu\nu}$, is defined as
\begin{align}
    \pi_{\mu\nu} &\ce \Sigma_{\mu\nu} - \frac13 h_{\mu\nu} h^{\alpha\beta} \Sigma_{\alpha\beta} \notag\\
    &= \left[h\ud{\alpha}{\mu} h\ud{\beta}{\nu} - \frac13 h_{\mu\nu} h^{\alpha\beta}\right] T_{\alpha\beta}\,,
\end{align}
where we used~\eqref{eq:DefsDecomp} and~\eqref{eq:TracePart}. We can now re-write $\Sigma_{\mu\nu}$ as
\begin{align}\label{eq:DecompSigma}
    \Sigma_{\mu\nu} &= p\,h_{\mu\nu} + \pi_{\mu\nu}\,.
\end{align}
By construction, we have
\begin{align}
    u^\mu \pi_{\mu\nu} &= 0 &&\text{and} & \pi\ud{\mu}{\mu} = g^{\mu\nu} \pi_{\mu\nu} = h^{\mu\nu}\pi_{\mu\nu} = 0\,.
\end{align}
In other words, $\pi_{\mu\nu}$ is a symmetric, spacelike, and traceless tensor field. By plugging~\eqref{eq:DecompSigma} into the decomposition~\eqref{eq:DecompT}, we conclude that the energy-momentum tensor $T_{\mu\nu}$ decomposed relative to an observer with four-velocity $u^\mu$ takes the form
\begin{align}
    T_{\mu\nu} &= \rho\, u_\mu u_\nu + q_\mu u_\nu + q_\nu u_\nu + p\, h_{\mu\nu} + \pi_{\mu\nu}\,.
\end{align}
For later convenience, we raise the first index using $g^{\mu\nu}$, which gives us the $(1,1)$-tensor field
\begin{align}\label{eq:ImperfectFluidT}
    T\ud{\mu}{\nu} &= \rho\, u^\mu u_\nu + q^\mu u_\nu + q_\nu u^\mu + p\, h\ud{\mu}{\nu} + \pi\ud{\mu}{\nu} \notag\\
    &= \left(\rho + p\right)\, u^\mu u_\nu + q^\mu u_\nu + q_\nu u^\mu + p\,\delta\ud{\mu}{\nu} + \pi\ud{\mu}{\nu}\,.
\end{align}
To arrive at the second line, we used the definition of $h_{\mu\nu}$ to rewrite $h\ud{\mu}{\nu}$ as $\delta\ud{\mu}{\nu} + u^\mu u_\nu$. Observe that if $q^\mu$ and $\pi_{\mu\nu}$ vanish, we are left with
\begin{align}
    T\ud{\mu}{\nu} = \left(\rho + p\right)\, u^\mu u_\nu + p\,\delta\ud{\mu}{\nu}\,,
\end{align}
which is the energy-momentum tensor of a perfect fluid as perceived by the observer with four-velocity $u^\mu$. This suggests the identifications
\begin{align}
    \rho &\equiv \text{energy density relative to $u^\mu$} \notag\\
    p &\equiv \text{isotropic pressure relative to $u^\mu$}\,.
\end{align}
We can also attach an interpretation to the spacelike vector field $q^\mu$ and the spacelike tensor field $\pi_{\mu\nu}$:
\begin{align}
    q^\mu &= \text{Heat flux within the fluid relative to $u^\mu$} \notag\\
    \pi_{\mu\nu} &= \text{anisotropic stress within the fluid relative to $u^\mu$}\,.
\end{align}
Thus, the second line of~\eqref{eq:ImperfectFluidT} represents the most general form of the energy-momentum tensor of an imperfect fluid. Schematically, we can think of this energy-momentum tensor as being represented by the matrix
\begin{align}
    T\ud{\mu}{\nu} =
    \begin{pmatrix}
        \rho + p & q_i \\
        q^j & p\, \delta\ud{i}{j} + \pi\ud{i}{j}
    \end{pmatrix}\,,
\end{align}
where $i,j$ range from $1$ to $3$. For the sake of completeness, we note that it is always possible to change frame of reference such that the heat flux $q^\mu$ vanishes. This is no surprise, since all quantities in~\eqref{eq:ImperfectFluidT} are defined relative to a four-velocity $u^\mu$. Hence, an observer who is stationary relative to $q^\mu$ would not perceive any heat flux.

The frame of reference in which $q^\mu$ vanishes is known as the Landau-Lifshitz frame, and it is defined by the equation
\begin{align}\label{eq:LandauLifshitzCondition}
    T\ud{\mu}{\nu} u^\nu = -\rho\, u^\mu\,.
\end{align}
Indeed, if we take~\eqref{eq:ImperfectFluidT} and we perform the contarction with $u^\mu$, we find
\begin{align}
    T\ud{\mu}{\nu} u^\mu = -\rho\, u^\mu - q^\mu\,,
\end{align}
where we used the normalization of $u^\mu$ and the orthogonality relations $q^\mu u_\mu = 0$ and $\pi_{\mu\nu} u^\mu = 0$. Thus, if $u^\mu$ satisfies the Landau-Lifshitz condition~\eqref{eq:LandauLifshitzCondition}, we find indeed $q^\mu = 0$.

In what follows, we shall assume a generic frame of reference, i.e., one which does \emph{not} satisfy the Landau-Lifshitz condition. We are interested in perturbing~\eqref{eq:ImperfectFluidT} to first order in $(\rho, p, q^\mu, \pi_{\mu\nu})$ and $u^\mu$ about the energy-momentum tensor of a perfect fluid. Specifically, we set
\begin{align}\label{eq:FirstOrderMatterPerturbations}
    \rho(\eta, \x) &= \bar{\rho}(\eta) + \delta\rho(\eta, \x) \notag\\
    p(\eta, \x) &= \bar{p}(\eta) + \delta p(\eta, \x) \notag\\
    q^\mu(\eta, \x) &= \bar{q}^\mu(\eta) + \delta q^\mu(\eta, \x) \notag\\
    \pi_{\mu\nu}(\eta, \x) &= \bar{\pi}_{\mu\nu}(\eta) + \delta \pi_{\mu\nu}(\eta, \x) \notag\\
    u^\mu(\eta, \x) &= \bar{u}^\mu + \delta u^\mu(\eta, \x)\,,
\end{align}
where a bar $\bar{}$ indicates background quantities and $\delta$ a small perturbation. Since the background energy-momentum tensor is the perfect fluid one, we have $\bar{q}^\mu = 0$ and $\bar{\pi}_{\mu\nu} = 0$. Rather than writing $\delta q^\mu$ and $\delta \pi_{\mu\nu}$, we simply denote the perturbations by $q^\mu$ and $\pi_{\mu\nu}$, to keep the notation lighter.

Before substituting these first order perturbations into the imperfect fluid energy-momentum tensor~\eqref{eq:ImperfectFluidT}, we need to examine the perturbed four-velocity more closely. Since the four-velocity of any observer has to be timeline and normalized, we have
\begin{align}
    g^{\mu\nu} u^\mu u^\nu &= g^{\mu\nu}\left(\bar{u}^\mu + \delta u^\mu\right) \left(\bar{u}^\nu + \delta u^\nu\right) \notag\\
    &= g^{\mu\nu} \bar{u}^\mu \bar{u}^\nu + 2 g^{\mu\nu} \bar{u}^\mu \delta u^\nu \notag\\
    &= -1 + 2 g^{\mu\nu} \bar{u}^\mu \delta u^\nu \overset{!}{=} -1\,,
\end{align}
where we used the normalizations of $\bar{u}^\mu$ and $u^\mu$, and where we dropped the second order term $\delta u^\mu \delta u_\mu$. It follows from the last line that
\begin{align}
    \bar{u}^\mu \delta u_\mu = 0\,.
\end{align}
Now, when we substitute~\eqref{eq:FirstOrderMatterPerturbations} into~\eqref{eq:ImperfectFluidT}, we obtain
\begin{align}
    T\ud{\mu}{\nu} = \bar{T}\ud{\mu}{\nu} + \left(\delta\rho + \delta p\right) \bar{u}^\mu \bar{u}_\nu + \bar{u}^\mu \left[(\bar{\rho} + \bar{p}) \delta u_\nu + q_\nu\right] + \left[(\bar{\rho} + \bar{p}) \delta u^\mu + q^\mu \right] \bar{u}_\nu + \pi\ud{\mu}{\nu}\,,
\end{align}
where
\begin{align}
    \bar{T}\ud{\mu}{\nu} = (\bar{\rho} + \bar{p}) \bar{u}^\mu \bar{u}_\nu + \bar{p}\, \delta\ud{\mu}{\nu}
\end{align}
denotes the background energy-momentum tensor. The perturbation $\tau\ud{\mu}{\nu}$ of the energy-momentum tensor can thus be written as
\begin{align}
    \tau_{\mu\nu} \ce T\ud{\mu}{\nu} - \bar{T}\ud{\mu}{\nu} = \left(\delta\rho + \delta p\right) \bar{u}^\mu \bar{u}_\nu + \bar{u}^\mu \left[(\bar{\rho} + \bar{p}) \delta u_\nu + q_\nu\right] + \left[(\bar{\rho} + \bar{p}) \delta u^\mu + q^\mu \right] \bar{u}_\nu + \pi\ud{\mu}{\nu}\,.
\end{align}
Since for the four-velocity $\bar{u}^\mu$ we have
\begin{align}
    \bar{g}_{\mu\nu}\bar{u}^\mu \bar{u}^\nu = a(\eta)^2 \eta_{\mu\nu}\bar{u}^\mu \bar{u}^\nu = -a(\eta)^2 \left(u^0\right)^2 = -1\,,
\end{align}
we must have
\begin{align}
    \bar{u}^\mu &= \frac{1}{a(\eta)}(1, 0, 0, 0) &\text{and} && \bar{u}_\mu &= a(\eta)(-1, 0, 0, 0)\,,
\end{align}
where we made use of the FLRW metric in conformal time~\eqref{eq:FLRWConfTime}. Given that $q^\mu$, $\pi_{\mu\nu}$, and $\delta u^\mu$ are spacelike, we have
\begin{align}
    q^\mu &= (0, q^{i})\,, & \pi\ud{0}{\nu} &= 0\,, & \pi\ud{\mu}{0} &= 0\,, &\text{and} && \delta u^\mu &= \frac{1}{a(\eta)}(0, v^{i})\,.
\end{align}
The factor $\frac{1}{a(\eta)}$ in $\delta u^\mu$ mirrors the structure of $\bar{u}^\mu$ and it has been introduced for convenience. It follows that the components of $\tau\ud{\mu}{\nu}$ can be written as
\begin{align}
    \tau\ud{0}{0} &= -\delta\rho(\eta, \x) \notag\\
    \tau\ud{0}{j} &= (\bar{\rho} + \bar{p}) v_j(\eta, \x) + \frac{1}{a(\eta)}q_j(\eta, \x) \notag\\
    \tau\ud{i}{0} &= -(\bar{\rho} + \bar{p}) v^{i}(\eta, \x) - a(\eta)\,q^{i}(\eta, \x) \notag\\
    \tau\ud{i}{j} &= \delta p(\eta, \x)\,\delta\ud{i}{j} + \pi\ud{i}{j}(\eta, \x)\,.
\end{align}
These are a slightly more generic expressions than the ones we introduced in~\eqref{eq:TauParametrization} to parametrize the perturbations of the energy-momentum tensor. The difference between~\eqref{eq:TauParametrization} and the $\tau$-components we obtained here is the presence of the heat flux vector $q^{i}$ in the latter.

Notice that we are working in a generic frame and that we are describing the matter perturbations in terms of two scalar fields ($\delta \rho$, $\delta p$), two three-dimensional vector fields ($v^{i}$ and $q^{i}$), and one symmetric, traceless, spacelike tensor field $\pi\ud{i}{j}$. In total, this gives us $2 + 6 + 5 = 13$ field components to parametrize the ten components of $\tau\ud{\mu}{\nu}$.

The resolve this tension, we note that we always have the freedom of fixing a specific frame $u^\mu$. Given that $u^\mu$ has to be normalized, this gives us three conditions, thus removing three excessive field components from our count. In particular, we can always use the Landau-Lifshitz frame, which eliminates $q^{i}$, giving us~\eqref{eq:TauParametrization} with its ten field components to parametrize $\tau\ud{\mu}{\nu}$.

\newpage
\bibliographystyle{JHEP}
\bibliography{Bibliography}

\end{document}